\documentclass{emulateapj}       

\newcommand{\about}{\mbox{$\sim$}}               
\newcommand{\alphaco}{\mbox{$\alpha_{\rm CO}$}}           
\newcommand{\Eu}{\mbox{$E_{\rm u}$}}                  
\newcommand{\frest}{\mbox{$f_{\rm rest}$}}      
\newcommand{\Lsol}{\mbox{$L_\odot$}}             
\newcommand{\Lsun}{\mbox{$L_\odot$}}            
\newcommand{\Mdyn}{\mbox{$M_{\rm dyn}$}}   
\newcommand{\Mmol}{\mbox{$M_{\rm mol}$}}    
\newcommand{\Msol}{\mbox{$M_\odot$}}            
\newcommand{\Msun}{\mbox{$M_\odot$}}           
\newcommand{\Tb}{\mbox{$T_{\rm b}$}}               
\newcommand{\uv}{\mbox{$u$--$v$}}                     
\newcommand{\Vsys}{\mbox{$V_{\rm sys}$}}        
\newcommand{\Xco}{\mbox{$X_{\rm CO}$}}           
\newcommand{\Xtwenty}{\mbox{$X_{20}$}}           
\newcommand{\hr}{\mbox{$^{\rm h}$}}                   
\newcommand{\mn}{\mbox{$^{\rm m}$}}                
\newcommand{\perbeam}{\mbox{beam$^{-1}$}}                         
\newcommand{\percubiccm}{\mbox{cm$^{-3}$}}                         
\newcommand{\kms}{\mbox{km s$^{-1}$}}                                    
\newcommand{\persquarepc}{\mbox{pc$^{-2}$}}                         
\newcommand{\peryr}{\mbox{yr$^{-1}$}}                                        
\newcommand{\unitofalpha}{\mbox{\Msol \persquarepc (K km s$^{-1}$)$^{-1}$}}   
\newcommand{\unitofX}{\mbox{cm$^{-2}$ (K km s$^{-1}$)$^{-1}$}}   

\newcommand{\plus}{\mbox{$+$}}    
\newcommand{\minus}{\mbox{$-$}}  
\newcommand{\twelveCO}{\mbox{$^{12}$CO}}                  
\newcommand{\thirteenCO}{\mbox{$^{13}$CO}}                
\newcommand{\CeighteenO}{\mbox{C$^{18}$O}}              
\newcommand{\Halpha}{\mbox{H$\alpha$}}                       
\newcommand{\HH}{\mbox{H$_2$}}                                      
\newcommand{\HthirteenCN}{\mbox{H$^{13}$CN}}          
\newcommand{\HCthreeN}{\mbox{HC$_{3}$N}}                 
\newcommand{\HCOplus}{\mbox{HCO$^{+}$}}                    
\newcommand{\propyne}{\mbox{CH$_3$CCH}}                 

\newcommand{\nd}{\nodata}

\newcommand{\citest}[1]{\citeauthor*{#1}}
\newcommand{\citesp}[1]{(\citeauthor*{#1})}

\shorttitle{NGC 3256 Molecular Outflows}
\shortauthors{SAKAMOTO et al.}

\begin{document}
\title{A Luminous Infrared Merger with Two Bipolar Molecular Outflows: \\ ALMA and SMA Observations of NGC 3256}

\slugcomment{submitted to ApJ on Mar. 04, 2014}

\author{Kazushi Sakamoto\altaffilmark{1}, 
Susanne Aalto\altaffilmark{2}, 
Francoise Combes\altaffilmark{3}, 
Aaron Evans\altaffilmark{4,5}, 
and
Alison Peck\altaffilmark{4}
}
\altaffiltext{1}{Academia Sinica, Institute of Astronomy and Astrophysics, Taiwan} 
\altaffiltext{2}{Department of Earth and Space Sciences, Chalmers University of Technology, Onsala Space Observatory, Onsala, Sweden} 
\altaffiltext{3}{Observatoire de Paris, 61 Av. de l'Observatoire, 75014, Paris, France}
\altaffiltext{4}{NRAO, Charlottesville, VA, U.S.A.}
\altaffiltext{5}{University of Virginia, Charlottesville, VA, U.S.A.}

\begin{abstract}
We report ALMA and SMA observations of 
the luminous infrared merger  NGC 3256, 
the most luminous galaxy within $z=0.01$.
Our data show that
both of the two merger nuclei separated by 5\arcsec\ (0.8 kpc) on the sky have a compact concentration
of molecular gas. 
We identify them as  nuclear disks with molecular gas surface densities over $10^3$ \Msol\persquarepc\ 
and determine that while one at the northern nucleus is nearly face-on 
the other at the southern nucleus is almost edge-on.
The northern nucleus is more massive and has molecular arcs and spiral arms around. 
The high-velocity molecular gas previously found in the system is resolved to two components.
They are two molecular outflows associated with each of the two nuclei.
The molecular outflow from the northern nuclear disk is part of a starburst-driven superwind seen nearly pole on.
It has a maximum velocity greater than 750 \kms\ and its mass outflow rate is estimated to be $\geq 60$ \Msol\peryr\
for a conversion factor $N_{\rm H_2}/I_{\rm CO(1-0)}$ of $1\times 10^{20}$ \unitofX. 
The outflow from the southern nucleus is a highly collimated bipolar molecular jet seen nearly edge-on.
Its line-of-sight velocity increases with distance out to 300 pc from the southern nucleus. 
The maximum de-projected velocity is on the order of 2000 \kms\ for the estimated inclination and 
should exceed 1000 \kms\ even allowing for its uncertainty.
The mass outflow rate is estimated to be $>50$ \Msol\peryr\ for this outflow.
There are possible signs that this southern outflow has been driven by a bipolar radio jet from an AGN that became inactive very recently. 
The sum of these outflow rates, although subject to the uncertainty in the molecular mass estimate, either exceeds or
compares to the total star formation rate in NGC 3256. 
The feedback from nuclear activities in the form of molecular outflows is therefore significant in the gas consumption budget, 
and hence evolution, of this luminous infrared galaxy.
\end{abstract}

\keywords{ 
        galaxies: active ---
        galaxies: interactions ---
        galaxies: individual (NGC 3256) ---        
        galaxies: ISM ---        
        ISM: jets and outflows
       }

\section{Introduction}  
\label{s.introduction}
NGC 3256 is an infrared-luminous merger with a bolometric luminosity of $L_{\rm bol}=4\times 10^{11} L_{\odot}$
($D = 35$ Mpc, see Table \ref{t.4418param} for other parameters).
Its two nuclei with a projected separation of 5\arcsec\ = 850 pc  \citep{Zenner93,Norris95} and 
two long tidal tails of stars and \ion{H}{1} gas \citep{English03}
indicate that the system is in a late stage of merging between two disk galaxies \citep{Toomre77}.
NGC 3256 belongs to the sequence of `most luminous galaxies within their distance ranges', which are, beyond the local group,
NGC 253, M82, NGC 1068, NGC 3256, Arp 299, and Arp 220 in the catalogue of \citet{Sanders03}.
It is therefore among the best targets to explore luminosity-related phenomena in local galaxies, although its location at
Dec. = $-43\arcdeg$ impeded its studies compared to other galaxies in the sequence.
\citet[hereafter SHP06]{Sakamoto06} made the first interferometric imaging of a CO line emission in NGC 3256 
soon after the commissioning of the Submillimeter Array (SMA)  and
discovered wide CO line wings underlying the much brighter narrow component in previous observations
\citep[e.g.,][]{Sargent89,Aalto02}.
The wing CO emission was attributed to a molecular outflow from the face-on merger.
The detection of a galactic molecular outflow from faint and wide CO line wings became possible at that time
owing in part to the new wide-band capabilities of the SMA.
Many extragalactic molecular outflows have been detected since then 
through broad CO line wings caught with wide-band spectrometers
\citep[e.g.,][]{Feruglio10,Chung11,Alatalo11}\footnote{Detection of molecular outflows from broad OH lines 
dates back much further \citep[and references therein]{Baan89}.
Galactic outflows of cold molecular gas have been also found 
from off-plane molecular gas of edge-on galaxies 
[e.g.,  \citet{Nakai87} toward M82 and \citet{Turner85}, \citet{GarciaBurillo00}, and \citet{Bolatto13a} toward NGC 253]
and 
from blueshifted molecular absorption lines against nuclear continuum
\citep[e.g.,][for Arp 220 and Mrk 231]{Baan89,Sakamoto09,Fisher10}. 
All galaxies mentioned here belong to the above-mentioned elite sequence of luminous nearby galaxies.}
Such molecular outflows coexist with outflows of ionized and atomic gas and are expected to
have significant impact on the luminosity-generation activities in galaxies and the evolution of galaxies themselves \citep[for reviews]{Veilleux05,Carilli13}.

We have used the new Atacama Large Millimeter/sub-millimeter Array (ALMA) 
in its first open-use (Cycle 0) to further study NGC 3256.
We aimed at the structure and properties of the molecular gas around the luminous merger nuclei
including the high-velocity molecular gas. 
Although the broad CO wings had been confirmed and found to be even broader 
in the ALMA commissioning and science verification data \citep{Sakamoto13a}
its structure was still largely unconstrained.
We therefore observed the galaxy in the 3 and 0.8 mm bands in ALMA Cycle 0 and  
also made supplemental 1.3 mm observations with the SMA.
These new observations provide much higher spatial resolution than before for the circumnuclear molecular gas,
up to about 1\arcsec\ for  CO(1--0),  0\farcs8 for CO(2--1), and 0\farcs5 for CO(3--2).
We also obtained high-resolution high-sensitivity data of CN(1--0), \propyne(6--5), \thirteenCO(2--1), \HCOplus(4--3), and 3 and 0.8 mm continuum.
In this paper, we report these new observations and 
give an overall account of the spatial and kinematical structure of the molecular gas in the center of NGC 3256.
We found the high-velocity gas to be two bipolar molecular outflows from the two nuclei
and that the two outflows have distinctively different properties from each other.

We describe our observations and data reduction in Section \ref{s.obs}
and present our observational results in Section \ref{s.result}.
We use the data in Section \ref{s.configuration} to constrain the merger configuration, 
which is critical to interpret the observed gas motion.
In Section \ref{s.twoOutflows} we present our two-outflow model for the observed velocity structure and gas distribution.
The one from the southern nucleus has remarkable properties in its velocity field, high velocity, high collimation, and large energy.
We discuss its driving mechanism in Section \ref{s.Snucleus}.
Section \ref{s.conclusions} compares our findings in NGC 3256 with similar objects and phenomena in galaxies
and then summarizes our conclusions.

\section{Observations}
\label{s.obs}
\subsection{ALMA} 
\label{s.obs-alma}
Our ALMA observations in Cycle 0 were made in 2011--2012 using
up to twenty-three 12 m-diameter antennas as summarized in Table \ref{t.obslog}.
We observed in the 3 mm band (Band 3) and the 0.85 mm band (Band 7) 
each in two array configurations jointly covering projected baselines between 15 m and about 370 m.
The Band 3 observations were for a single pointing at a position between the two nuclei. 
The primary beam of the ALMA 12 m antennas has a full width at half maximum (FWHM) of 53\arcsec\ 
at the frequency of the redshifted CO(1--0) line\footnote{The FWHM size of the primary beam is assumed
to be $1.17 (\lambda/12)$ for ALMA and $1.15 (\lambda/6)$ for SMA, where $\lambda$ is the wavelength in meters.}.
In Band 7 we made a seven-point hexagonal mosaic with the same central position and a 7\farcs3 spacing 
between adjacent pointings.
The FWHM of the individual primary beam is 18\arcsec\ and that of the mosaicked primary beam is about 25\arcsec\ 
for the redshifted CO(3--2) line.  
We used a correlator setup having 0.488 MHz channel spacing and about 3.5 GHz continuous coverage in each sideband.

We also combined with our Band 3 data an earlier ALMA dataset obtained 
through the Commissioning and Science Verification (CSV) program carried out by the Joint ALMA Observatory.
The CSV observations, also listed in Table \ref{t.obslog},  had about the same on-source integration time of 3 hr as
our Cycle 0 observations albeit with 7 or 8 antennas. 
They provide dense sampling of short projected baselines between 12 m and 90 m.
The CSV observations were made toward a slightly offset position (4\farcs2 from our Cycle 0 observations)
with a correlator setup of 15.6 MHz channel spacing
and with almost the same frequency coverage as our Cycle 0 observations (Table \ref{t.freqCoverage}). 
The CSV and Cycle 0 data were combined as a mosaic because of the pointing offset.

All ALMA data were calibrated from the raw data\footnote{in ALMA Science Data Model format} 
in a uniform manner using the CASA\footnote{Common Astronomy Software Applications} reduction package versions 4.0 and 4.1. 
Most notably, we used the `Butler-JPL-Horizons 2012' model for Titan and Mars in our flux calibration, measured
and accounted for the spectral slopes of our bandpass and gain calibrators in our calibration, and checked
the flatness of our spectral bandpass by looking at the spectrum of the gain calibrator after all calibrations.
Data showing non-linear baselines were flagged. 
The versions of CASA that we used do not allow to flag only one of the two linear polarizations on individual baselines
because the two polarizations share a flagging variable in the data structure.
Therefore, when one of the two linear polarizations on a baseline was found faulty the remaining one was copied over 
the corrupted one and both the original and copied visibilities were down-weighted to conserve the net weights of the
rescued data.  

Imaging and basic data analysis were also made in CASA. 
For lines we binned our data to spectral resolutions of 4, 10, and 20 MHz for Cycle 0 Band 3, 15.6 MHz for the combined
CSV\plus Cycle 0 data in Band 3, and 10 and 30 MHz for Band 7.
Table \ref{t.linelist} lists the six lines detected in our ALMA data as well as two notable non-detections.
We made continuum data after carefully inspecting the full widths of these lines and by summing up line-free channels. 
The continuum has been subtracted from our line data in the \uv\ domain for the CSV\plus Cycle 0 data 
and in the image domain for the rest.
This is because we are most interested in weak and broad line emission near the phase center in the former dataset
while better subtraction across the imaging area is more desired for other datasets.

Parameters of our reduced data are summarized in Tables \ref{t.data_cont_properties}  and \ref{t.data_line_properties}. 
Compared to our previous SMA observations the new ALMA observations improved 
spatial resolution by about a factor of three and 
sensitivity in line brightness temperature by about an order of magnitude. 
Our CO(1--0) data cubes made from the Cycle 0 data alone recovered 76--87\% of the single-dish flux
measured with a 43\arcsec\ beam (FWHM)  by \citet{Aalto95}.
The fraction is highest in the data cube made with lower weights to longer baselines.
We recovered 91--97\% of the single-dish CO(1--0) flux in the cubes made from the CSV\plus Cycle 0 data.
We expect similar or higher recovery rates for other Band 3 lines and continuum 
but the recovery rate must be lower for emission in Band 7
because the central hole in the \uv\ plane is larger in Band 7.

\subsection{SMA} 
\label{s.obs-sma}
We added to our SMA 1.3 mm observations reported in \citest{Sakamoto06} new data taken in 2008, 
increasing the maximum projected baseline from 179 m to 509 m and doubling the total on-source time
from 6.9 hr to 12.9 hr. 
The new observations in two nights had 7 antennas and excellent weather with the 220 GHz zenith opacity between 0.04 and 0.06.
We observed the same position as in our previous observations (as well as our ALMA observations) using the
tuning for the same three $J=2$--1 lines as before, namely, \twelveCO, \thirteenCO, and \CeighteenO\
although only the first two were bright enough to be imaged at high angular resolutions.
The primary beam of the SMA 6-m antennas has a FWHM size of 52\arcsec\ at the frequency of the redshifted CO(2--1) line.
The data were reduced with the same steps as before using the MIR reduction package.

\subsection{Conventions}
The offset coordinates in this paper are with respect to our SMA and ALMA Cycle 0 phase-tracking center
in Table \ref{t.4418param}.
We adopt radio positions for the two merger nuclei, namely,
R.A. = 10\hr27\mn51\fs23,  Dec. = \minus43\arcdeg54\arcmin14\farcs0 (J2000) for the northern (N) nucleus
and 
R.A. = 10\hr27\mn51\fs22,  Dec.~=~\minus43\arcdeg54\arcmin19\farcs.2 (J2000) for the southern (S) nucleus
\citep{Neff03}.
Our phase tracking center is the midpoint of these nuclei with the last RA digit rounded up.
We use radio-defined velocity with respect to the Local Standard of Rest (LSR) throughout this paper
(LSRK in the ALMA terminology).
We adopt 2775 \kms\ (radio, LSR) for the systemic velocity of the galaxy and measure offset
velocities from this \Vsys\ (e.g., in presenting channels maps). 
Our previous SMA observations found this to be a good fiducial velocity not only for the whole system but also
for individual nuclei because they almost align on the kinematical minor axis of the merger \citesp{Sakamoto06}.

\section{Observational Results}
\label{s.result}

\subsection{Continuum} 
\label{s.result-continuum}
\begin{figure}[t]
\epsscale{0.3}
\plotone{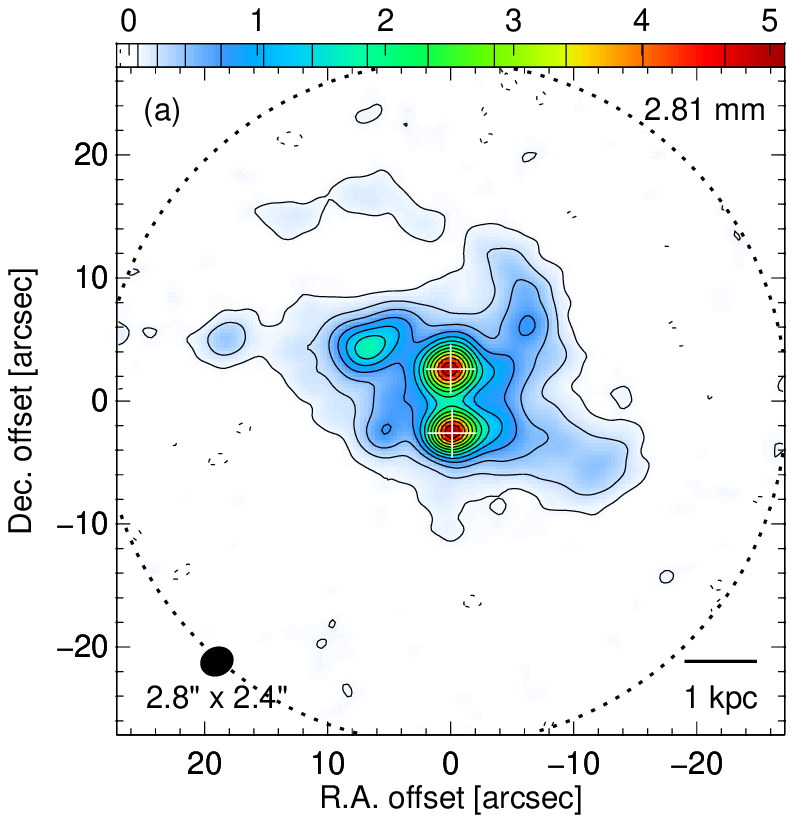} 
\plotone{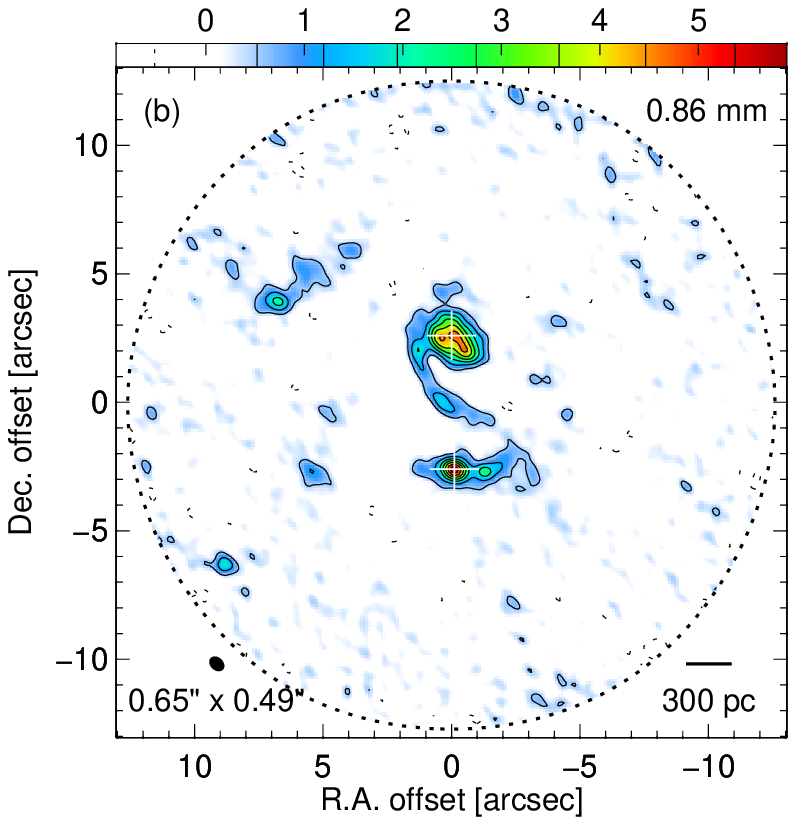} 
\plotone{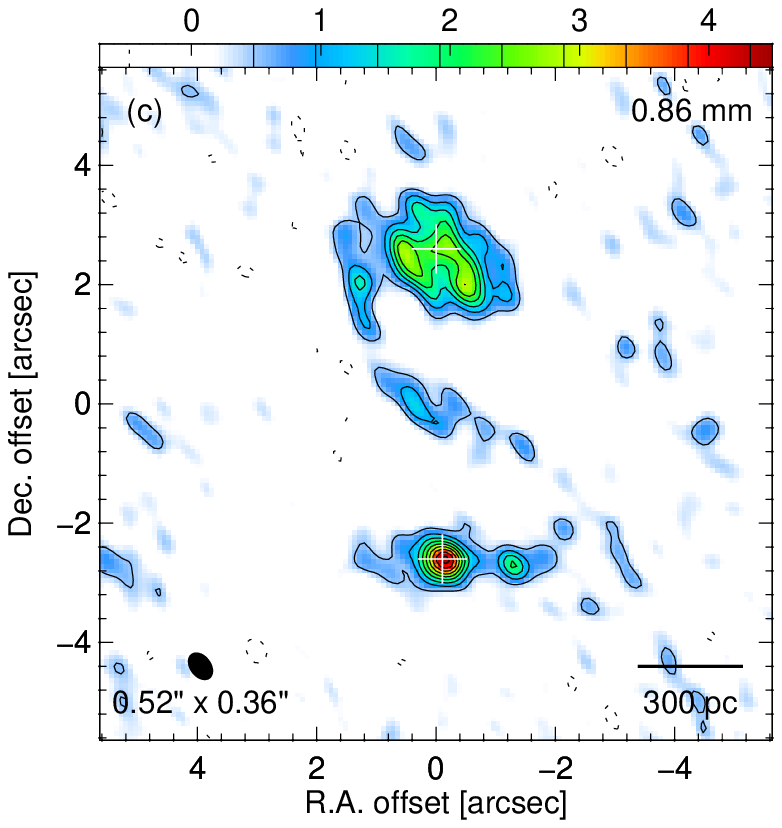} 
\caption{ \label{f.contmaps}
NGC 3256 continuum at $\lambda$ \about\ 2.8 mm (left) and 0.86 mm (middle and right). 
The two plus signs are at the positions of the cm-wave radio nuclei in \citet{Neff03}.
The offset coordinates are measured from the ALMA Cycle 0 pointing position in Table \ref{t.4418param}. 
The maps are corrected for the attenuation of the (mosaicked) primary beam response and are truncated
at the 50\% of their peaks.
The $n$th contours are at $\pm4n^{1.7}\sigma$,  $\pm4n^{1.2}\sigma$, and $\pm3n\sigma$ in (a), (b), and (c), respectively, with
zero contours omitted and negative contours dashed.
The rms noise $\sigma$ are the ones measured before the primary-beam correction and are given in Table \ref{t.data_cont_properties}.
The intensity unit of the color bars at the top is mJy \perbeam.
The peaks in (a), (b), and (c) are 5.1 (0.08), 5.9 (0.19), and 4.5 (0.24) mJy \perbeam\ (K), respectively.
}
\end{figure}

The 3 mm continuum emission shown in Fig. \ref{f.contmaps} (a) peaks at the two nuclei
and
is extended to a radius of at least 20\arcsec\ (3 kpc) with arcs and arm-like features in the region.
Millimeter continuum at 1.3 mm also peaks at the two nuclei \citesp{Sakamoto06}.
The nuclear peaks and the extended emission at 3 mm are morphologically similar to those
previously observed at 6 and 3.6 cm \citep[in their Fig. 1]{Norris95,Neff03}.  
Both nuclei are resolved in our 0.86 mm continuum images in Fig. \ref{f.contmaps} (b), (c).
The northern nucleus has a high-intensity plateau with a diameter of  about 2\arcsec\ (0.3 kpc)
and the southern nucleus has a compact (\about0\farcs5, 80 pc) peak 
with a linear feature elongated by about 3\arcsec\ (0.5 kpc) in the east-west direction through the nucleus.
The extent of the northern nucleus agrees with that in X-rays \citep[FHWM \about1\farcs5  in 0.5--10 keV measured by][]{Lira02}. 
Our highest-resolution continuum image in Fig. \ref{f.contmaps} (c) hints at a (broken) ring
in the plateau around the northern nucleus. 
It is comparable in size to an optical ring-like structure noted by \citet{Laine03}.
There is also conspicuous bridge-like emission between the two nuclei.
It emanates from the circumnuclear region of the northern nucleus and curves toward the western side
of the elongated continuum emission across the southern nucleus.
This feature and the near linear emission across the southern nucleus, 
are present in the  3.6 cm continuum data of \citet[their Figs. 1, 2]{Neff03}.
The peak brightness temperatures in Fig. \ref{f.contmaps} are between 0.08 and 0.24 K.
The compact southern nucleus shows higher peak brightness temperatures 
when the northern circumnuclear plateau is spatially resolved.

We measured the spectral slope of the continuum emission at 2.8 mm and 0.86 mm by comparing the data in
the upper and lower sidebands separated by about 12 GHz from each other.
We used for this single-sideband continuum images that have a common spatial resolution and
were made with a common \uv\ baseline lengths; the shortest baselines in the LSB and the longest in the USB were
flagged for this.
Unfortunately, we cannot reliably compare the 2.8 mm and 0.86 mm data to estimate the spectral index between them
because the difference in their \uv\ coverages is too large.
The spectral index $\alpha$ of the continuum emission (for $S_\nu \propto \nu^\alpha$ where $\nu$ is frequency
and $S_\nu$ is flux density)
is found to be  $-0.1$ at 2.8 mm  and $+3$ at 0.86 mm in the central 20\arcsec.
The spectral indexes at the individual nuclei are also measured and listed in Table \ref{t.contFluxSpix}.
It is found that $\alpha$ is significantly larger at 0.86 mm than at 2.8 mm also for each nucleus.
Moreover, $\alpha$ is found larger at the northern nucleus than at the southern nucleus clearly at 0.86 mm and also at 2.8 mm.
The spectral index $\alpha$ is 3--4 for thermal dust emission that is optically thin because
dust mass opacity coefficient has a power law index of 1--2.
Free-free emission from thermal electrons often has a spectral index around $-0.1$.
Synchrotron emission from galaxy nuclei often has a spectral index of about $-1$ ($\pm 0.5$).
The smaller $\alpha$ of the southern nucleus is consistent with the nucleus having a larger fraction
of free-free or synchrotron emission and less of dust thermal emission than the northern nucleus.

\subsection{Line} 
\label{s.result-line}

Line maps are shown in Figures \ref{f.maps.CO102132} and \ref{f.maps.nonCO} for the eight lines that we imaged.
Plots shown are: the integrated intensity, intensity-weighted mean velocity, intensity-weighted velocity dispersion, and
peak brightness temperature.
Also shown are line channel maps in Figs \ref{f.chans.CO10.br}, \ref{f.chans.CN10.br10MHz}, \ref{f.chans.CH3C2H.Na10MHz},
\ref{f.chans.CO32.br}, and \ref{f.chans.HCOp43.na30MHz}.
They are of low velocity resolutions for space reasons although we also made data cubes with higher velocity resolutions.
Note that contour levels are switched between channels with and without strong signals in the CO channel maps
in order to display both faint high-velocity emission and strong emission near the systemic velocity.

\subsubsection{Spatial Distribution}
All the molecular lines have emission peaks at or near the two nuclei, as does the continuum emission.
The degree of concentration and the relative strengths of the two nuclei vary among the lines.
The  bridge-like feature between the northern and southern nuclei is also visible in line emission, 
most clearly in CO(2--1), (3--2) and \HCOplus(4--3) integrated intensity images.
There are other arc features; some are seen in continuum and some are visible only in the line data.
The near-linear feature crossing the southern nucleus in the east-west direction is also visible
in line emission, most clearly in CO(3--2) and \HCOplus(4--3).

\subsubsection{Velocity Field}
\label{s.result.line.velocity_field}
\begin{figure}[t]
\epsscale{0.3}
\plotone{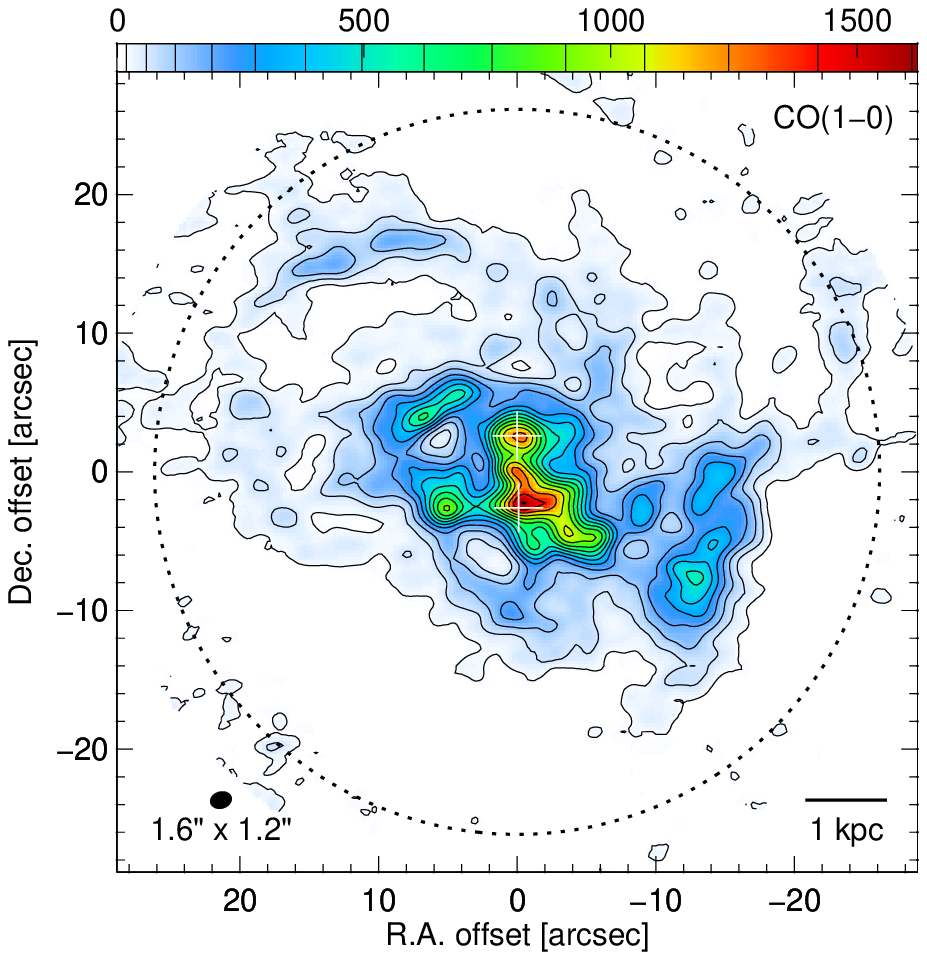} 
\plotone{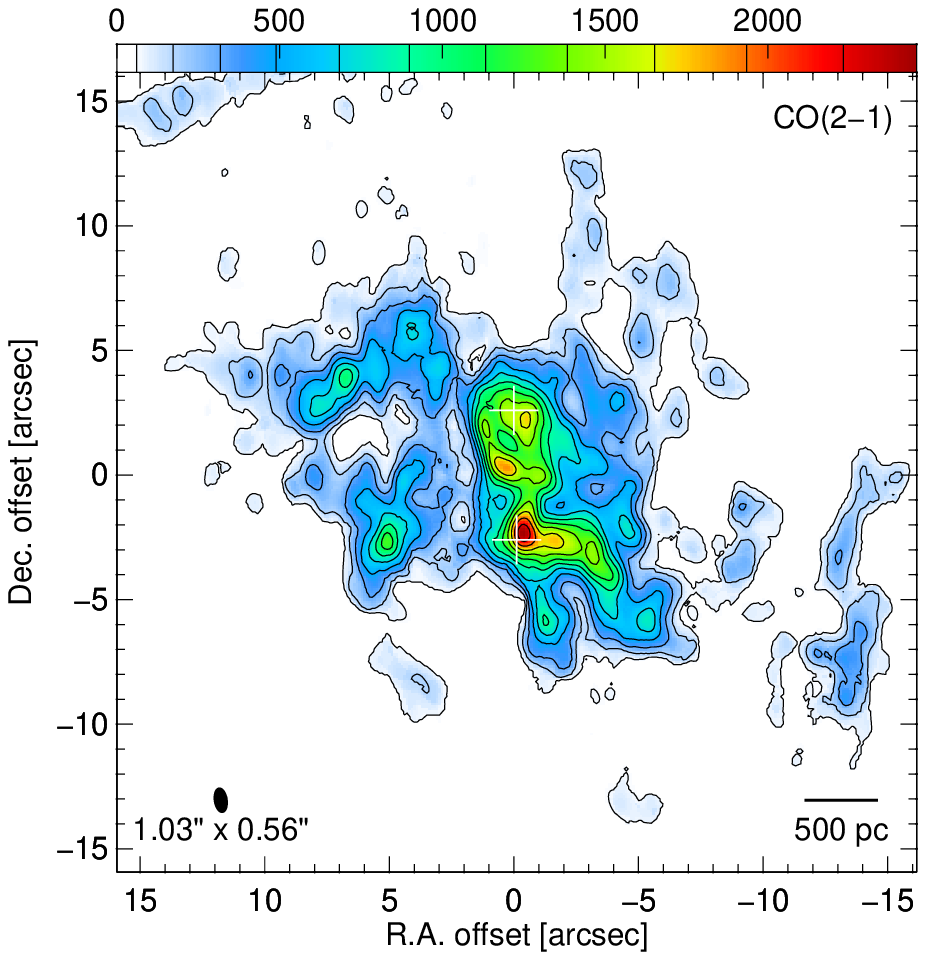} 
\plotone{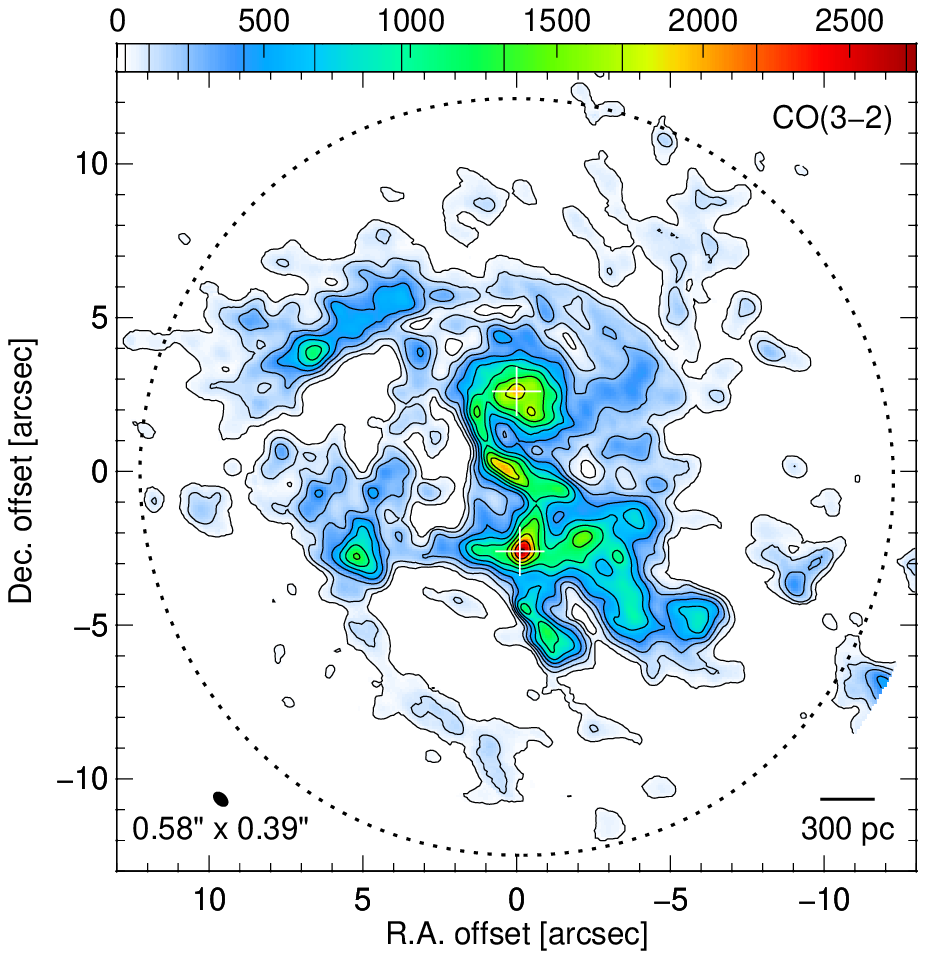}\\ 
\plotone{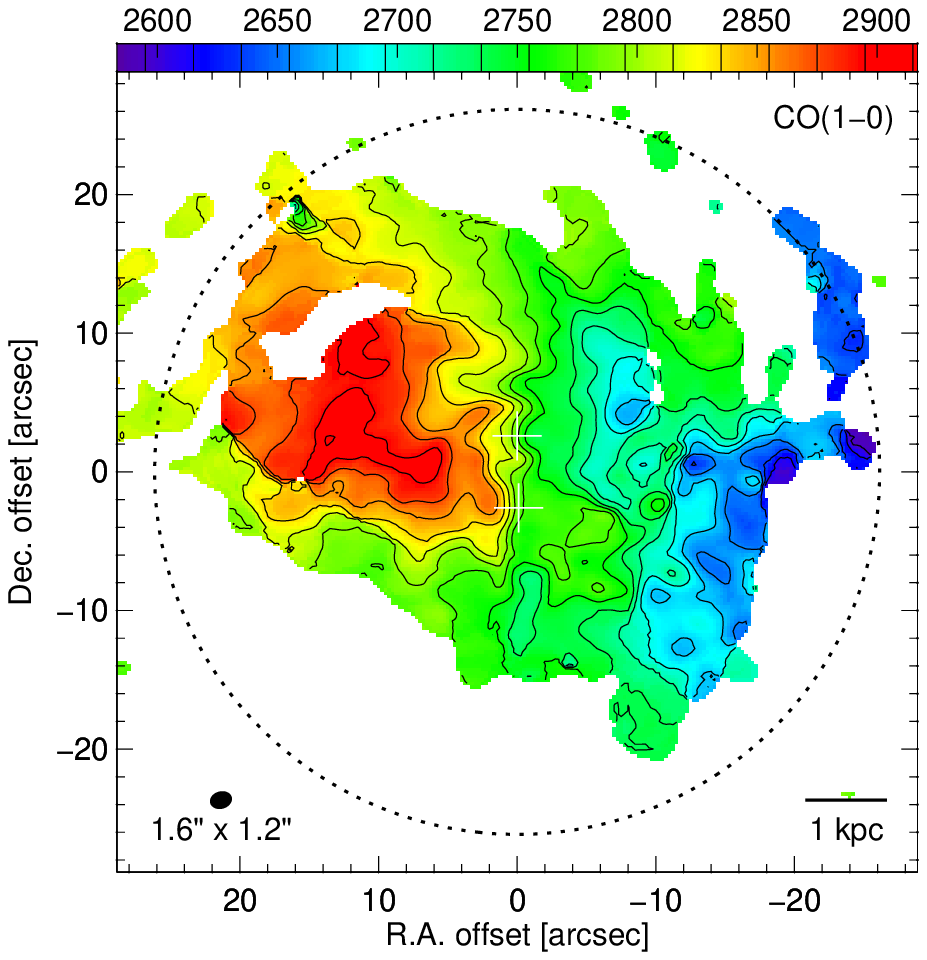} 
\plotone{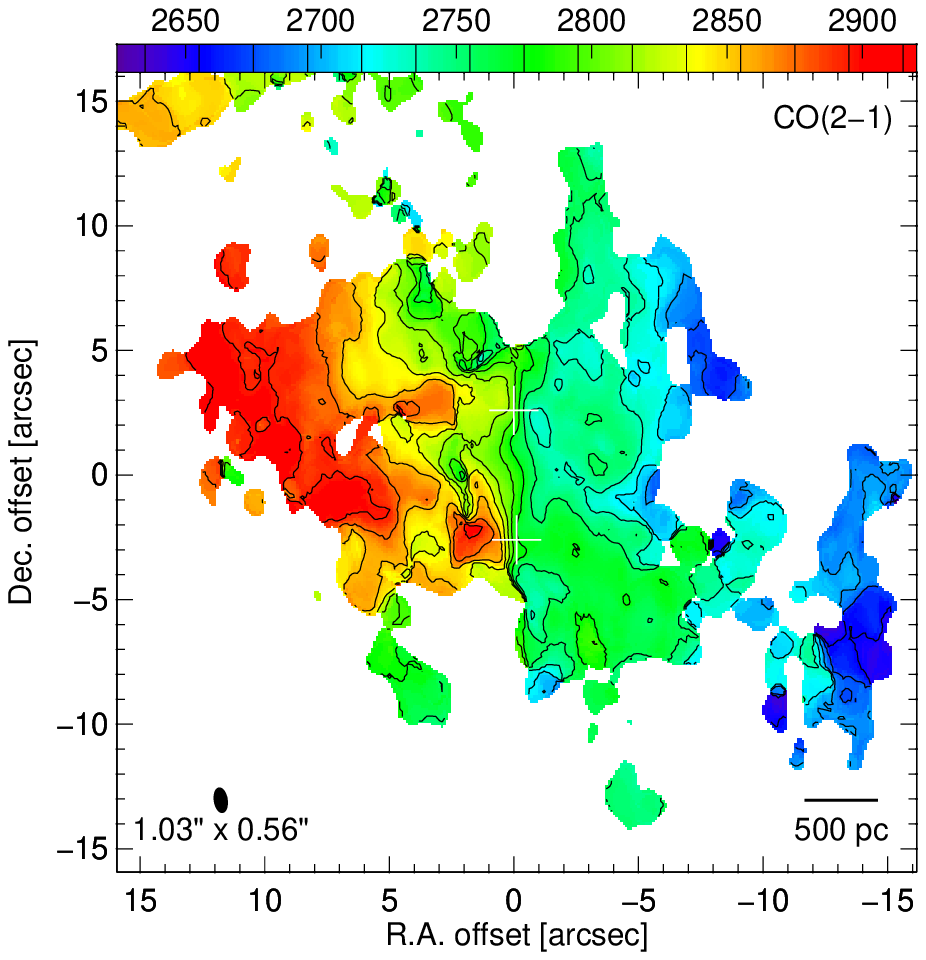} 
\plotone{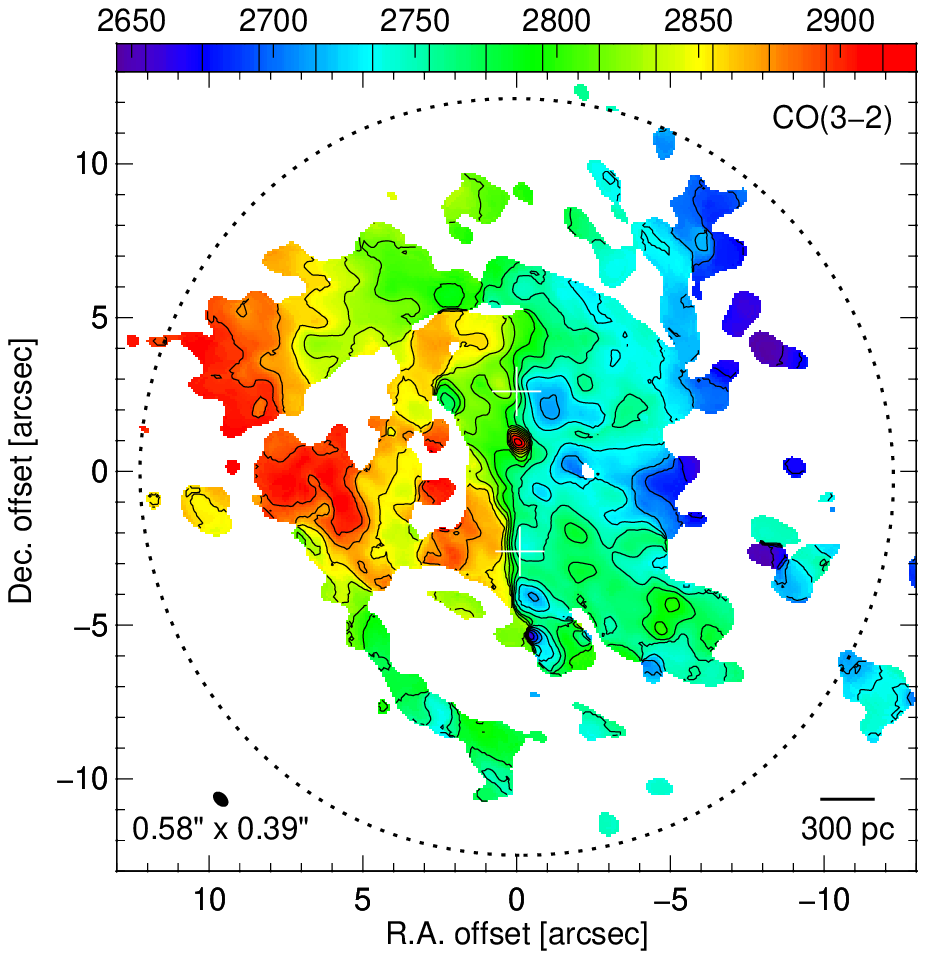}\\ 
\plotone{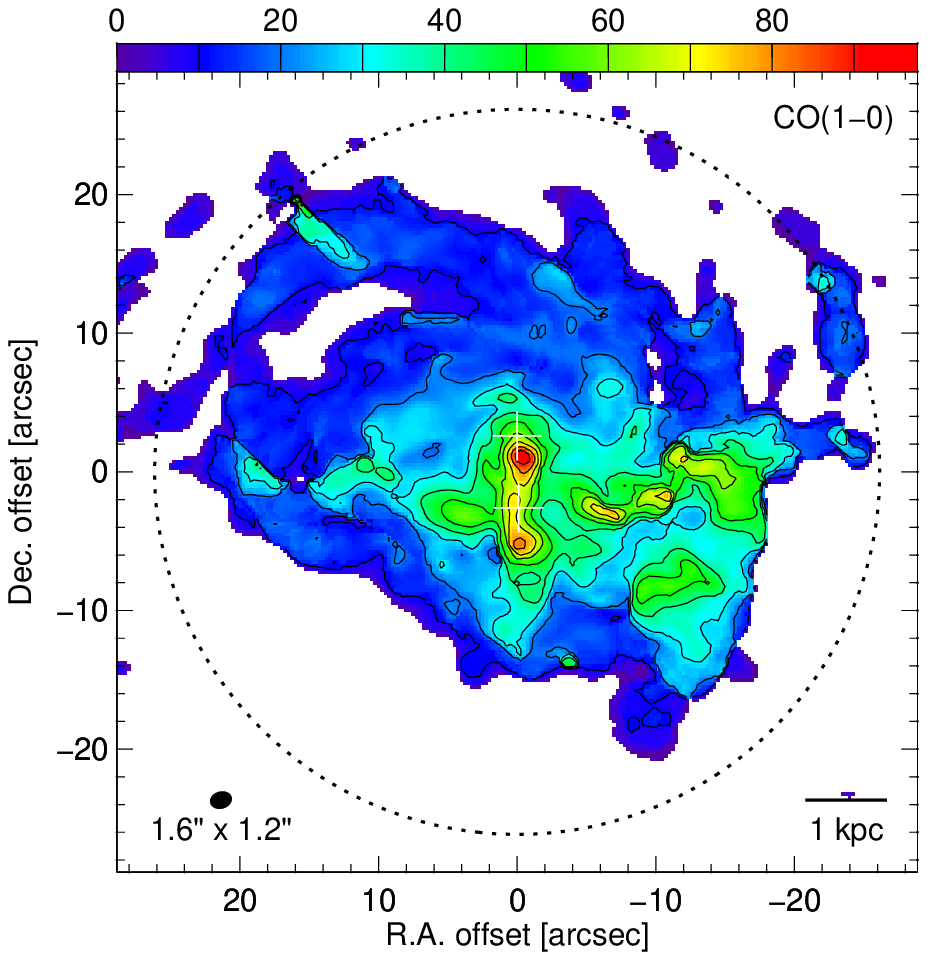} 
\plotone{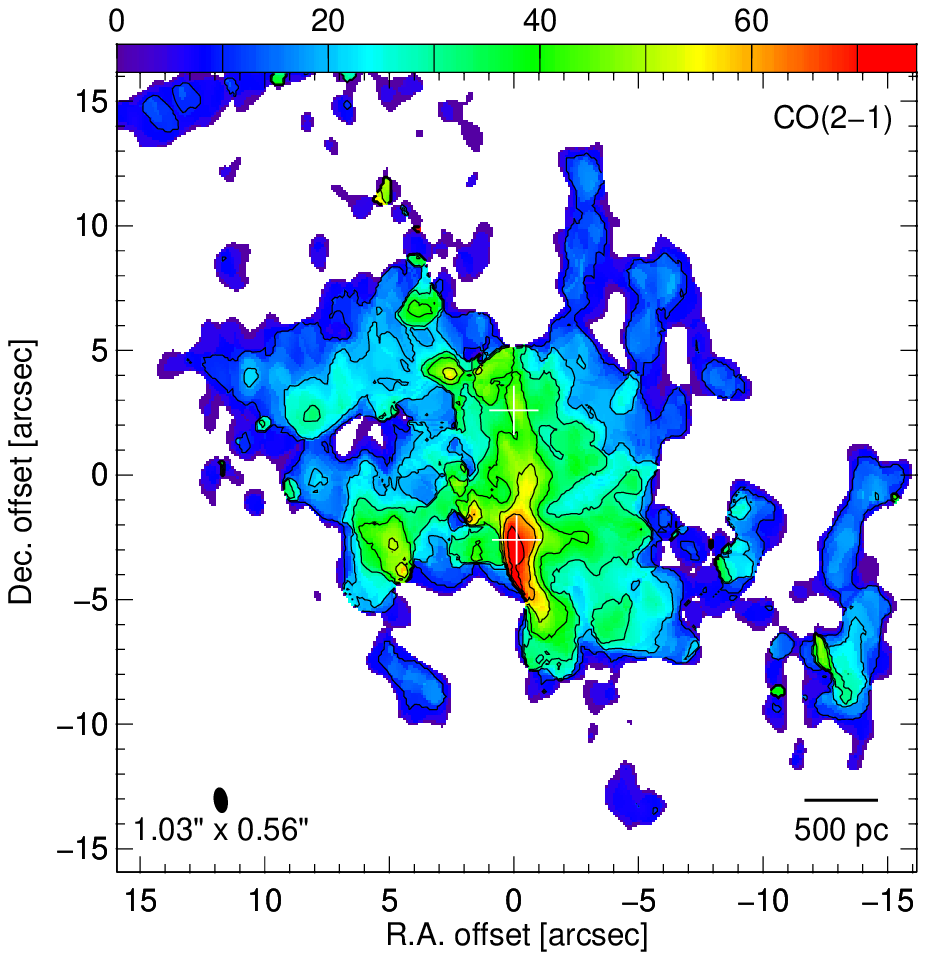}  
\plotone{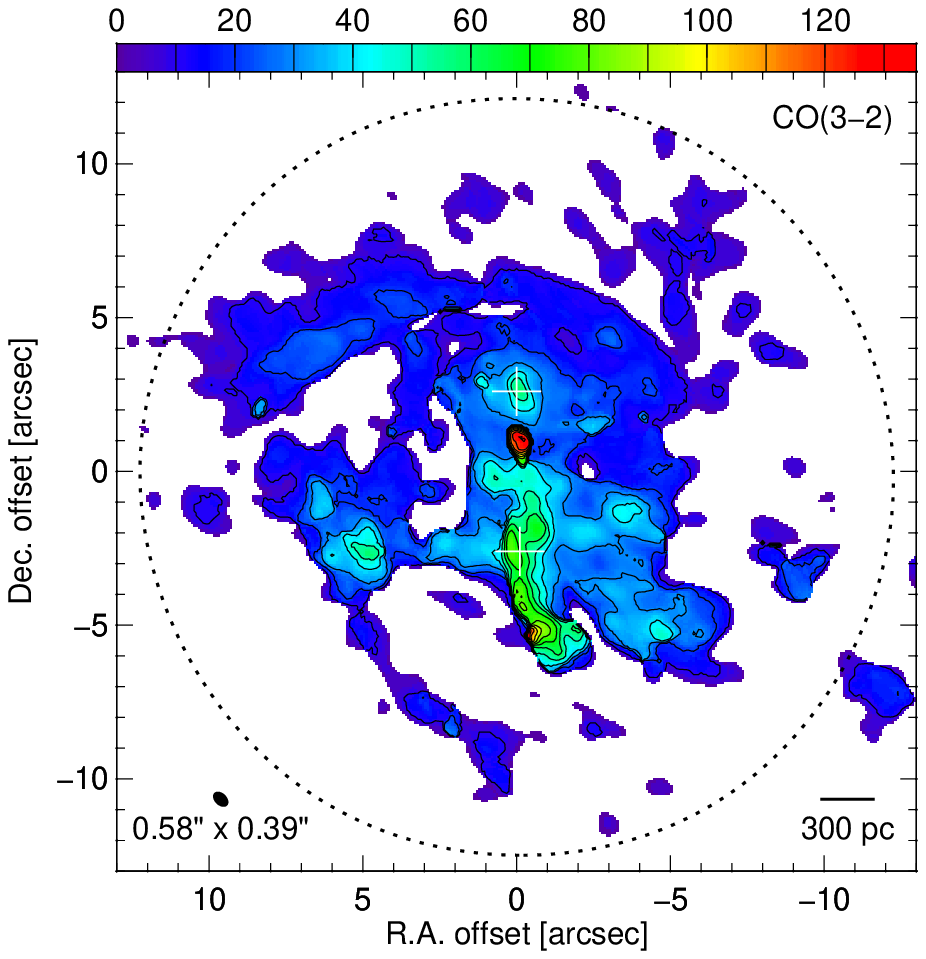}\\ 
\plotone{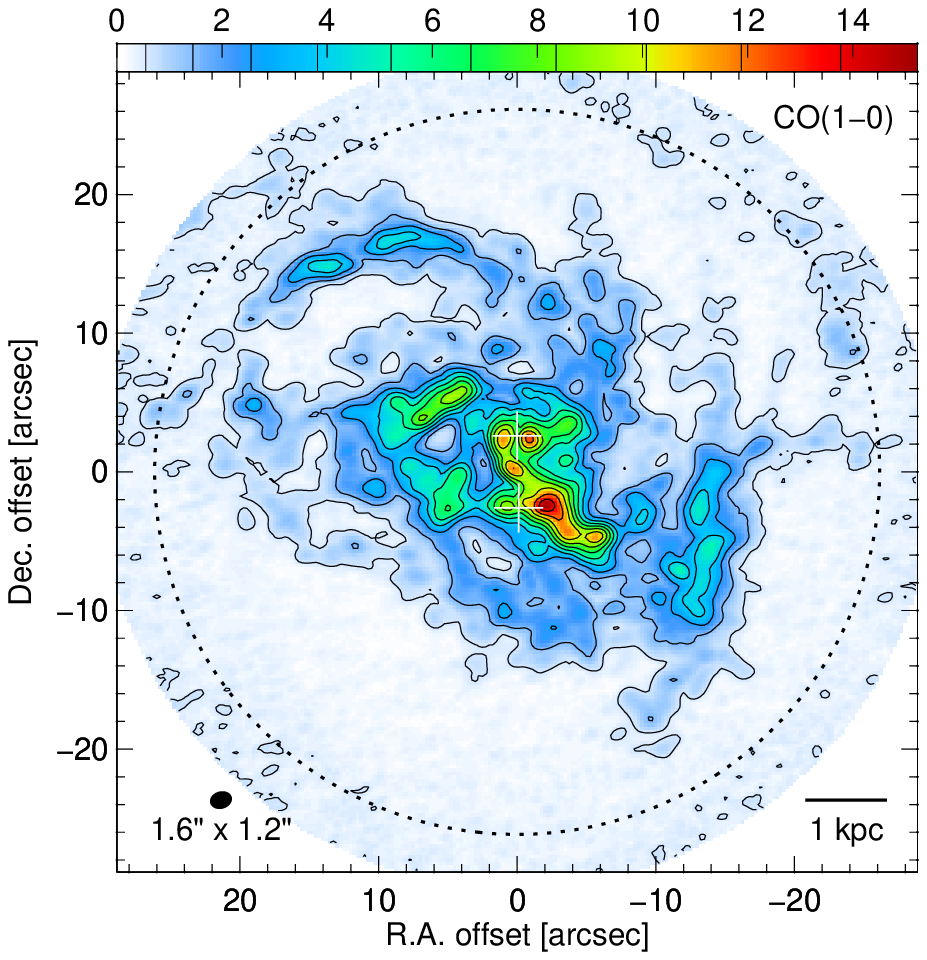} 
\plotone{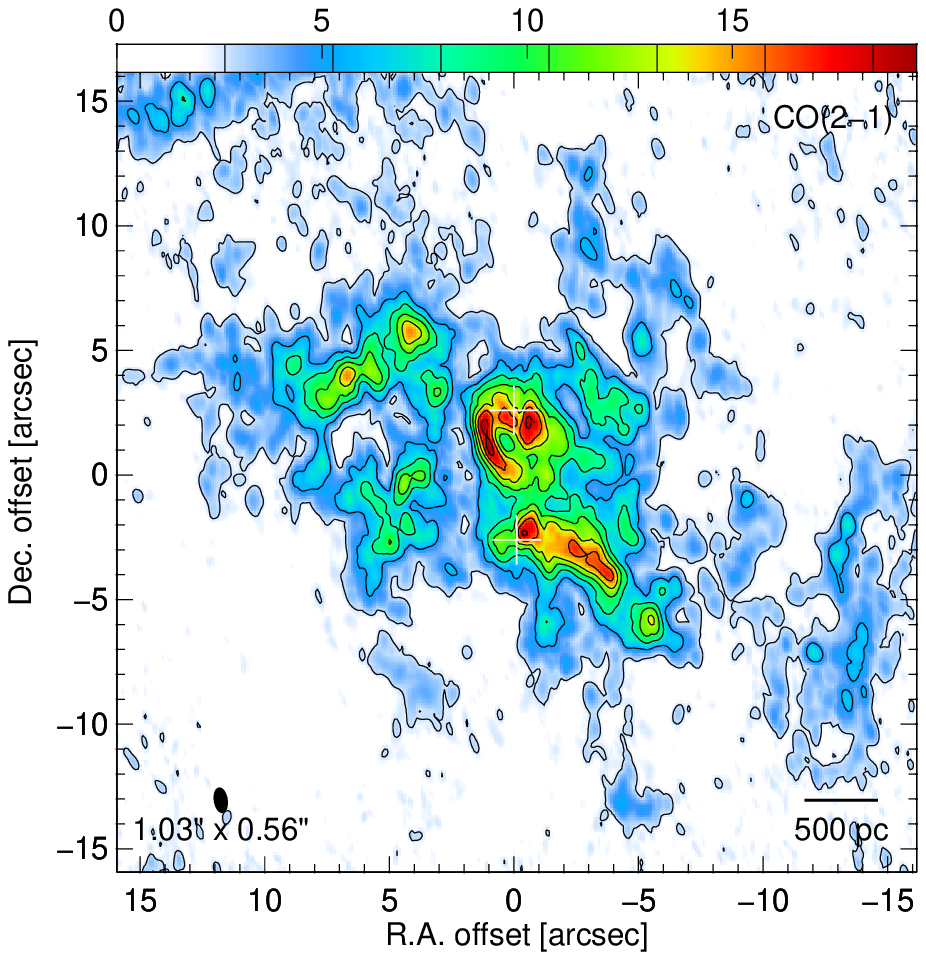}  
\plotone{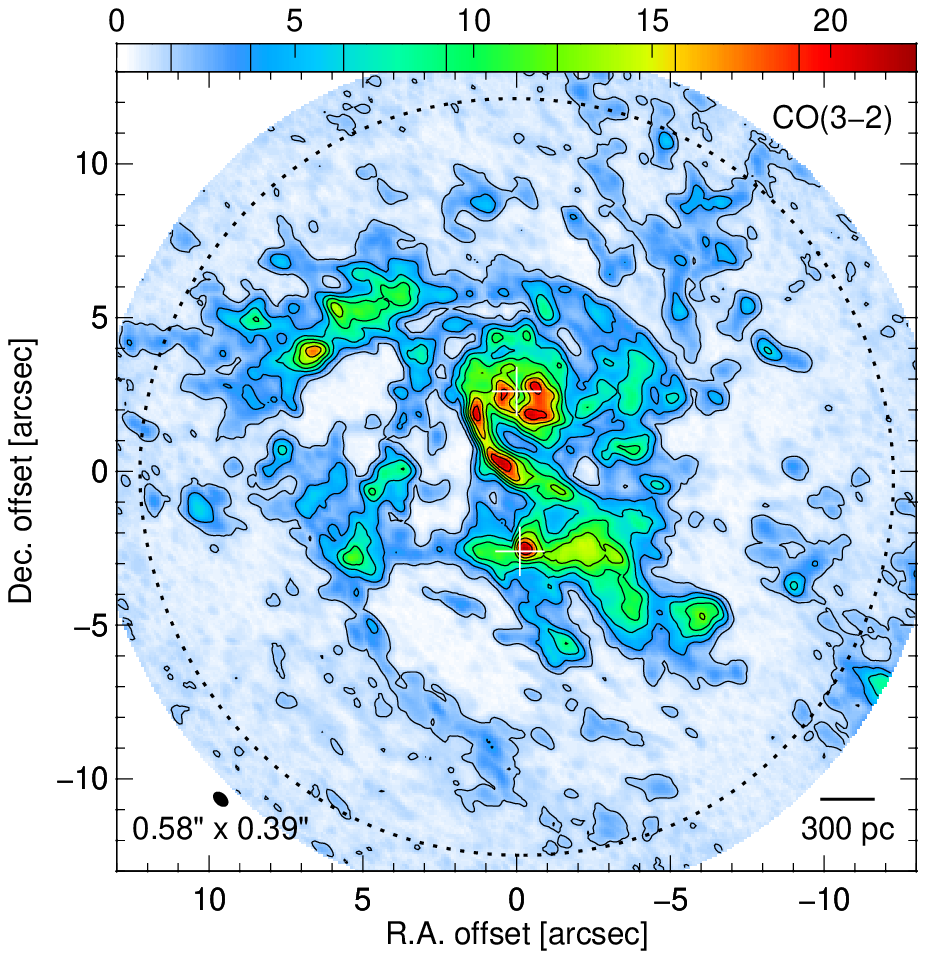}\\ 
\caption{\scriptsize \label{f.maps.CO102132}
\twelveCO(J=1--0) (left), \twelveCO(J=2--1) (middle),  and \twelveCO(J=3--2) (right) images of NGC 3256.
The four rows are, from the top, moment 0, 1, 2 maps and  peak \Tb\ maps with the data units
of K \kms, \kms, \kms, and K, respectively.
The two plus signs are at the positions of the cm-wave radio nuclei.
The offset coordinates are measured from the common ALMA Cycle 0 and SMA pointing position in Table \ref{t.4418param}. 
Intensity maps (the top and the bottom rows) are corrected for the (mosaicked) primary beam responses; 
dotted lines show them at 50\% of their peaks.
The $n$th contours in the moment 0 maps are at $cn^p$ K\kms, where $(c, p) = (18, 1.7), (61, 1.5),$ and $(27, 2)$ for CO(1--0), (2--1), (3--2), 
respectively.  
Contours in the moment 1 maps are at every 20 \kms\ including 2775 \kms, and those in the moment 2 maps
at every 10 \kms\ starting from 10 \kms.
The $n$th contours in the peak \Tb\ maps are at $dn^q$ K, where $(d, q) = (0.55, 1.4), (2.5, 1.0),$ and $(1.5, 1.3)$ for CO(1--0), (2--1), (3--2), respectively. 
The maximum brightness temperatures in the maps are 15.2, 19.5, and 22.4 K for  CO(1--0), (2--1), (3--2), respectively.
Synthesized beams are shown with their FWHM sizes at the bottom-left corners of the panels.
}
\end{figure}

{\em Large scale:}
The CO(1--0) velocity map in Fig. \ref{f.maps.CO102132} shows
overall rotation in the central \about5 kpc with the receding major axis at position angle \about70\arcdeg.
Significant deviations from circular motion at this scale are visible mostly at the locations of the arm-like features.
The apparent kinematical major axis is at p.a. \about90\arcdeg\ within about 1 kpc from the nuclei.
Both nuclei are therefore approximately on the apparent kinematical minor axis at this scale, as seen in
e.g., the 1st moment maps of CO(1--0), CO(2--1) and CN(1--0).
These large scale kinematics of molecular gas are consistent with those in \citest{Sakamoto06}.

\begin{figure}[t]
\begin{center}
\includegraphics[height=35mm]{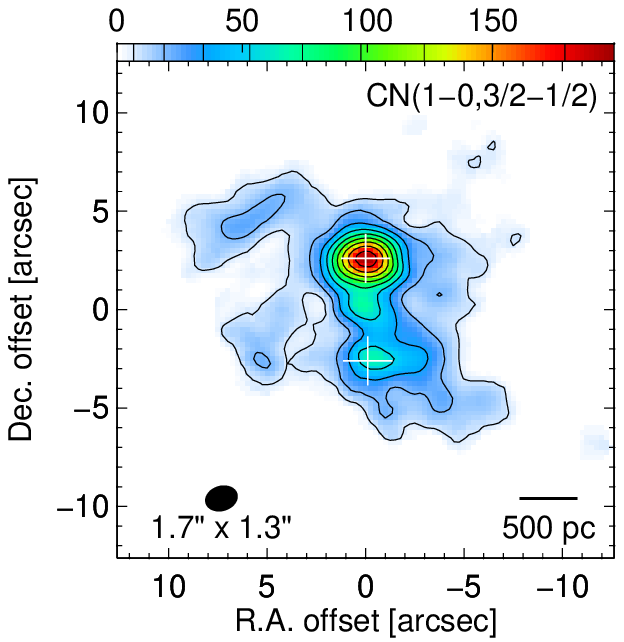} 
\includegraphics[height=35mm]{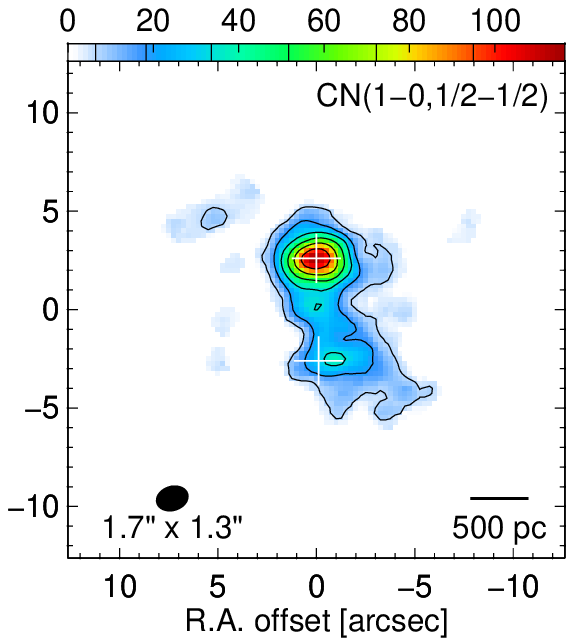} 
\includegraphics[height=35mm]{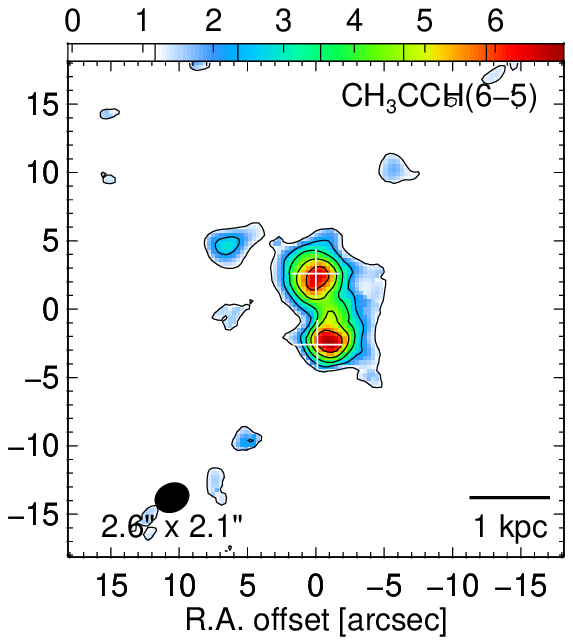} 
\includegraphics[height=35mm]{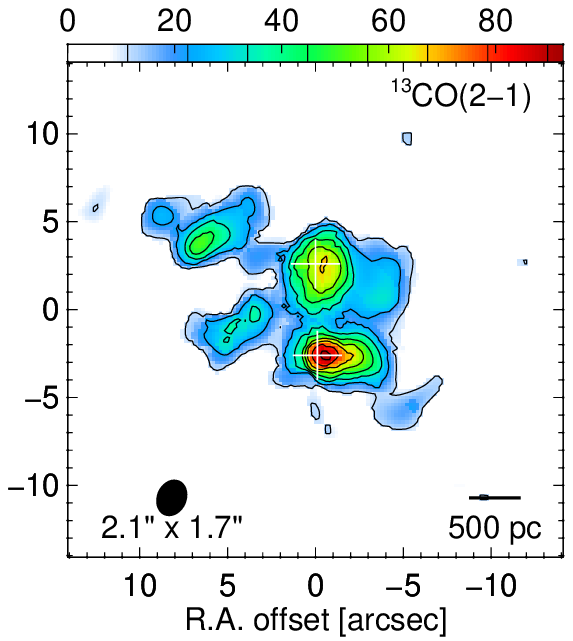} 
\includegraphics[height=35mm]{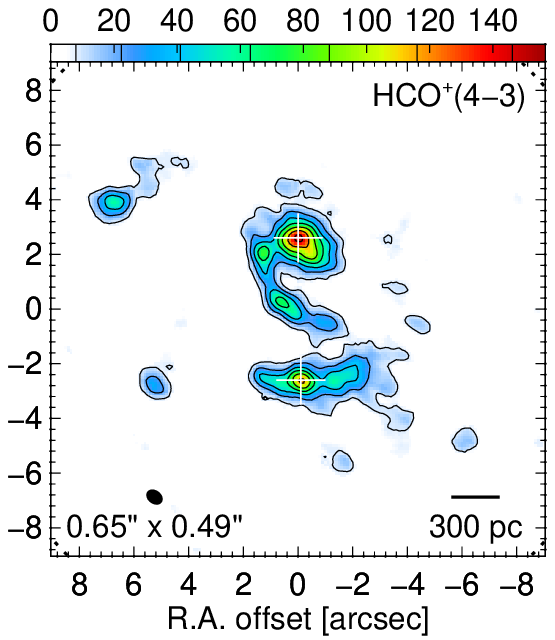}\\ 
\includegraphics[height=35mm]{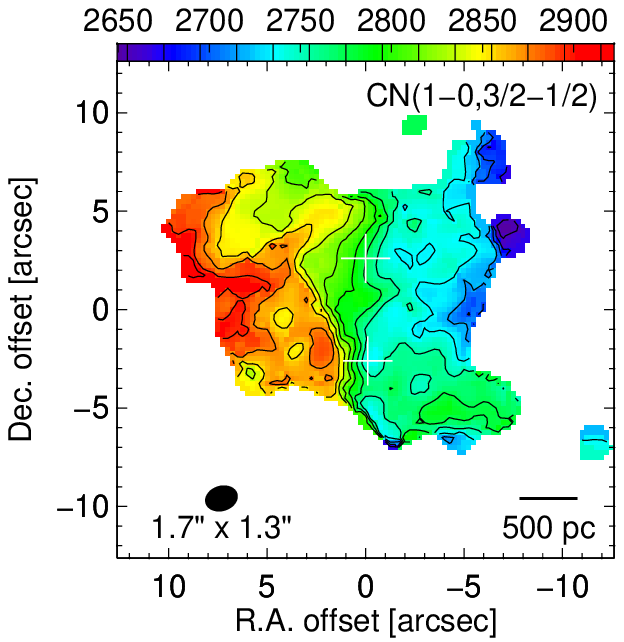} 
\includegraphics[height=35mm]{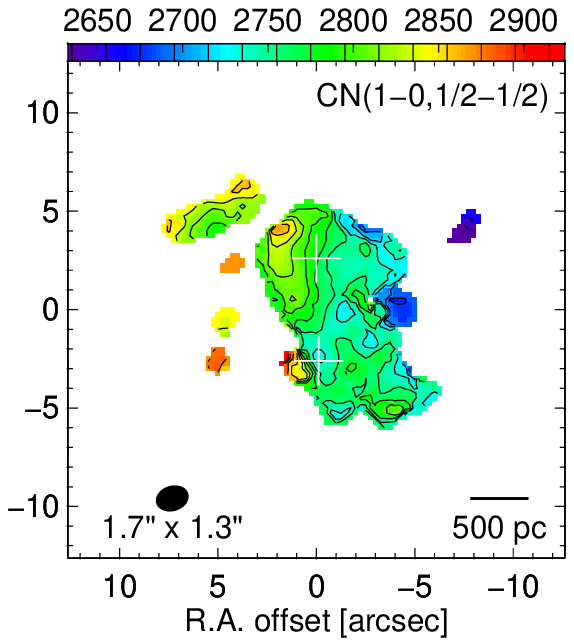} 
\includegraphics[height=35mm]{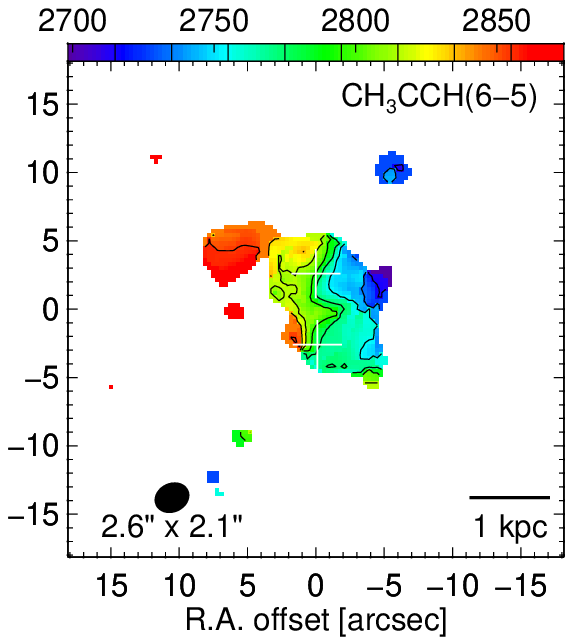}  
\includegraphics[height=35mm]{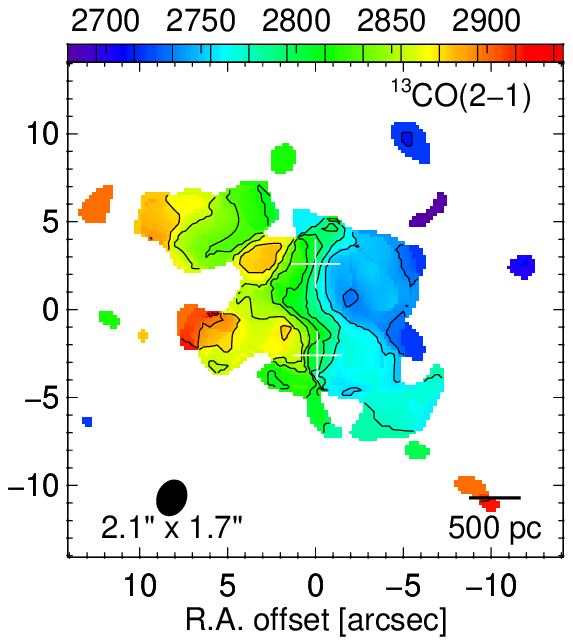} 
\includegraphics[height=35mm]{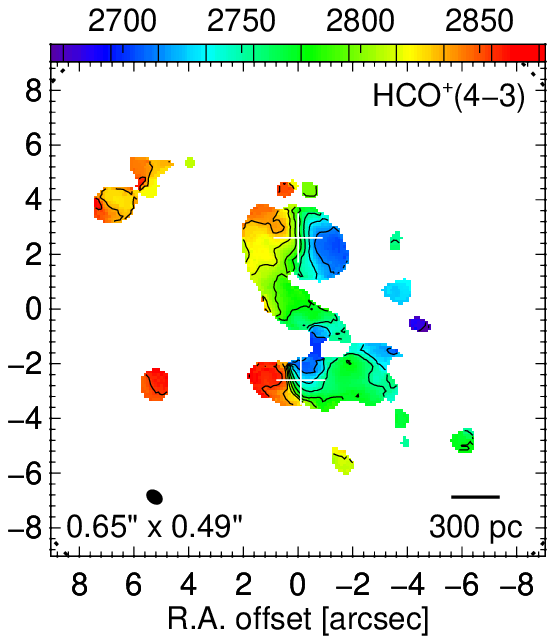}\\ 
\includegraphics[height=35mm]{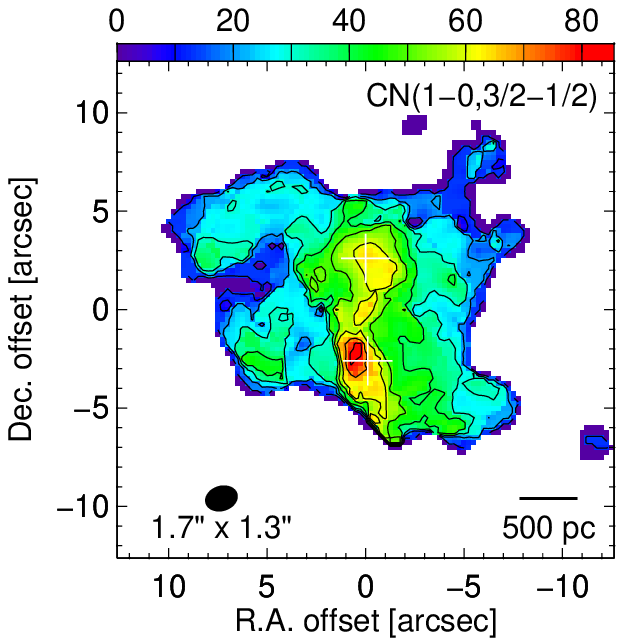} 
\includegraphics[height=35mm]{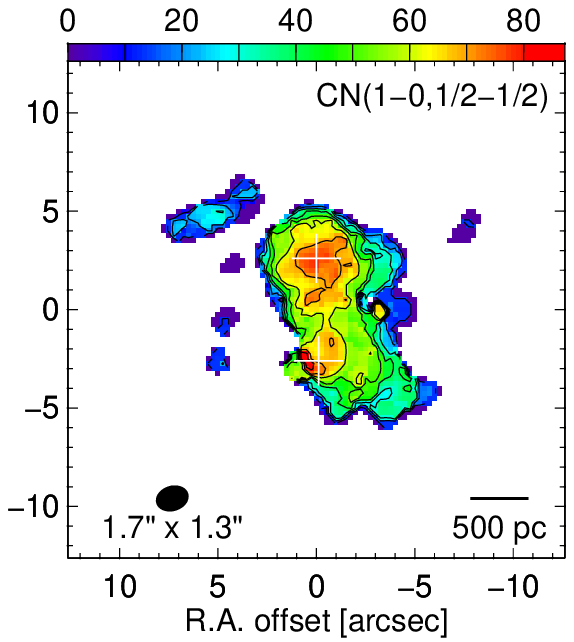} 
\includegraphics[height=35mm]{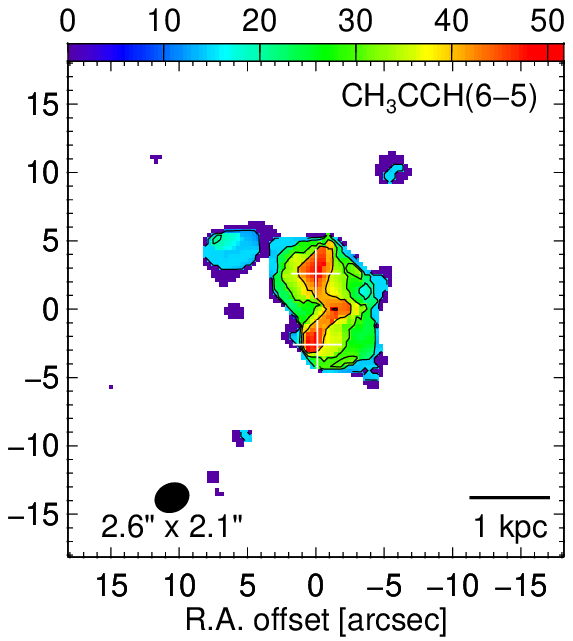} 
\includegraphics[height=35mm]{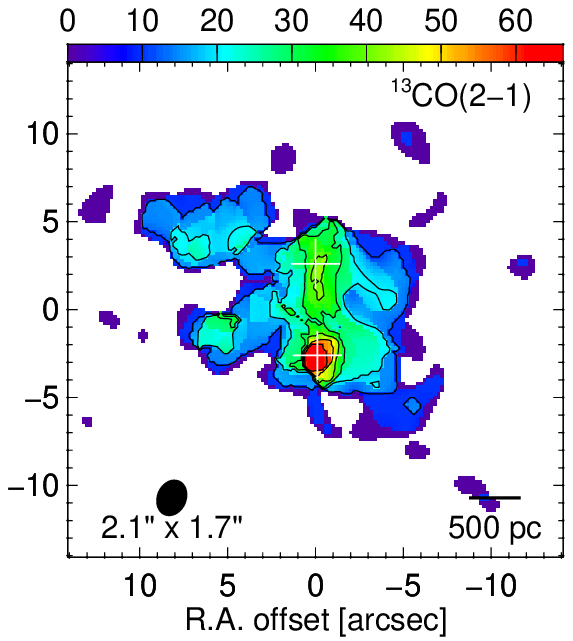}  
\includegraphics[height=35mm]{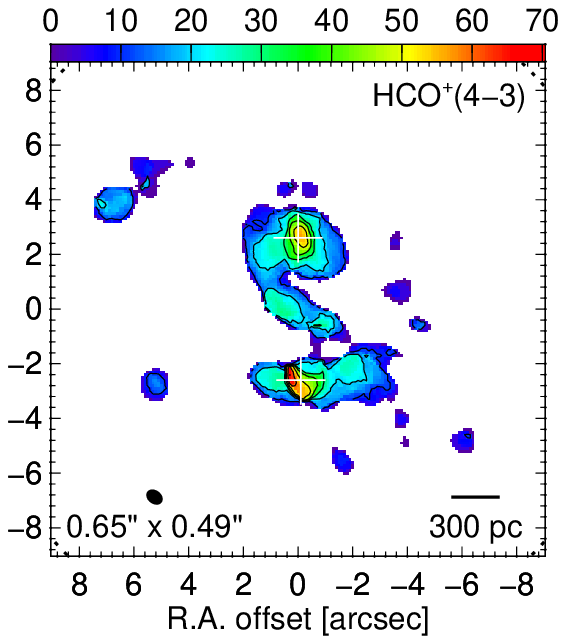}\\ 
\includegraphics[height=35mm]{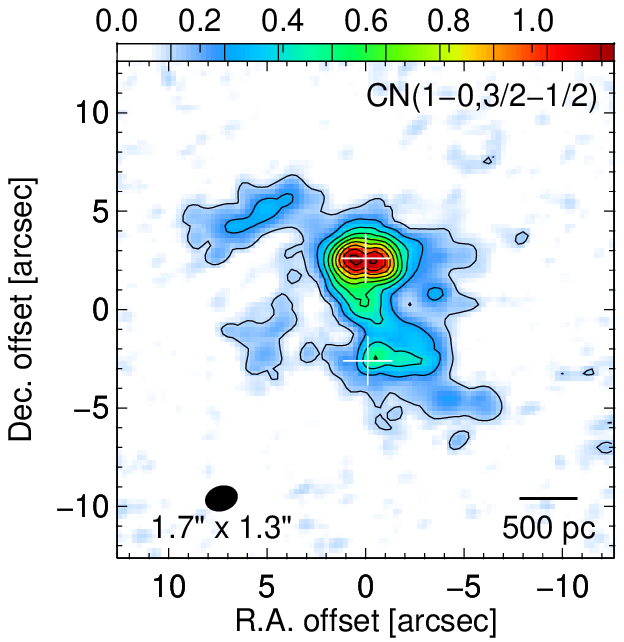}
\includegraphics[height=35mm]{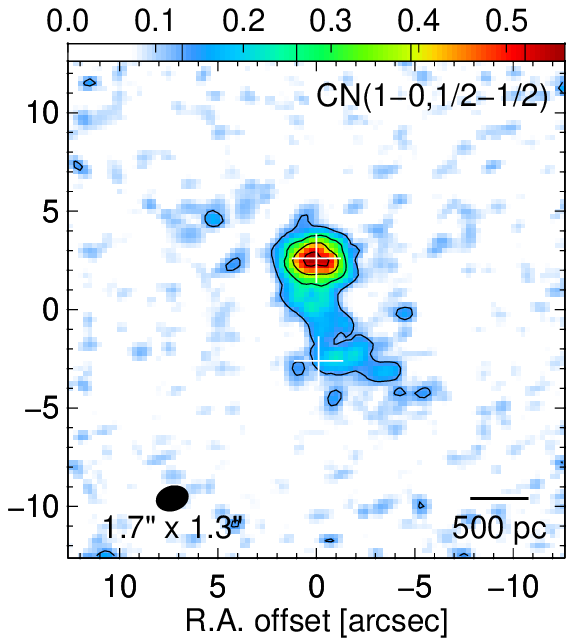} 
\includegraphics[height=35mm]{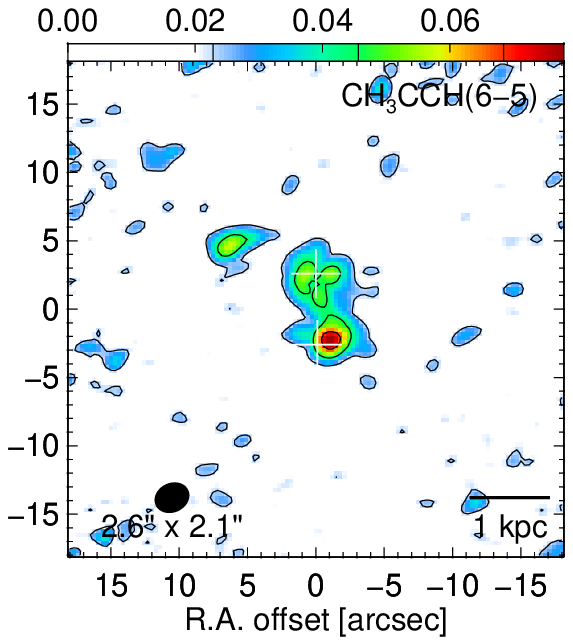} 
\includegraphics[height=35mm]{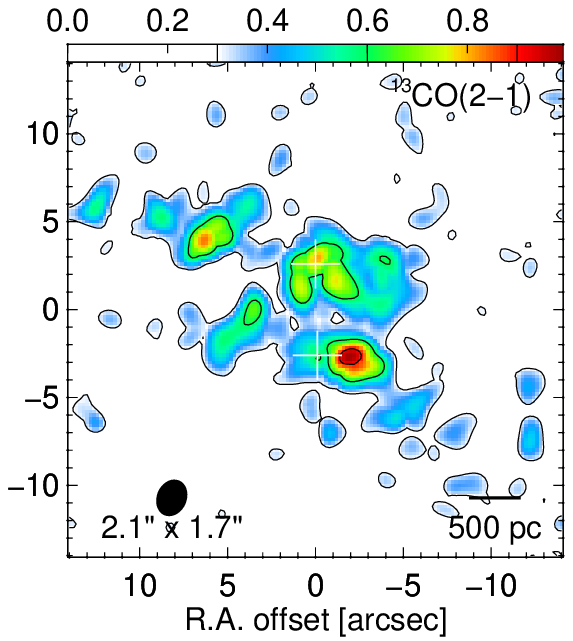} 
\includegraphics[height=35mm]{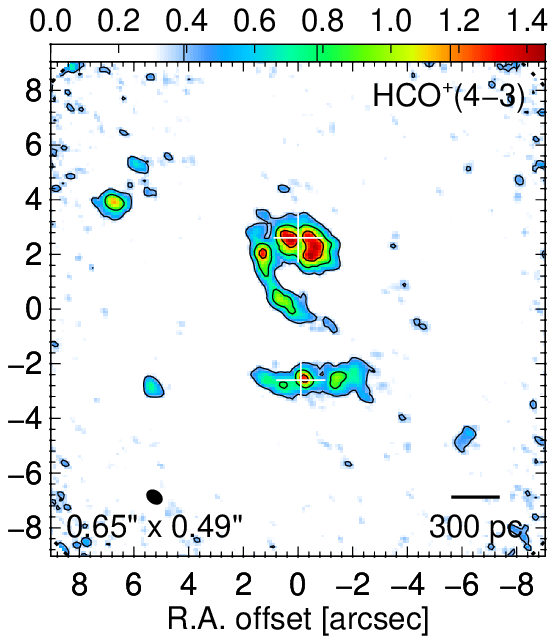}\\ 
\end{center}
\caption{ \label{f.maps.nonCO}
Molecular line maps of NGC 3256.
The lines are, from left to right,
CN(N=1--0; J=3/2--1/2), 
CN(N=1--0; J=1/2--1/2), 
\propyne(J=6--5),
\thirteenCO(J=2--1),
and  
\HCOplus(J=4--3).
The four rows are, from the top to the bottom, the moment 0, 1, 2 maps and  peak \Tb\  map in units
of K \kms, \kms, \kms, and K, respectively.
Intensity maps at the top and the bottom rows are corrected for the sensitivity patterns of 
the primary beams; all plots are within 50\% of their peaks.
The $n$th contours are at $cn^p$ K\kms\ in the moment 0 maps  and $dn^q$ K in the peak \Tb\ maps 
with $(c, p) = (6.6, 1.5), (6.5, 1.5), (1.2, 1), (11.2, 1) $ and $(7.9, 1.5)$       
and $(d, q) = (0.13, 1), (0.13, 1), (0.023,1), (0.30, 1) $ and $(0.39, 1)$,     
respectively, from the left-most column to right.
}
\end{figure}

\begin{figure}[t]
\epsscale{1}
\plotone{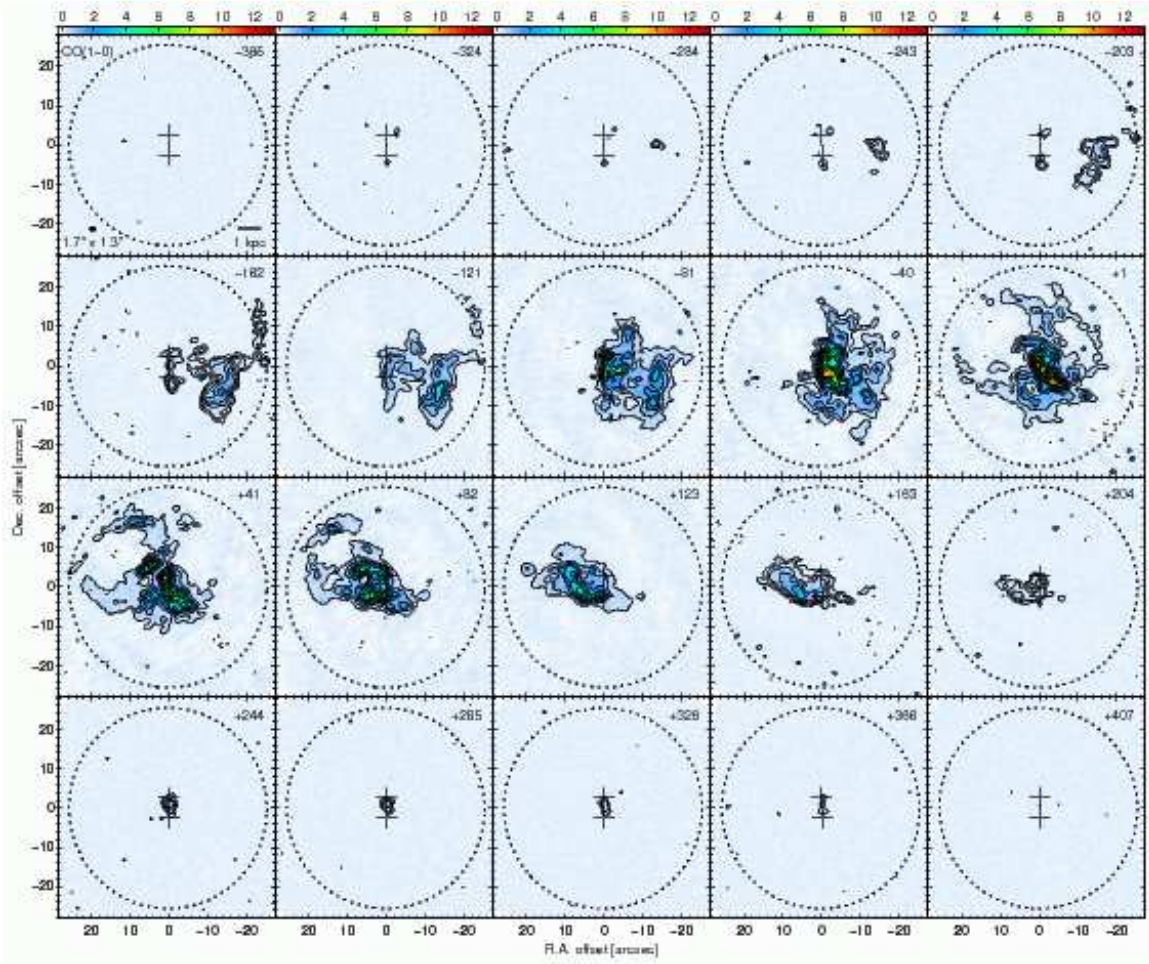}
\caption{ \label{f.chans.CO10.br}
CO(1--0) channel maps. 
Offsets from 2775 \kms\ are at the top-right corners. 
Contours are at $\pm7n^{1.8} \sigma$ $(n=1,2,3,\cdots)$ in channels from $-121$ to $+123$ \kms,
whose maxima exceed $100\sigma$,
and at $\pm3.5\times 2^{m} \sigma$ $(m=0,1,2,\cdots)$ in the other channels.
The rms noise is $\sigma=29$ mK. 
Negative contours are dashed.
This plot is not corrected for the primary-beam response, whose 50\% contours are the dotted circles.
The peak intensity of CO(1--0) is 13.3 K in this plot.
The two plus signs in each panel are at the two nuclei. 
The black ellipse in the top-left panel shows the FWHM size of the observing beam.
}
\end{figure}

\begin{figure}[t]
\epsscale{1}
\plotone{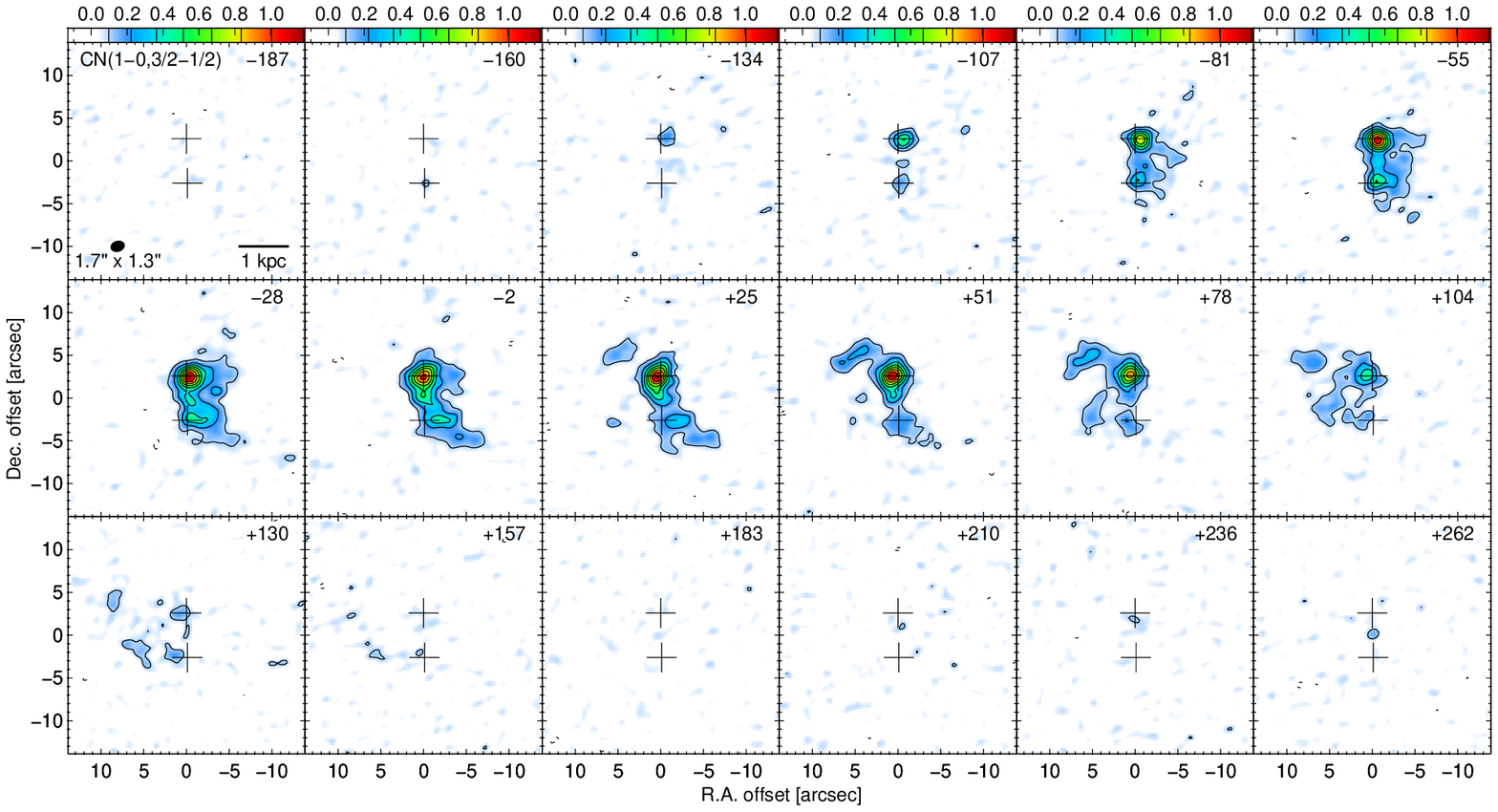} 
\plotone{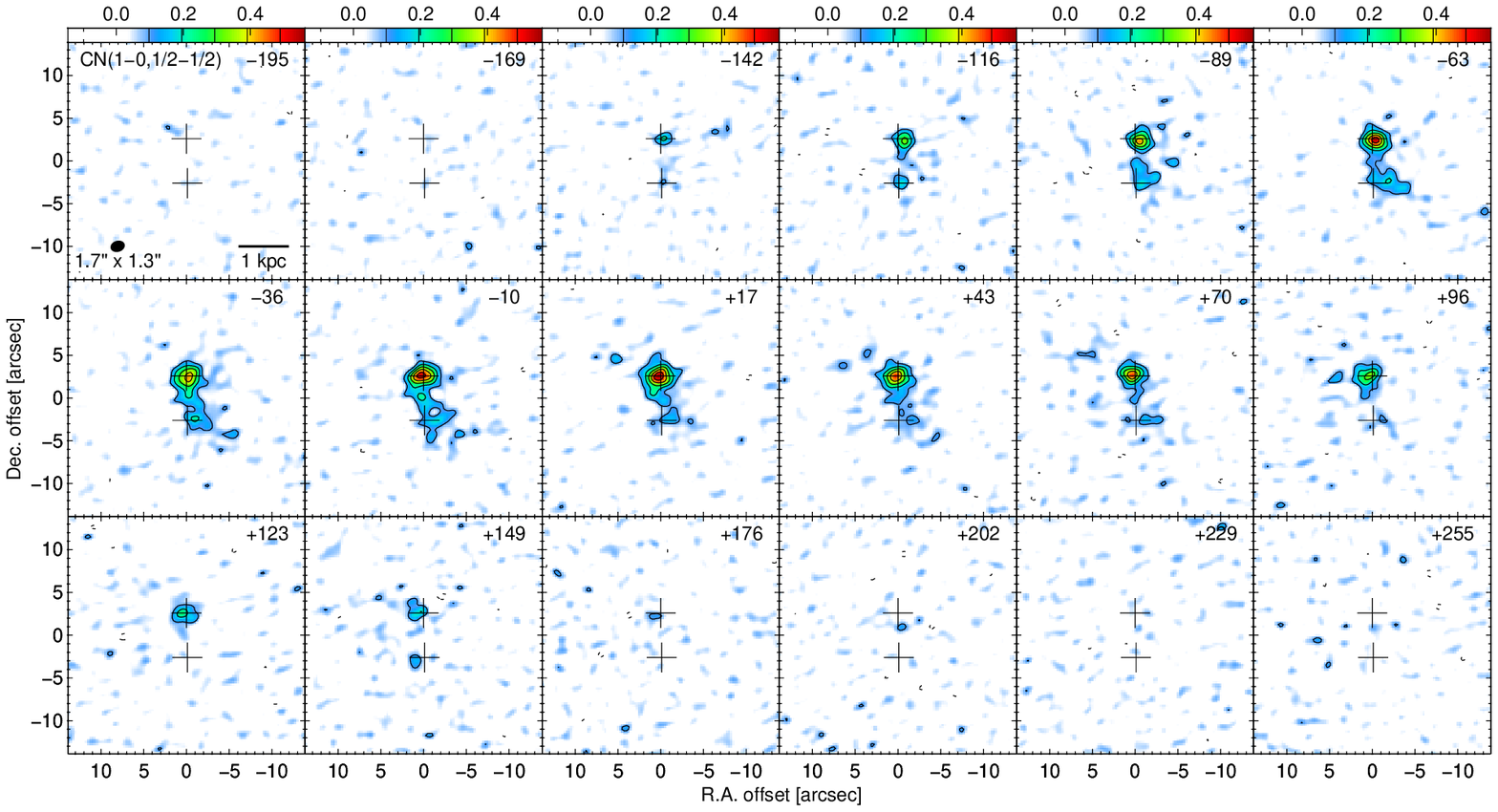} 
\caption{ \label{f.chans.CN10.br10MHz}
CN(N=1--0) channel maps. 
(top) J=3/2--1/2 line at \frest=113.4949 GHz.          
(bottom) J=1/2--1/2 line at \frest=113.1688 GHz.   
Offsets from 2775 \kms\ are at the top-right corners.
Contours are at $\pm 3 n^{p} \sigma$ $(n=1,2,3,\cdots)$, where the power $p$ and the rms noise $\sigma$ are
$p=1.2, \sigma=32$ mK for J=3/2--1/2 and $p=1,  \sigma=33$ mK for J=1/2--1/2.
Negative contours are dashed.
The data are not corrected for the primary beam response, whose FWHM is 54 arcsec.
The peak line intensity in these plots are 1.2 K and 0.56 K, respectively, for  J=3/2--1/2 and 1/2--1/2.
The two plus signs in each panel are at the two radio nuclei. 
A black ellipse in the first panel shows the FWHM size of the observing beam.
}
\end{figure}

\begin{figure}[t]
\epsscale{1}
\plotone{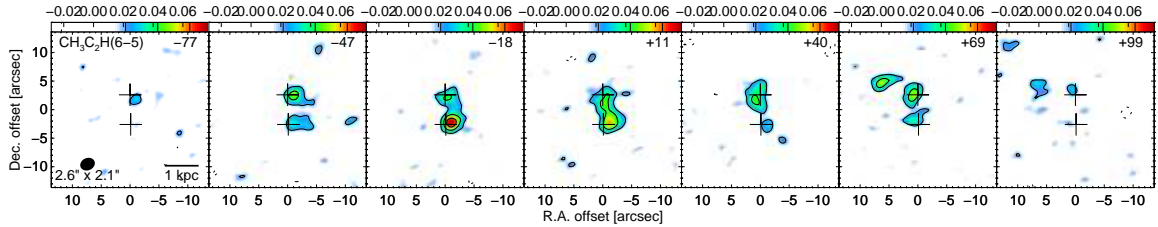} 
\caption{ \label{f.chans.CH3C2H.Na10MHz}
\propyne(6--5) channel maps.
Offsets from 2775 \kms\ are at the top-right corners.
Contours are at $\pm 3 n \sigma$ $(n=1,2,3,\cdots)$, where the rms noise is $\sigma=7.2$ mK.
Negative contours are dashed.
The data are not corrected for the primary beam response, whose FWHM is 59 arcsec.
The peak line intensity here is 77 mK.
The two plus signs in each panel are at the two nuclei. 
A black ellipse in the first panel shows the FWHM size of the synthesized beam.
}
\end{figure}

{\em N nucleus:}
Further inside and around the northern nucleus, our data show rotation within about 300 pc from the nucleus. 
This clearly appears as a butterfly-like pattern of isovelocity contours  
in the mean velocity maps of \HCOplus(4--3) (Fig. \ref{f.maps.nonCO}) and CO(3--2)  (Fig. \ref{f.maps.CO32nuc} b), 
the latter of which was made only with brighter circumnuclear emission.  
We fitted the velocity field to estimate the kinematical major axis to be at ${\rm p.a.} \approx 75\arcdeg$
and the disk inclination to be $ i \approx 30\arcdeg$ for a region with a 3\arcsec\ major axis diameter.
This kinematical major axis reasonably agrees with the morphological major axis of the circumnuclear high-intensity region
in CO, \HCOplus, and 0.86 mm continuum emission around the northern nucleus.
The kinematical major axis gradually changes its position angle in the sense
that it is smaller, about 60\arcdeg, at larger radii and is about 90\arcdeg\ closer to the nucleus.
This may be due to warp of the northern nuclear disk or non-circular motions of the gas in the disk.

{\em S nucleus:}
The southern nucleus has in its vicinity a velocity gradient in the east-west direction (${\rm p.a.} \approx 90\arcdeg$)
in the mean velocity maps.
The isovelocity contours, however, do not show a clear butterfly pattern there.
Also, the largest gradient of mean velocity is at about 0\farcs5 east of the southern radio nucleus (white plus sign).
It is in contrast to the peaks of line integrated intensity that are often at slightly west or northwest, by about 0\farcs3 -- 0\farcs5, 
from the nucleus (e.g., in CO, \thirteenCO, CN, but not in \HCOplus).
We are going to model in the following the near-linear feature running east-west across the S nucleus as 
a near edge-on circumnuclear disk of radius \about300 pc.
The lack of clear butterfly pattern around the southern nucleus is attributed to the edge-on viewing angle.

{\em Between the Nuclei:}
Conspicuously, the most redshifted CO(3--2) emission is located by about 2\arcsec\ south of the northern nucleus, 
as seen in the CO(3--2) mean velocity map in Fig. \ref{f.maps.CO102132}.  
This is due to the high-velocity wing of CO emission and is the reason for the very large line width at the same location in the 
CO(3--2) line-width map.
This feature does not show up in Fig. \ref{f.maps.CO32nuc}b because the wing emission is faint and below the cutoff used for the moment analysis.
The high-velocity emission is separately described in \S\ref{s.result.highV} along with the line width information
in Fig. \ref{f.maps.CO102132} and \ref{f.maps.nonCO}.

\subsubsection{Peak \Tb\ and Integrated Intensity}
The three \twelveCO\ lines have peak integrated intensities on the order of $2\times 10^3$ K \kms\ 
and maximum brightness temperatures of about 20 K, both at about 1\arcsec\ resolution.
The maxima are 22.4 K and 2730 K \kms\ in CO(3--2) at our highest spatial resolution
($0\farcs58\times0\farcs39 \approx 80$ pc).
The peaks of line emission are in the vicinity of the two nuclei, 
the spiral feature running between the two nuclei,
and in the linear feature across the western nucleus, particularly in its western side.
At least for these regions, our data do not show 
significant decline of (integrated) intensity in higher transitions that would suggest significantly subthermal excitation
or 
significant increase in (integrated) intensity arising from optically-thin emission from thermalized warm molecular gas.

Other lines are much weaker than the \twelveCO\ lines, having peak brightness temperatures at about 1 K or lower.
Possible reasons for this include that these lines are optically thin, 
have lower excitation temperatures than \twelveCO\ (i.e., subthermally excited), 
and are emitted from smaller regions than the \twelveCO\ lines.

\subsubsection{Line Flux, Gas Mass, and Surface Density }
\label{s.obs.line.flux}
The total flux of CO line emission is measured to be $1.0\times10^3$, $3.0\times10^3$, and $5.7\times10^3$ Jy \kms\ 
for J=1--0, 2--1, and 3--2 transitions, respectively,
in a 20\arcsec\ diameter aperture centered at the midpoint of the two nuclei.
The CO(1--0) flux in the concentric 40\arcsec\ diameter aperture is  $1.6\times10^3$ Jy \kms.
These are corrected for the primary beam (and mosaic) responses but not for any missing flux in the interferometric data.
The fluxes above are measured in data cubes with resolutions 
\about2\farcs7, \about0\farcs6, and \about0\farcs6 for CO(1--0), (2--1), and (3--2), respectively.
The CO(2--1) flux in the same 20\arcsec\ aperture is measured to be $4.0 \times10^3$ Jy \kms\ in a \about3\farcs0 resolution
data cube.

We note that the CO(2--1) to CO(1--0) flux ratio
at about 3\arcsec\ resolution, 4.0 with \about10\% calibration uncertainty, 
is what is expected for the thermalized optically thick gas at $\gtrsim 30$ K.
The two data sets have about the same ranges of baseline length in units of wavelength and hence 
the ratio must be affected little by missing flux.
The CO(3--2) to (2--1) ratio at about 0\farcs6 resolution is 1.9 in flux and 0.86 in brightness temperature.
It is fully compatible with thermalized optically thick CO at $\gtrsim30$ K  considering the calibration uncertainties and 
the probably larger missing flux in the CO(3--2) data.
On the whole, the data are consistent with the CO being thermalized at least up to $J = 3$ and optically thick.

We estimate the mass of molecular gas using the CO(1--0) to \HH\ mass
conversion factor $\Xco \equiv N_{\rm H_2}/I_{\rm CO(1-0)} = 1 \times 10^{20}$ \unitofX\ and
36\% mass contribution from He.
We do not have the true \Xco\ in NGC 3256 nor do we have a strong reason to believe that \Xco\ is constant across the galaxy.
Therefore we give our molecular gas masses with the parameter \Xtwenty\ that is \Xco\ in units of $1 \times 10^{20}$ \unitofX.
While \Xtwenty\ is unity for our assumed (i.e., fiducial) conversion factor, 
this parameterization allows our mass estimates to be easily rescaled when a more plausible value of \Xco\ is given.
The conversion factors estimated with various methods in galaxies at solar metallicities or higher are usually
in the range of \Xtwenty = 0.3 -- 3 with high values for `normal' galaxies such as our Galaxy in its disk 
and low values for luminous infrared galaxies \citep[see][for a review]{Bolatto13b}.
\citet{Bolatto13b} recommend \Xtwenty=0.4 with an uncertainty of 0.5 dex for  luminous starburst galaxies
and \citest{Sakamoto06} obtained a value within 10\% of it for the central 3 kpc of NGC 3256 after
averaging various estimates.
In this paper, however, we adopt the normalization with \Xtwenty=1 partly for simplicity 
and also because the gas to dynamical mass ratios
that we later calculate for the nuclei and for a larger area appear more reasonable with \Xtwenty=1.
In any case, we expect a factor of 3 uncertainty in our adopted \Xtwenty\ of 1.0 and therefore \Xtwenty=0.4 is within the uncertainty.
The conversion factor between CO(1--0) integrated intensity and molecular gas surface density is 
$\alphaco \equiv \Sigma_{\rm mol} / I_{\rm CO(1-0)} = 2.2 \Xtwenty$ \unitofalpha.

The molecular gas mass estimated from our CO(1--0) line flux is 
$\Mmol(r \leq 10\arcsec ) = 7\times10^{9} \Xtwenty\, \Msol$ in the central 20\arcsec\ diameter aperture
and
$\Mmol(r \leq 20\arcsec ) = 1\times10^{10} \Xtwenty\, \Msol$ for the central 40\arcsec\ (7 kpc).
The scaling parameter \Xtwenty\ should be read as the average value for each region in consideration. 
The peak molecular gas surface densities toward individual nuclei are 
$4\times 10^3 \Xtwenty $ and $3 \times 10^3 \Xtwenty$ \Msol \persquarepc\   
for the northern and southern nuclei, respectively,
at \about1\farcs4 (240 pc) resolution on the basis of the CO(1--0) data in Fig. \ref{f.maps.CO102132}.
The southern nucleus has the highest CO integrated intensity in the merger and its peak gas column density is
$\Sigma_{\rm mol}({\rm S}) = 6\times 10^3 \Xtwenty$ \Msol \persquarepc\  
in our \about0\farcs5 (80 pc) resolution CO(3--2) data in Fig. \ref{f.maps.CO102132}.
Here we do not correct for the different transition because of the CO excitation inferred above.
Converting this peak molecular gas column density to the peak hydrogen and proton column densities,
we obtain toward the southern nucleus
$\log (N_{\rm H, equiv.}/{\rm cm^{-2}})  = 23.9$
and 
$\log (N_{\rm p}/{\rm cm^{-2}})  = 23.8$.
The former converts \HH\ and He with hydrogen atoms of equivalent mass
and the latter gives proton column density.
Both have 0.5 dex uncertainty inherited from the uncertainty in \Xtwenty.

\subsection{High Velocity Emission}
\label{s.result.highV}
We detected wide faint line wings in our data, most clearly in CO(1--0) and (3--2) and also in CN(1--0).
The new sensitive ALMA data not only confirm the previous detection of \citest{Sakamoto06} 
but also better constrain  the velocity extent and spatial distribution of the high velocity gas.

\subsubsection{Channel Maps}
\label{s.result.highV.channelMaps}
Figures \ref{f.HVchans.CO10.tp.VHV} and \ref{f.HVchans.CN10.tp.VHV} 
show our CO(1--0) and CN(1--0) channel maps, respectively, 
made with \uv\ tapering (i.e., spatial smoothing) and wide channel widths to better detect high-velocity emission.
These Band 3 images use the CSV and Cycle 0 data combined to maximize sensitivity. 
Continuum was determined more than 750 (400) \kms\ away from \Vsys\ for the CO (CN) lines
and has already been subtracted from each data cube.
The CO(1--0) data show $>3 \sigma$ emission from $-650$ \kms\ to $+650$ \kms\ around the northern nucleus,
up to about 500 \kms\ from \Vsys\ between the two nuclei, 
and up to \about$\Vsys \pm400$ \kms\ around the southern nucleus.
At the offset from \Vsys\ of about 300 \kms, redshifted emission is stronger than the blueshifted and the former
peaks between the two nuclei.
The same is observed in CN(1--0) and was also the case in the CO(2--1) observations of \citest{Sakamoto06} in which
the wing emission was first found up to $\Vsys \pm 300$ \kms.  

Figure \ref{f.HVchans.CO32.RedBlueGray} shows CO(3--2) channel maps displaying blueshifted and redshifted emission
on the same panel for the same absolute offset from \Vsys. The background image in gray scale is continuum.
The upper panels (Fig. \ref{f.HVchans.CO32.RedBlueGray}a) are our 1\farcs1 resolution data.
Emission stronger than $4\sigma$ is detected up to 450 \kms\ from \Vsys\ in both blueshifted and redshifted velocities.
The northern nucleus has emission up to this largest offset velocity
and the centroid of the blueshifted emission is on the northwestern side of the nucleus 
while the redshifted emission centroid is on the southeastern side.
Around the southern nucleus, redshifted and blueshifted emission are roughly symmetrical about the nucleus, redshifted
to the north and blueshifted to the south, except for the redshifted emission extending east from the southern nucleus
at the leftmost channel.
It is also notable that emission more than about 300 \kms\ from \Vsys\ is clearly detached from the southern nucleus
unlike the high velocity emission around the northern nucleus.
The lower panels (Fig. \ref{f.HVchans.CO32.RedBlueGray}b) are our 0\farcs5 resolution channel maps for the high velocity emission.
They more clearly show the symmetry around the southern nucleus.
Notable new observations here are that the high velocity gas has clumps in the extended structures
and 
that the blueshifted gas slightly curves toward west at larger distances from the southern nucleus.
The highest velocity emissions are again clearly detached, by about 1\farcs8 (310 pc), from the southern nucleus. 
Little CO emission is detected around the northern nucleus in these higher resolution data 
indicating that the high velocity CO(3--2) emission around the
northern nucleus is more extended than that around the southern nucleus.
The extent of the high-velocity blueshifted gas is larger around the northern nucleus than around the southern nucleus 
also in CO(1--0) as seen in Fig. \ref{f.HV.CO10.RedBlues}a.

\subsubsection{High-Velocity Line Flux}
\label{s.result.line.hv.flux}
The flux of the high velocity CO emission has been measured by only integrating the high velocity channels.
Fig. \ref{f.HV.CO10.RedBlues} shows CO(1--0) maps integrating about 530 \kms-wide ranges 
offset by about 220--750 \kms\ from our fiducial velocity (\Vsys) 2775 \kms. 
The CO(1--0) flux in our 2\farcs7 resolution data (Fig. \ref{f.HV.CO10.RedBlues}b) is 
8.9, 3.2, and 1.2 Jy \kms\ 
for the redshifted emission,
blueshifted emission associated with the northern nucleus, and 
blueshifted emission associated with the southern nucleus,
respectively. 
The response of the primary beam was corrected for these measurements and the blueshifted emission about 15\arcsec\ east of 
the nuclei excluded as it is associated with an arm there and is detected only down to about $\Vsys -300$ \kms.
In total this high velocity emission has 1.3\% of the total CO(1--0) flux detected in the central 20\arcsec\ diameter aperture given in
\S \ref{s.obs.line.flux}.
The flux of CO(3--2) emission integrated in the same velocity ranges in our 0\farcs6 resolution data are
40, 11, and 21 Jy \kms\ 
for the redshifted emission, 
blueshifted emission associated with the northern nucleus, and 
blueshifted emission associated with the southern nucleus, 
respectively.
In total this high velocity emission has 1.2\% of the total CO(3--2) flux in the central 20\arcsec\ given in \S \ref{s.obs.line.flux}.

\subsubsection{High-Velocity Gas Mass}
\label{s.result.line.hv.mass}
The total mass of the high-velocity molecular gas is calculated to be
$\Mmol(223\, \kms \leq |V-\Vsys| \leq 752\, \kms) = 8.8\times10^7 \Xtwenty\, \Msol$   
from the high-velocity CO(1--0) line flux. 
The mass of high-velocity molecular gas associated with each nucleus is estimated to be 
$6.3\times10^7 \Xtwenty\, \Msol$ for the northern nucleus and  
$2.5\times10^7 \Xtwenty\, \Msol$ for the southern nucleus         
under the assumption 
that the redshifted high-velocity gas is composed of gas associated with the two nuclei with the same fractions
as in the blueshifted high-velocity gas (i.e., 72\% to the north and 28\% to the south.)
We use the ratio in our CO(1--0) data and not the ratio of ${\rm N}:\rm{S}=34:66$ in our CO(3--2) data.
This is  because the former data suffer less from missing flux.
Also CO(1--0) is less affected by any difficulty in CO excitation in the high velocity gas.

We assume, unless otherwise noted, 
that the \Xtwenty\ value is unity for the high-velocity gas as we did for the bulk CO emission of NGC 3256.
This is partly motivated by our observation
that the fraction of the high-velocity flux with respect to the total flux is almost the same in CO(1--0) and CO(3--2).
This can be, though not uniquely so, 
because the physical properties of the high-velocity gas and those of the gas at lower velocities are
not drastically different from each other.
Our choice is also because we have insufficient information to specify a different value.

A possible alternative choice of \Xco\ for the high-velocity gas, which we suggest below to be high-velocity outflows,
is the one for optically thin CO emission.
This is possible because the peak CO brightness temperature of the high-velocity emission is only
on the order of 0.5 K for CO(1--0) and 1.5 K for CO(3--2) in Figs. \ref{f.chans.CO10.br} and \ref{f.chans.CO32.br}.
In beam-matched data of 1\farcs6 $\times$ 1\farcs2 resolution, 
the CO(3--2) to CO(1--0) ratios of peak brightness temperatures around $\Vsys \pm 200$ \kms\ are mostly $\sim1 \pm0.5$ 
for the high-velocity emission associated with the southern nucleus. 
Taken at face value, i.e., assuming little effect of CO(3--2) missing flux to this ratio because the high-velocity gas
around the southern nucleus is relatively compact, the ratio can be not only due to optically thick emission from thermalized CO
but also due to optically thin CO emission.
In the latter case, the ratio corresponds to the excitation temperature of $12 \pm 3$ K in LTE. 
If the CO excitation is not in LTE then CO(1--0) can have a higher excitation temperature than this but
the CO(3--2) excitation temperature must be much lower than that.
The conversion factor for optically thin CO(1--0) emission is on the order of $\Xtwenty = 0.1$ 
in both of these LTE and non-LTE cases for a CO abundance of $[{\rm CO}/\HH]=10^{-4}$.
The non-LTE conversion factor for optically thin CO(1--0) depends little on gas temperatures above \about15 K 
provided that CO molecules are well excited only to J=2 but not to 3 and beyond.  
The CO level population is determined by the level statistical weights in such a case.
As the observed line ratio is consistent with multiple gas conditions 
we keep in mind that the conversion factor for the high-velocity gas can be an order of magnitude lower 
than our fiducial value of unity.

\subsubsection{Spectra}
\label{s.result.line.hv.spectra}
Figure~\ref{f.spectra.nuclei} shows spectra at the two nuclei.
Each line is fitted with a Gaussian to help highlight the high-velocity wings,
i.e., emission in excess of the Gaussian fit at large offset velocities.
At both nuclei, line centroids are within about 10 \kms\ from our fiducial velocity of 2775 \kms\ 
and line widths are about 150--200 \kms\ in FWHM.

The CO(1--0) data have the highest signal-to-noise ratio and show a clear redshifted wing at \about3\% level
in the spectrum toward the northern nucleus. There are also high-velocity wings at the level of 1\% of the peak
or less in both redshifted and blueshifted velocities.
The full width at zero intensity of the emission is about 1600 \kms\ toward the northern nucleus.
The southern nucleus also has blue and red-shifted wings visible at the level of  1--2\% of the main line; the 
red wing is again stronger.
The fraction of the wing component to the main line is probably larger when the wing features are interpolated 
to the systemic velocity.
The full width at zero intensity (FWZI) for the southern nucleus is about 1200 \kms.
Our observations in the spectra are consistent with what we saw above in channel maps (Fig. \ref{f.HVchans.CO10.tp.VHV}) in that
the full line widths exceed 1000 \kms, the line is wider toward the northern nucleus, and the high-velocity emission
is stronger in redshifted velocities at around $\left| V-\Vsys \right| = 300$ \kms.

The CO(3--2) spectra also show the high-velocity wings.
While more emission is in the redshifted wing in the aperture containing the northern nucleus,
fainter and broader wings than this are seen in both blue and redshifted velocities toward both nuclei.
The full width of the CO(3--2) line is about 1000 \kms\  (i.e., $\pm500$ \kms) in our data,  consistent with
our observation in channel maps (Fig.~\ref{f.HVchans.CO32.RedBlueGray}a).
Line full width depends on sensitivity because noise can mask faint and wide high-velocity emission.
The smaller full line width in CO(3--2) than in CO(1--0) must be partly due to the lower signal-to-noise ratio (S/N)
in the former data.

In \HCOplus(4--3) we did not detect high-velocity emission. 
This may be mostly because \HCOplus(4--3) has the lowest S/N among the lines shown in Fig. ~\ref{f.spectra.nuclei}
and also because the J=4 excitation of \HCOplus\ has a high critical density of $10^7$ \percubiccm.
The \HCOplus\ line profiles have double peaks (or a dip near the line center) on both nuclei.
This is also seen in CO(3--2) toward the southern nucleus with a smaller 2\arcsec\ diameter aperture.

\subsubsection{CN in the High-Velocity Gas}
\label{s.result.line.hv.CN}
The CN(1--0) spectra in Fig.~\ref{f.spectra.nuclei} show the redshifted wing at about the same level as in the CO(1--0) data.
Although each of the CN lines consists of a group of hyperfine lines, the redshifted wing is not due to the line distribution
because if it were then the redshifted emission in Fig.~\ref{f.HVchans.CN10.tp.VHV} should have peaked on the nuclei
and not between them.
Thus both CN and CO red wings are probably from the same high-velocity gas.
The flux ratio of the two CN lines is 1.8 on both nuclei, 
only slightly less than the ratio of 2 from optically thin lines \citep{Turner75}. 
The CN emission is therefore mostly optically thin; the opacity of the brighter line is calculated to be 0.4 from
$(1-e^{-\tau})/(1-e^{-\tau/2}) = 1.8$. 
Under a safe assumption that the low-velocity CO(1--0) with peak \Tb\ $\gtrsim$ 10 K
is optically thick and has a higher optical depth than high-velocity CO emission, the fraction of the high-velocity emission
to the main low-velocity component should be larger in CN than in CO after the CO opacity correction. 
This suggests enhanced CN abundance or excitation in the high velocity gas.
If the CN enhancement is due solely to collisional excitation then the high-velocity gas is denser than the low-velocity gas
because the critical density for CN(1--0) is $10^6$ \percubiccm\ and is $10^3$ times higher than that for CO(1--0).
This CN detection as well as enhancement in the high-velocity gas is noteworthy because the line has not been detected 
in galactic molecular outflows before.

\subsubsection{Robustness of the Detection}
We regard our detection of these high-velocity emission components as robust for the following reasons.
Firstly, 
the faint and broad wing emission cannot be errors in continuum subtraction,
because continuum in each channel is only at the levels of 30$\sigma$ and 20$\sigma$ in the CO(1--0) and (3--2) data, 
respectively, while our passband calibration is much  more accurate than 1/30 = 3\%
as seen in the flatness of our spectra sufficiently away from the line in Fig. \ref{f.spectra.nuclei}.
Moreover, much of the high-velocity emission peaks slightly offset from the nuclei where any passband error
could make the largest artifact.
Secondly, 
it is unlikely that the high-velocity emission is due to line-blending, i.e., from lines other than the target line,
in part because of the offset of the high-velocity emissions from the nuclei where all lines peak 
and also because of the lack of molecules that may plausibly contribute to the observed emission.
Individual line wings of a single CO transition sometimes have possible alternative sources, 
such as \HCthreeN(38--37) and \HthirteenCN(4--3) on the red (i.e., low-frequency) side of CO(3--2). 
However, these molecules cannot explain the redshifted emission of CO(2--1) or CO(1--0) 
because their lower transitions are not adjacent to these CO transitions.
In addition, the peak of the emission on the red side of CO(3--2) is not exactly 
at the redshifted frequencies of \HCthreeN(38--37) and \HthirteenCN(4--3).
Fig. \ref{f.spec.noblend} shows this in the spectrum sampled at the midpoint of the two nuclei. 
The peak of the redshifted component clearly does not coincide with the expected frequencies of \HCthreeN(38--37) and \HthirteenCN(4--3).
Therefore their contribution to the high-velocity emission should be small, if any.
Finally,
the line wings are unlikely due to the response pattern of the spectral correlator to a strong narrow line,
as this would appear symmetric about the line center.

\subsubsection{Position-Velocity Diagrams}
Figure \ref{f.COpv} shows CO position-velocity diagrams across the nuclei.
 The upper panels are for CO(1--0) and the lower for CO(3--2).
The three columns are, from left to right, 
p.a.=270\arcdeg\ cuts through the N nucleus, 
p.a.=270\arcdeg\ cuts through the S nucleus,
and
p.a.=0\arcdeg\ cuts through the midpoint of the N and S nuclei.
The position angle for the northern nucleus is along the kinematical major axis at the center of the northern circumnuclear disk.
The p.a. for the southern nucleus is because there is a structure extending across the nucleus in p.a.$\approx$ 90\arcdeg. 
The PV diagrams along p.a.=0\arcdeg\ are for the high velocity emission that showed symmetrical velocity structure 
around the southern nucleus approximately along this axis.  
Rotation of the circumnuclear disk  is evident around the northern nucleus in the panels (a) and (d). 
The high-velocity emission at the northern nucleus is also clear in these PV diagrams.
Gas motion around the southern nucleus is more complex, in particular in the CO(3--2) data in panel (e), but
an overall velocity gradient within about 5\arcsec\ from the nucleus and presence of high velocity gas at the nucleus are
consistent with what we see in the channel maps.

The most interesting of the PV diagrams are the cuts in the north-south direction across the two nuclei, panels (c) and (f).
There we clearly see two components of high velocity gas. 
One is on the northern nucleus and shows little positional shift with velocity.
The other is symmetric about the southern nucleus, blueshifted to the south (left in the plot) and redshifted to the north
within about 4\arcsec\ from the southern nucleus. 
The terminal velocity increases with the distance from the southern nucleus up to about the offset of 2\arcsec.
It is also notable that the range of emission velocities at each position is large, about 500 \kms, across this region of
a north-south velocity gradient.

\subsection{Comparison with Other Observations}
\label{s.result.comparison}
\subsubsection{HST Optical Images}
\label{s.result.comparison.HST}
Figures \ref{f.hst.L} and \ref{f.hst.M} compare our CO images with multi-color HST images of NGC 3256.
The merger has many dark lanes, particularly  in its southern part, as shown in Fig.~\ref{f.hst.L}(a).
Comparison of Fig.~\ref{f.hst.L}(a) with the color excess image in  Fig.~\ref{f.hst.L}(b) shows that 
the dark lanes are generally redder in color than their adjacent areas.
This suggests that the dark lanes are due to higher dust extinction.
As seen in Fig.~\ref{f.hst.L}(c), there is overall match between these dark lanes in the optical and the CO(1--0) distribution
shown in Fig.~\ref{f.hst.L}(d).
This is what is expected when the dark lanes are due to obscuration by the interstellar dust.
In addition, there is an interesting match between the dark lanes (i.e., optical color excess) and the CO line widths
as shown in Fig.~\ref{f.hst.L}(b). 
Both are enhanced in a roughly triangular area on the south-western side of the binary nucleus.  
The similarities between dark (dust) lanes and CO emission are also seen in our
higher resolution CO(3--2) data in Fig. \ref{f.hst.M} (c).
At this higher resolution, however, it becomes evident that the optical color excess (i.e., reddening) and the
CO integrated intensity are not strictly proportional. 
Also the matching is poor between the high color excess regions and regions of large line widths in
the vicinity of the two nuclei (see Fig. \ref{f.maps.CO102132} for our CO(3--2) 2nd moment map), 
as was already the case in our CO(1--0) comparison in Fig.~\ref{f.hst.L}(b).

\subsubsection{VLA Radio Continuum Images}
\label{s.result.comparison.VLA}
%
There are remarkable correlations between our ALMA data and VLA radio continuum data in \citet{Neff03}.
The spatial distribution of  6 and 3.6 cm continuum in their Fig.~1 matches quite well with 
that of the sub/millimeter continuum shown in our Fig. \ref{f.contmaps}. 
The agreement includes not only the two nuclei and the overall shape of the diffuse emission
but also a short arc (arm) about 5\arcsec\ northeast of the northern nucleus, 
a spot about 20\arcsec\ west of the northern nucleus, 
the bridge-like arm emanating from the northern nuclear disk to south,
and
a linear feature across the southern nucleus.
The 3.6 cm image also shows a faint spur that emanates from the southern nucleus to south and slightly curves toward west. 
It has a counterpart in our CO data. 
The blueshifted emission in the $|V - \Vsys| = 200$ \kms\ channel of Fig. \ref{f.HVchans.CO32.RedBlueGray} (b)
coincides with the radio spur.

Figure \ref{f.vla-almaHV} compares a higher resolution 3.6 cm image, in Fig. 2 of \citet{Neff03}, with our ALMA data.
The 3.6 cm continuum in black contours and 860 \micron\ continuum in gray scale again show very good correlation.
In the radio emission there is a pair of narrow spurs that emanate from the southern nucleus to north and south; 
the one to the south is probably a part of the spur mentioned above.
Although we did not detect this feature in submillimeter continuum, our CO data have their counterparts.
The highest velocity CO(3--2) emission shown in red and blue contours are at the tips of these
radio continuum spurs. 
We are going to discuss these observations in \S \ref{s.outflow.southern.driver}.

\subsubsection{Spitzer Infrared Images}
\label{s.result.comparison.Spitzer}
Figure.~\ref{f.spitzer} shows archival infrared images of NGC 3256 
taken with the Spitzer Space Telescope Infrared Array Camera (Program ID. 32).
We show in each panel the same area as in Fig.~\ref{f.contmaps} (a) for our 2.8 mm continuum 
and use the same linear intensity scale. The infrared images have 1\arcsec--2\arcsec\ resolutions.
Similarities between the infrared and millimeter continuum distributions are 
not only the two bright nuclei but extended features around them
including the spiral arm to the north of the nuclei, 
two bright areas about 5\arcsec\ and 20\arcsec\ east of the northern nucleus, 
and a linear feature that protrudes west from the central region by about 15\arcsec\ at about the latitude of the southern nucleus.
The northern nucleus is brighter than the southern nucleus in the Spitzer images
(except at 4.5 \micron\ not shown here) and even more so at 11.5 \micron\ \citep{Lira08}.
This is also the case in millimeter emission.  
The northern nucleus has comparable or more integrated flux density than the southern nucleus in 1\arcsec--3\arcsec\ apertures
(see Table \ref{t.contFluxSpix}), although the southern nucleus is more compact and has comparable 
or higher peak brightness than the northern nucleus at $\lesssim2\arcsec$ resolutions.
It is very likely that the northern nucleus has larger flux densities also between 11.5 \micron\ and 860 \micron\
and hence a larger bolometric luminosity than the southern nucleus.

\section{Merger Configuration}
\label{s.configuration}
We suggest the merger configuration in Fig. \ref{f.illust} for the reasons given in this section.
There are two nuclei as in the model in \citest{Sakamoto06}. 
Their identification as the nuclei of two merging galaxies is strongly supported 
by the peaks of line and continuum emission at the two dominant radio sources 
and by our detection of large velocity gradients there
(Figs.~\ref{f.contmaps} and \ref{f.maps.CO32nuc}).
Parameters estimated in this section are summarized in Table \ref{t.4418measured.param}.

\subsection{NGC 3256N}
The northern nucleus has a nuclear gas disk that has a low inclination and nearly circular rotation,
showing a clear butterfly pattern in the velocity field (Fig. \ref{f.maps.CO32nuc}). 
We measured in \S \ref{s.result.line.velocity_field} that the disk major axis is at ${\rm p.a.(N)} \approx 75\arcdeg$
and inclination is $i_{\rm N} \approx 30\arcdeg$.
The molecular spiral arms around the northern nucleus, those shown in gray in Fig. \ref{f.illust}, 
must be also nearly face-on based on their morphology. 
Since they emanate from the northern nuclear disk or its vicinity the arms most likely belong to the northern galaxy
and are coplanar with the northern nuclear disk.
The near side of the northern nuclear disk is then its southeastern side assuming 
that the molecular spiral arms are trailing.

\subsection{NGC 3256S}
\label{s.configuration.n3256s}
The southern nucleus must be in front of the northern galaxy disk and nearly edge-on.
This deeply obscured nucleus cannot be much behind the northern galaxy disk because if so we should have seen 
the foreground northern galaxy disk at its location.
The extinction peak toward the southern nucleus must be due mostly to the southern galaxy itself, because even if
the southern nucleus were slightly behind the northern galaxy disk its surface gas density, hence extinction, does not peak
at the radius of the southern nucleus. 
A nearly edge-on configuration is therefore suggested for the obscured southern nucleus and the southern galaxy.
The high inclination is supported by the shape of the region that has both large optical extinction (reddening) 
and large CO line width in Fig.~\ref{f.hst.L}(b).
This region is extended in the east-west direction across the southern nucleus
as expected when the foreground (part of the) southern galaxy has 
a high inclination and a major axis at ${\rm p.a.(S)} \approx 90\arcdeg$.
Such a configuration also explains the distribution of large line widths 
as due to the overlap of the two galaxies 
and also due to the nearly edge-on geometry of the southern disk. 
This argument disfavors the possibility that the southern nucleus is on or slightly behind the northern disk 
because we do not see any high velocity-dispersion region with little reddening 
(i.e., gas behind the northern galaxy disk) in the central few kpc of the southern galaxy. 
Further outskirts of the southern galaxy appear already strongly disturbed and leaving their original orbital plane
judging from the large scale distribution of color excess in Fig.~\ref{f.hst.L}(b).
Closer to the center, there is a bar-like distribution of molecular gas and dust 
across the southern nucleus (Fig. \ref{f.maps.CO32nuc}, HCN in Fig. \ref{f.maps.nonCO}, and Figs. \ref{f.contmaps} b and c ). 
This is the edge-on southern nuclear disk in our interpretation.
The lack of clear butterfly pattern in its velocity field (Fig. \ref{f.maps.CO32nuc}) is consistent with
the proposed large inclination.
For the reasons given in \S \ref{s.outflow.southern}, the near side of the nearly edge-on southern nuclear disk 
must be its northern side and the disk inclination is constrained to be 
$70\arcdeg < i_{\rm S} \lesssim 85\arcdeg$.

\subsection{Mass Ratio}
\label{s.configuration.massRatio}
The northern nucleus is probably a few times more massive than the southern nucleus.
The ratio of the CO(3--2) line width (FWHM) at the northern nucleus to that at the southern nucleus is 0.77
for the 4\arcsec\  aperture used in Fig. \ref{f.spectra.nuclei}.
This ratio, after correction for the inclinations, reflects the mass ratio of the nuclei at 0.7 kpc scale
because the broad emission wings at the nuclei are too faint to affect FWHM.
For $i_{\rm N} \approx 30\arcdeg$ and $i_{\rm S} \approx 80\arcdeg$ and ignoring the effect of any difference in gas radial distributions, 
the mass ratio $M_{\rm N}/M_{\rm S}$ is 2.3 for the line width ratio of 0.77;
the mass ratio is between 2.1 and 2.4 for $70\arcdeg < i_{\rm S} < 90\arcdeg$.
The ratio of line FWHM appears to increase to about 1 when the sampling area increases; the mass ratio
would be 3.9 for the FWHM ratio of 1 and $i_{\rm S}$ of 80\arcdeg. 
This trend can be due to different degrees of mass concentration between the two nuclei 
but can be also due to more contamination to the southern nucleus from the northern galaxy disk.
With these uncertainties in mind
we suggest $\Mdyn({\rm N})/\Mdyn({\rm S}) \sim 2.5$ with an error up to $\pm 1$ for 1 kpc diameters.

\subsection{Merger Orbit}
The orbital plane of the two nuclei is probably close to the disk plane of the northern galaxy.
In other words, not only must the southern nucleus be in front of the northern galaxy as argued above
it is probably near the northern galaxy plane.
This is deduced from two observations. 
One is that the most prominent molecular arm emanating from the northern nuclear disk extends in the direction of the
southern nucleus as if bridging the two nuclei.
The other is that other molecular arms at larger radii are around the two nuclei; 
the most notable is the arm starting from the northern nuclear disk and running east of the binary nuclei 
by almost 180\arcdeg\ in our CO(3--2) map in Fig. \ref{f.maps.CO102132}. 
These are expected to be so if the southern galaxy has been close to the disk plane of the northern galaxy and exerting
its gravitational force to the disk gas in the direction nearly within the disk plane.  
Because the northern galaxy was estimated to be nearly face-on the merger orbital plane is also close to face-on.

The southern galaxy has a high inclination angle with respect to the orbital plane in the configuration we suggested above.
This high inclination is consistent with much of the gas in the outer disk of the southern galaxy leaving its original galactic plane 
because for the southern galaxy the perturber is on a nearly polar orbit.
Direct contact of the gas in the two disks is another plausible reason for the disturbance although this works for both disks.
Fig. \ref{f.hst.L} (b) and (c) show a one-arm reddening and CO feature that starts at about 30\arcsec\ east of the
two nuclei and spirals into the southern nucleus after a 270\arcdeg\ clockwise turn. 
This may well be material stripped from the southern galaxy tracing its past trajectory 
around the center of mass near the northern nucleus.
If this is the case, the southern nucleus must be currently moving from west to east (right to left on our maps).

The right panel of Fig. \ref{f.illust} shows the two nuclear disks viewed from above the merger orbital plane.
As was in the sky-projection in  Fig. \ref{f.illust} (left), the northern nuclear disk is close to face-on and the southern nuclear disk 
is close to edge-on because the orbital plane is estimated to have a low inclination ($\lesssim$30\arcdeg) with respect to our sight line.
However, the sense of rotation of the southern nuclear disk is opposite between the sky-projection and orbital-plane-projection 
in our model.
It is certainly possible in our model, with the southern nucleus in front of the northern disk, 
that both of the two nuclear disks have prograde rotation with respect to the orbital motion of the two nuclei.
Whether this is not only possible but is indeed so is not certain from our argument above, but the prograde-prograde
configuration has been suggested to explain the long tidal tails seen in the optical and \ion{H}{1}  \citep{Toomre72,English03}.

\subsection{Merged Gas Disk}
The gas presumably stripped from the southern galaxy and the gas from the northern galaxy appear to be 
forming, from larger radii, a merged gas disk that is connected to the northern galaxy disk.
The overall CO(1--0) velocity field in Fig. \ref{f.maps.CO102132} is largely consistent with that of the northern nuclear disk
regarding the kinematical major axis and an apparently low inclination.
The stripped gas that we inferred above from the color index image does not stand out in our CO mean velocity field.
In our proposed configuration, this is mainly because the gas on the large-scale is settling to the merger orbital plane 
that is close to the plane of the northern galaxy. 
Such a merged gas disk was proposed in \citest{Sakamoto06}.
It is expected to form because gas is not collisionless unlike stars and hence cannot remain on the original disks 
at the larger radii where the two disks have already collided with each other.
Although the small visible perturbation in the observed CO velocity field can be also because the northern galaxy, 
whose nucleus we found to be more massive than the southern one,
had a dominant fraction of gas in the system, the presence of two \ion{H}{1} tidal tails makes it unlikely that the large scale gas
disk is only from the northern galaxy.

\subsection{Comparison with Other Estimates}
The configuration suggested above is consistent with what 
\citet{English03} estimated from their \ion{H}{1} imaging of the merger.
They suggested on the basis of the wide \ion{H}{1} tidal tails that the merger orbital plane is almost face-on.
They further attributed the different shapes of the two tails to different inclinations of the progenitor galaxy disks
with respect to the orbital plane.
The spin of each galaxy was estimated to be prograde with respect to the binary orbital motion as mentioned above.
The consistency of these estimates by \citet{English03} from \ion{H}{1} observations 
and ours from molecular gas and optical data adds credence to our model in Fig. \ref{f.illust}.
In addition, \citet{Trancho07} made a notable observation from their optical spectroscopy 
of young star clusters that while the majority of the clusters follow the rotation of the
main (i.e., northern) gas disk some clusters about 20\arcsec\ west of the southern nucleus do not. 
They deduced that the former belong to the northern galaxy and the latter either belong to the
other (i.e., southern) galaxy or may have formed in tidal-tail gas falling back to the system.
The locations of the out-of-rotation clusters are consistent with them belonging to the nearly edge-on southern galaxy.
The dominance of the northern galaxy over the southern one in the number and motion of the clusters
is consistent with the presumably larger mass of the northern progenitor.

\subsection{\Mdyn\ and Gas-to-Dynamical Mass Ratios}
\label{s.configuration.Mdyn_MgasMdyn}
We here calculate the dynamical masses of 
the two nuclear disks and an area encompassing the two nuclei
and compare them to our gas mass estimates to see whether the gas masses are reasonable.
We estimate the dynamical mass of the northern nucleus to be on the order of
$\Mdyn(r_{\rm N} \leq {\rm 200\, pc}) \sim 4\times10^9$ \Msol\ 
using the line-of-sight rotational velocity of 150 \kms\ inferred from the CO position-velocity plot (Fig \ref{f.COpv} d) 
and $i_{\rm N} \approx 30\arcdeg$ measured above. 
We also estimate the dynamical mass of the southern nucleus to be $\Mdyn(r_{\rm S} \leq {\rm 200\, pc}) \sim 2\times10^9$ \Msol\ 
using the line-of-sight rotational velocity of 200 \kms\ inferred from the CO position-velocity plot (Fig \ref{f.COpv} e) 
and $i_{\rm S}$ of 80\arcdeg.
The ratio between the two dynamical masses within 400 pc diameters is 2.2, 
consistent with the ratio of $2.5\pm1$ in 1 kpc diameters estimated in \S \ref{s.configuration.massRatio}.
These dynamical masses have large uncertainties 
because we cannot accurately measure the rotational terminal velocity of each nuclear disk in the PV diagrams
contaminated by the faint and broad line wings that we attribute to outflow in the next section.
We also crudely estimate the dynamical mass in the central 20\arcsec\ of the merger to be
$\Mdyn(r \leq {\rm 1.7\, kpc}) \sim 6\times10^{10}$ \Msol\ from a rotational line-of-sight velocity of about 200 \kms\ inferred from the
CO(1--0) channel maps (Fig. \ref{f.chans.CO10.br}) and an inclination of 30\arcdeg. 
Although the high-velocity emission near the nuclei does not contaminate rotation at this large scale 
this estimate still has a large uncertainty due to the assumed inclination and the possibility 
that some gas at this radius may not be on a merged disk.

The molecular gas masses for the same regions are estimated from CO(1--0) to be 
$\Mmol(r_{\rm N} \leq {\rm 200\, pc}) \sim 3\times10^8$ \Msol,
$\Mmol(r_{\rm S} \leq {\rm 200\, pc}) \sim 2\times10^8$ \Msol\,
and
$\Mmol(r \leq {\rm 1.7\, kpc}) \sim 6\times10^9$ \Msol\
 for $X_{20} = 1$.
 The gas to dynamical mass ratios are therefore about 6\%, 12\%, and 9\% for the northern nuclear disk, southern nuclear disk, and
 the the merger in its central 3.4 kpc.
 These ratios inherit the uncertainties of the adopted \Xco\ and any of its spatial variation and
 any error in the dynamical masses.
 The reasonable gas mass fractions on the order of 10\%, however, suggest 
 that the gas masses above are probably not very wrong.
 The 0.5 dex uncertainty for the adopted $X_{20}=1$ seems reasonable for the bulk (though not all)
 of molecular gas in the observed region.

\section{Two Outflows}
\label{s.twoOutflows}
We argue from our observations of high-velocity molecular emission 
(in particular Figs. \ref{f.HVchans.CO10.tp.VHV}, \ref{f.HVchans.CO32.RedBlueGray}, and \ref{f.HV.CO10.RedBlues})
that each of the two nuclei has its own bipolar molecular outflow.
In our model illustrated in Fig. \ref{f.illust},
activities in the northern nucleus and its low-inclination nuclear gas disk 
are driving a bipolar outflow with a wide opening angle in the direction perpendicular to the northern nuclear disk.
This causes the high-velocity molecular line emission observed around the northern nucleus.
The southern nucleus drives a more collimated bipolar outflow perpendicular to the southern nuclear disk, i.e.,
in the north-south direction on the sky. 
The high velocity CO emission along ${\rm p.a.} \sim 0\arcdeg$ and 180\arcdeg\ are due to this outflow.
The redshifted gas of the two outflows overlap on the sky between the two nuclei
causing the peak of redshifted high velocity CO found in \citest{Sakamoto06}.
Outflow parameters derived in this section are summarized in Table~\ref{t.4418measured.param}.

Before elaborating on the two outflows we briefly mention two conceivable alternatives 
for the southern outflow and why we do not favor them.
An alternative interpretation of the gas motion around the southern nucleus is that the north--south velocity gradient is
due to rotation around the nucleus. 
If so the projected rotation axis of this hypothetical southern nuclear disk is along ${\rm p.a. } \approx 90\arcdeg$.
Then the continuum and line emission features along this p.a. ($\approx$ 90\arcdeg),  
e.g., in Figs. \ref{f.contmaps} (b), (c), \ref{f.maps.CO32nuc} (a), and the leftmost channels in Fig. \ref{f.HVchans.CO32.RedBlueGray},
would be polar structures, plausibly a bipolar outflow.
The optical color-excess region across the southern nucleus would also be a polar structure for the southern galaxy.
This model is not favored because it makes the bipolar structures much larger than the base nuclear disk.
Another alternative interpretation of the high-velocity gas around the southern nucleus is that it may be
a merger-driven tidal feature rather than a bipolar outflow. 
The tidal force exerted on the southern nucleus by the northern galaxy is along the north-south direction, i.e., the major axis
direction of the high-velocity gas.
We note, however, that the blueshifted high-velocity gas comes out almost directly from the southern nucleus 
in Fig.~\ref{f.HVchans.CO32.RedBlueGray} (b).
If the tidal force were strong enough to strip gas in the nucleus from such a small radius
then the gas elongated in the east-west direction across the southern nucleus (Fig.  \ref{f.maps.CO32nuc} a) would not be there.
Also, we estimated in the previous section that the merger orbital plane is close to face on.
Since the tidal force vector is along the orbital plane the force cannot give large line-of-sight velocities to the tidally stripped gas.
We therefore regard this alternative as equally unlikely.

\subsection{Northern Outflow: Uncollimated Bipolar Wind}
\label{s.outflow.northern}
\subsubsection{Evidence, Geometry, Driving Mechanism}
The following observations in \S \ref{s.result.highV}
are the pieces of evidence for a bipolar outflow with a wide opening angle from the northern nuclear disk.
CO(1--0) emission is detected ($\gtrsim 4\sigma$) around this nucleus up to $|\Delta V| = 650$ \kms\ from systemic   
in Fig.~\ref{f.HVchans.CO10.tp.VHV} and the full extent of the line to zero intensity is about 1600 \kms\ (Fig. \ref{f.spectra.nuclei}).
The high velocity gas in CO(3--2) is detected on the northern nuclear disk 
with its blueshifted emission slightly shifted to northwest and its redshifted counterpart biased toward southeast 
in Fig.~\ref{f.HVchans.CO32.RedBlueGray} (a).
These spatial shifts of blueshifted and redshifted high-velocity gas are 
along the minor axis of the northern nuclear disk 
and the shift of the blueshifted gas is toward the far-side of the nuclear disk. 
These observations are consistent with the high-velocity emission being an outflow from the nucleus in the direction
perpendicular to the northern nuclear disk  (see Fig.  \ref{f.illust}).

Since the northern nuclear disk is nearly face-on the outflow axis is close to our line of sight.
This pole-on viewing angle is consistent with the small spatial offset between the blueshifted and redshifted emission.
This northern outflow must be extended, i.e., must have a wide opening angle, because it is better detected at lower resolution 
(in Figs. \ref{f.HVchans.CO32.RedBlueGray} and \ref{f.HV.CO10.RedBlues}). 
In particular, the blueshifted emission in Fig. \ref{f.HV.CO10.RedBlues}(a) directly shows that the high-velocity gas
that we attribute to outflows is more extended around the northern nucleus than around the southern nucleus.

We note that the large extent is another reason, in addition to the velocity gradient along the minor axis, why
the northern high-velocity gas is unlikely to be due to rotation. 
If the high velocities were rotational the enclosed dynamical mass would be unrealistically large, 
although we do not exclude a small fraction of rotational high-velocity gas very close to the dynamical center.
We also note why we model the high velocity gas as outflow rather than inflow.
It would be too much of a coincidence to have polar inflow from both sides of the northern nuclear disk at the same time.

The most plausible driver for the northern molecular outflow is starburst in and around the northern nuclear disk.
The current data do not suggest the outflow originates from a particular single point, such as an active galactic nucleus (AGN), within the nuclear disk.

\subsubsection{Northern Outflow Parameters}
\label{s.outflow.northern.parameter}
We estimate the outflow rate from the northern nucleus to be  $\dot{M}_{N} \approx 60 \Xtwenty$ \Msol\ \peryr.
We assumed for this the outflow axis to have the same inclination as the northern nuclear disk, i.e., $i_{\rm N,outflow} \approx i_{\rm N} \approx 30\arcdeg$.
The extent of the outflow along its outflow axis is estimated to be 0.8 kpc from this inclination and the 2\farcs4 offset between the
peak of the blueshifted emission and the northern nucleus in Fig.  \ref{f.HV.CO10.RedBlues}(b). 
The outflow velocity along the outflow axis is $650/\cos(30\arcdeg) = 750$ \kms\ for the largest velocity in Fig. \ref{f.HVchans.CO10.tp.VHV}.
The outflow timescale is therefore 1 Myr. 
(This is not necessarily the age of the outflow because the extent of the outflowing molecular gas may be limited by
interaction with ambient gas, 
dissociation of the molecules, 
gravity of the galaxy,
and our sensitivity.)
Dividing the mass of the high velocity gas around the northern nucleus (\S\ref{s.result.line.hv.mass}) 
with this timescale gives the outflow rate above.
This outflow rate is a lower limit because it does not account for the mass that is in the outflow 
but has lower line-of-sight velocities than the 224 \kms\ cutoff in our flux measurement for the high-velocity emission.

The kinetic luminosity of the outflow is on the order of
$L_{\rm kin, N} \about 4\times 10^{8} \Xtwenty\, \Lsol $ (=$2 \Xtwenty \times 10^{35}$ W)
where we use the mass of the northern high-velocity gas in \S\ref{s.result.line.hv.mass},
300 \kms\ for a characteristic outflow velocity (i.e., 260 \kms\ along our sightline), and the characteristic timescale of 1 Myr.
The lower velocity gas excluded from our outflow mass adds to the luminosity but less so than to the outflow rate.
The outflow kinematic luminosity is about 10\Xtwenty \% of the mechanical luminosity from 
a half of the star formation  in NGC 3256 (25 \Msun\peryr),  $\sim2\times 10^{36}$ W \citep{Leitherer99}.
Thus the northern outflow can reasonably be driven by the starburst.

The gas depletion time from the northern nucleus, a 300 pc diameter region centered at the nucleus, 
is calculated to be $3\chi_{N}$ Myr on the basis of our observations.
The parameter $\chi_{N}$ is the ratio of CO to \HH\ conversion factor 
for the nuclear disk and that for the high-velocity gas, i.e., $\Xco({\rm nucleus})/\Xco({\rm outflow})$,
for the northern nucleus.
It is one in our default assumption but it can be \about10 if the outflow CO emission is optically thin.

%
\subsubsection{Comparison with Previous Outflow Observations}
Outflow of ISM around the northern nucleus has been reported and the parameters measured in several previous works
besides our own detection of high-velocity molecular gas in \citest{Sakamoto06}.
\citet{Scarrott96} found with optical imaging polarimetry a dust reflection nebula extending out to 7 kpc (40\arcsec) from the galactic center
and attributed it to dust entrained to the halo by a starburst-driven superwind.
\cite{Moran99} made optical slit spectroscopy across NGC 3256N and 
found LINER-like emission line ratios off the nucleus (up to 30\arcsec\ from the center) coupled
with large line widths (FWHM up to 400 \kms). 
They attributed these to shock-induced kinematics and ionization and 
concluded the presence of a starburst-driven superwind.
\citet{Heckman00} detected Na~D absorption lines of 550 \kms\ width and 309 \kms\ blueshift
and concluded the presence of an outflowing superwind.
\citet{Lipari00} found blue wings of \Halpha\ and [\ion{N}{2}] lines in their spectroscopy toward the northern nucleus
and deduced an outflow with a velocity of \about350 \kms\ and line width \about130 \kms.
Notably, the minor axis of their outflow at ${\rm p.a.} \approx 70\arcdeg$ agrees with that of our molecular outflow
and so does their wide outflow opening angle (140\arcdeg).
\citet{Leitherer13} detected in their UV spectroscopy blueshifted line absorption of C and Si.
They detected three velocity components at $-126, -447$, and $-867$ \kms\ with the bulk velocity of $-461$ \kms\
at the position of their observations 2\arcsec\ northeast of the northern nucleus.
Our molecular outflow from the northern nucleus agrees with these observations regarding the
magnitude of the outflow velocity, outflow direction and opening angle, and that the outflow has a large spatial extent.
Therefore previous observations reporting an outflow/superwind in NGC 3256 are probably observations of 
various aspects of this northern outflow,
although a minor contribution from the southern outflow is likely in \citest{Sakamoto06}.

\subsection{Southern Outflow: Molecular Bipolar Jet}
\label{s.outflow.southern}

\subsubsection{Morphology}
\label{s.outflow.southern.morph}
The southern bipolar outflow is clearly seen as bisymmetric high-velocity emission around the southern nucleus.
It is along a ${\rm p.a.} \sim 0\arcdeg$ and is redshifted to the north and blueshifted to the south of the nucleus 
(Fig. \ref{f.HVchans.CO32.RedBlueGray}).
The outflow axis projected onto the sky is orthogonal to the nearly edge-on southern nuclear disk.
We therefore assume that the outflow is along the rotation axis of the southern nuclear disk. 
The near side of the southern nuclear disk is then estimated to be its northern side 
because the outflow toward us (i.e., blueshifted outflow) is on the south of the nucleus (see Fig. \ref{f.illust}).

\begin{figure}[t]
\epsscale{1}
\plotone{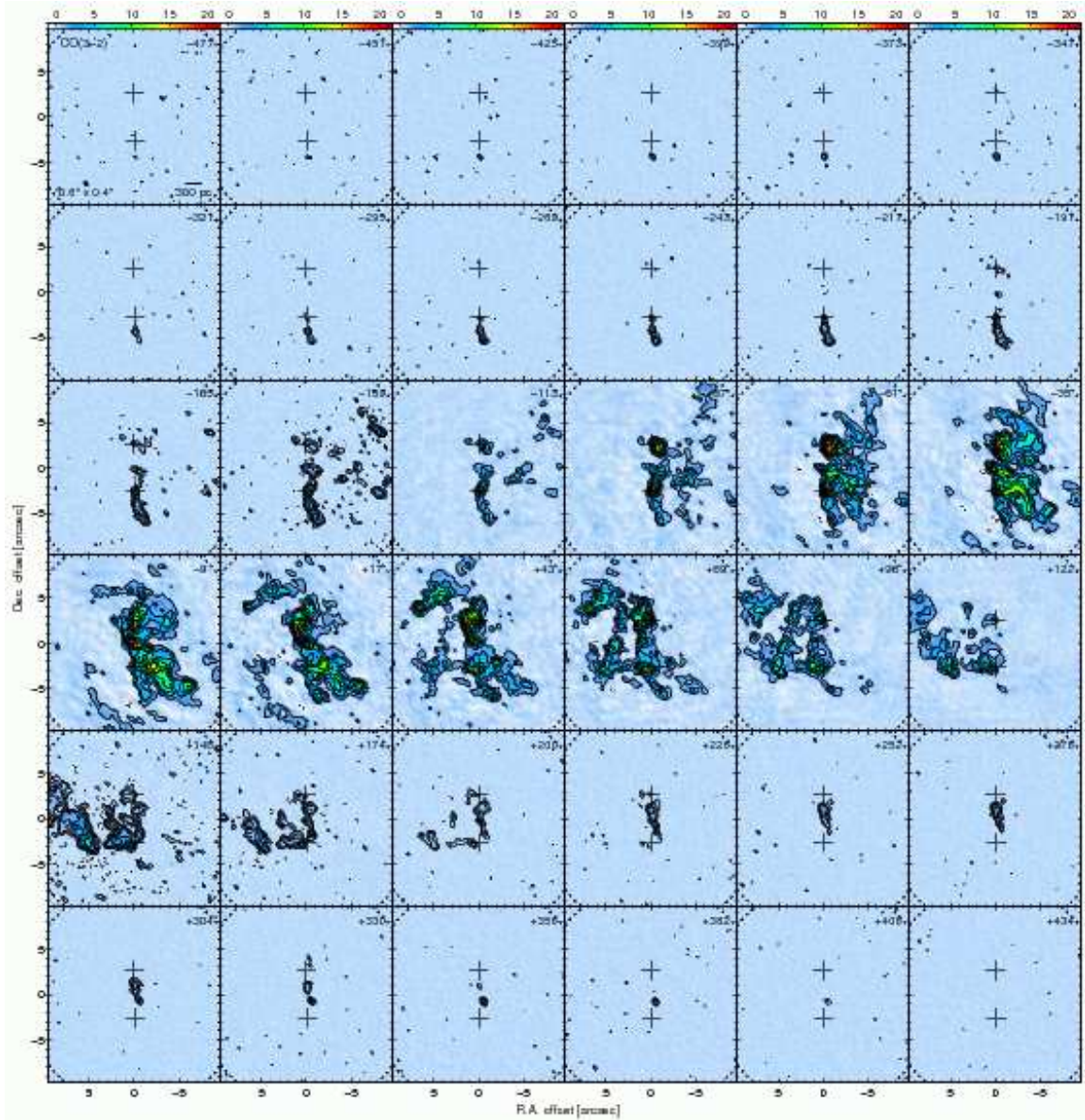}
\caption{ \label{f.chans.CO32.br}
CO(3--2) channel maps. 
Offsets from 2775 \kms\ are at the top-right corners of individual panels.
Contours are at $\pm10n^{1.5} \sigma$ $(n=1,2,3,\cdots)$ in channels from $-113$ to $+122$ \kms, 
whose maxima exceed 100$\sigma$,
and at  $[-6,-3,3,6,12,24,48,96]\sigma$ in the rest.
The rms noise is $\sigma=88$ mK.
Negative contours are dashed.
The data are not corrected for the mosaicked primary-beam response, whose 50\% contours are the dotted circles
(visible only at the corners).
The peak intensity of the line is 21.7 K in this plot whereas that of the 0.88 mm continuum already subtracted here is 0.22 K
at this resolution.
The two plus signs in each panel show the locations of the two nuclei. 
The black ellipse in a corner of the first panel shows the FWHM size of the observing beam.
}
\end{figure}

\begin{figure}[t]
\epsscale{1}
\plotone{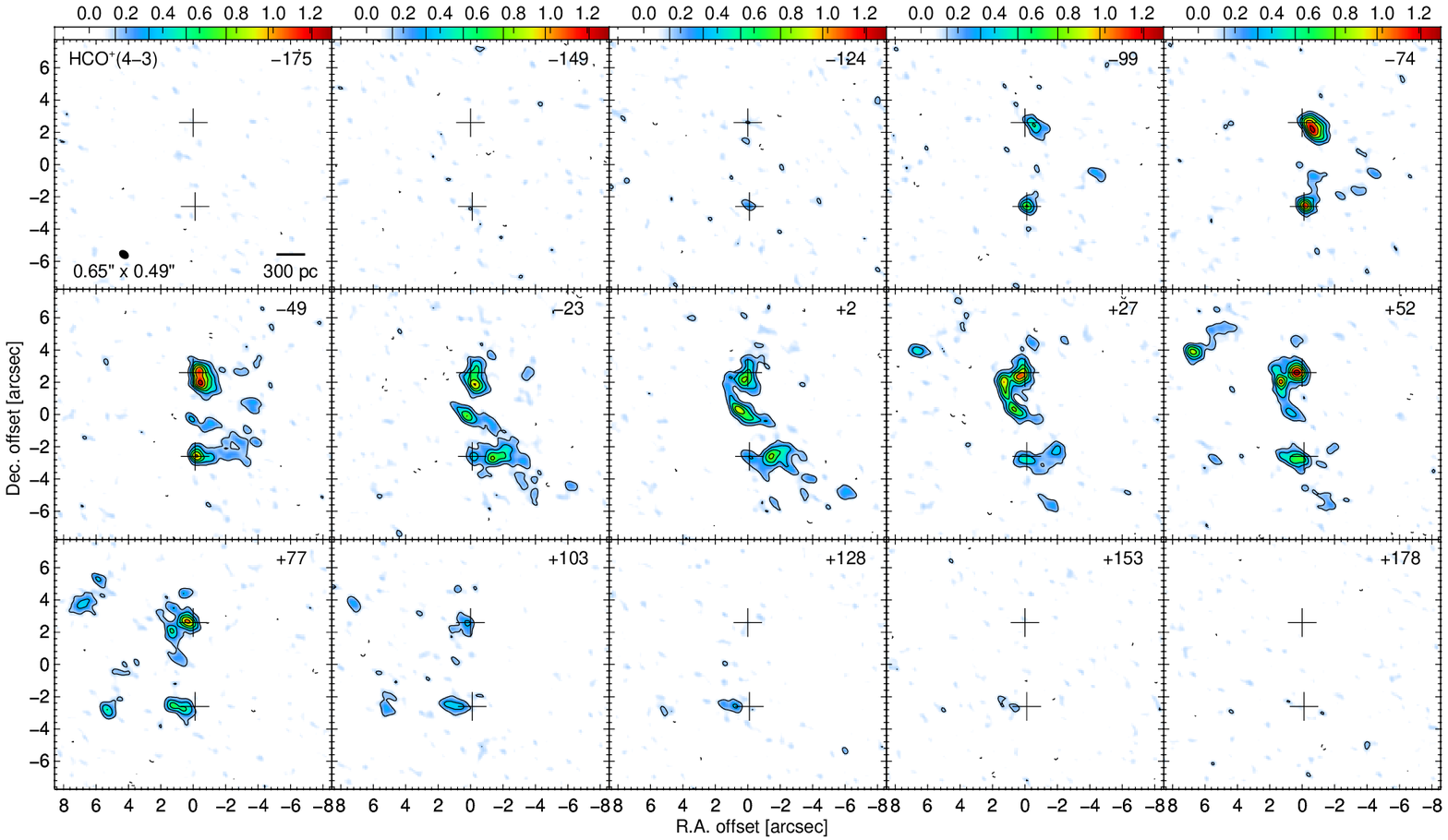} 
\caption{ \label{f.chans.HCOp43.na30MHz}
\HCOplus(4--3) channel maps. 
Offsets from 2775 \kms\ are at the top-right corners.
Contours are at $\pm 3 n^{1.2} \sigma$ $(n=1,2,3,\cdots)$, where
the rms noise is $\sigma=47$ mK.
Negative contours are dashed.
The data are not corrected for the mosaicked primary-beam response, whose FWHM is about 24 arcsec.
The peak line intensity is 1.3 K in this plot.
The two plus signs in each panel are at the two nuclei. 
A black ellipse in the first panel shows the FWHM size of the observing beam.
}
\end{figure}

\begin{figure}[t]
\begin{center}
\epsscale{0.85}
\plottwo{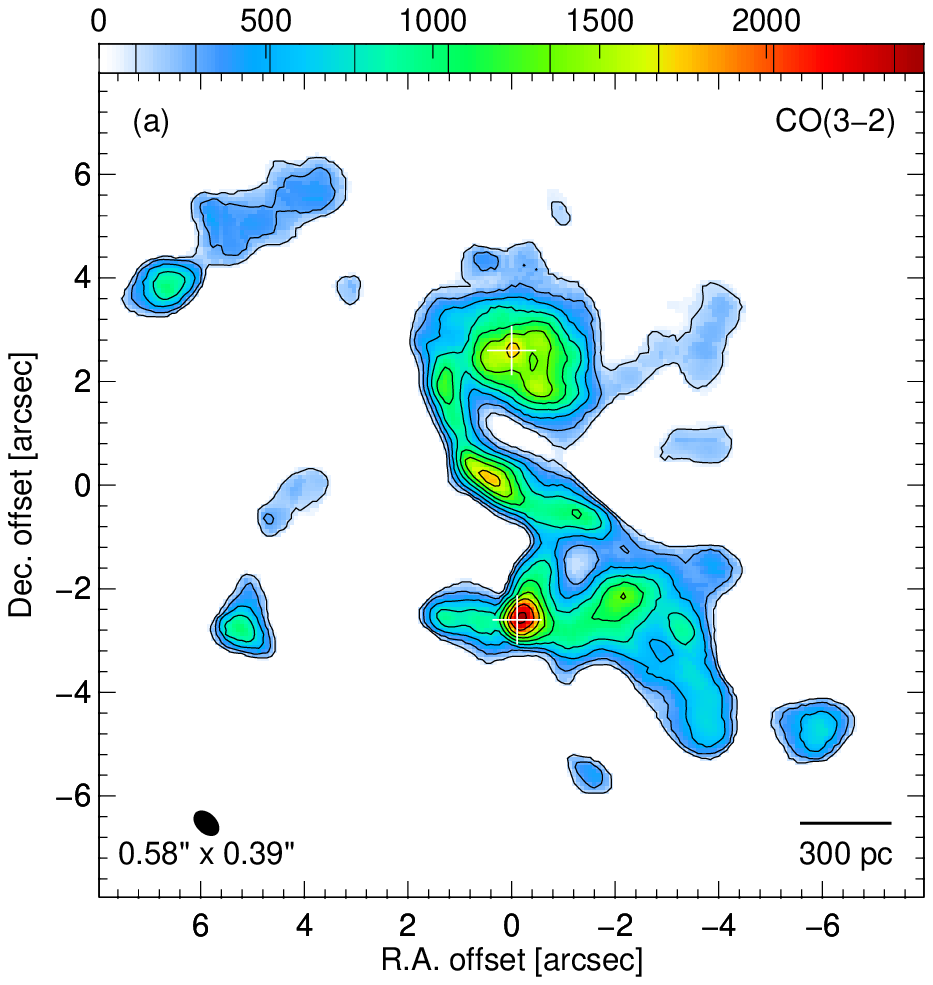}{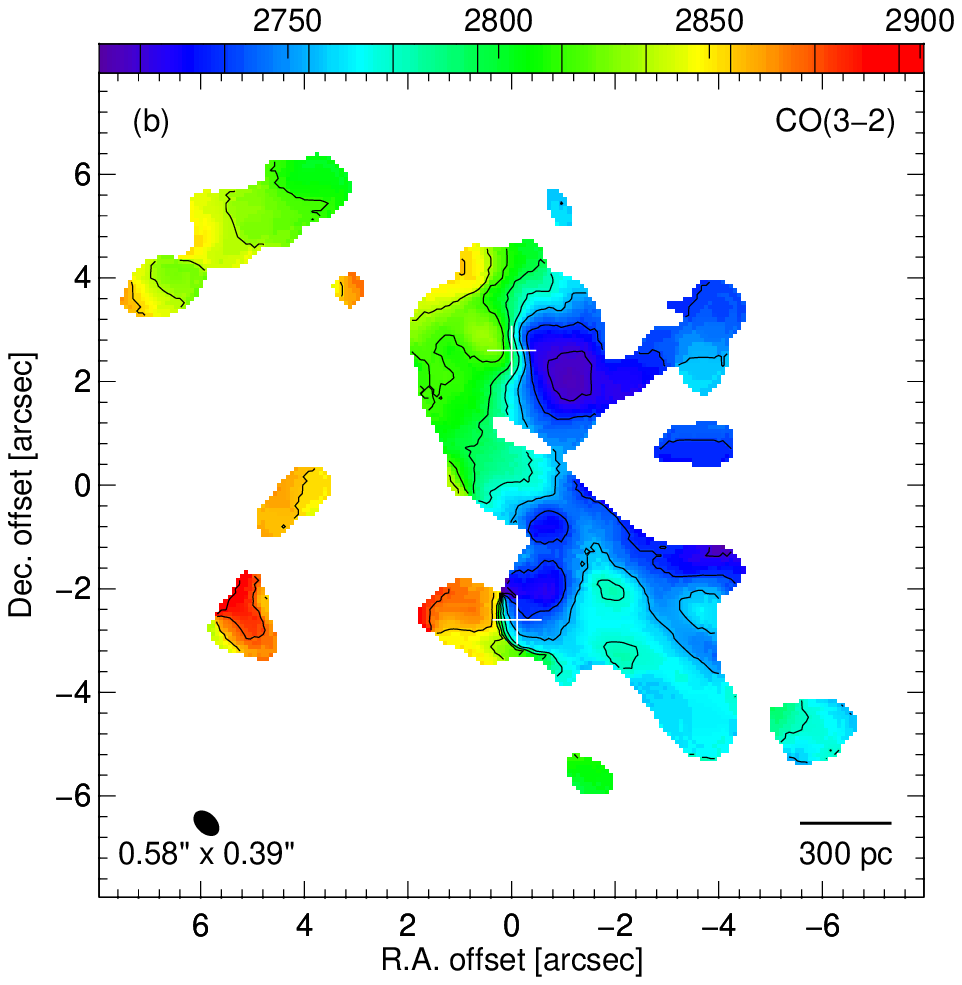} 
\end{center}
\caption{ \label{f.maps.CO32nuc}
CO(3--2) moment 0 and 1 maps (i.e., integrated intensity and mean velocity maps)
made with a high cutoff for the moment analysis to emphasize the brightest circumnuclear emission.
The intensity map (a) is corrected for the mosaic sensitivity patterns. 
The $n$th contours are at $110 n^{1.4}$ K \kms.
The velocity contours in (b) are at 2775 \kms\ and every 20 \kms\ from it.
The two white plus signs are at the position of the cm-wave nuclei. 
}
\end{figure}

\begin{figure}[t]
\epsscale{0.70}
\plotone{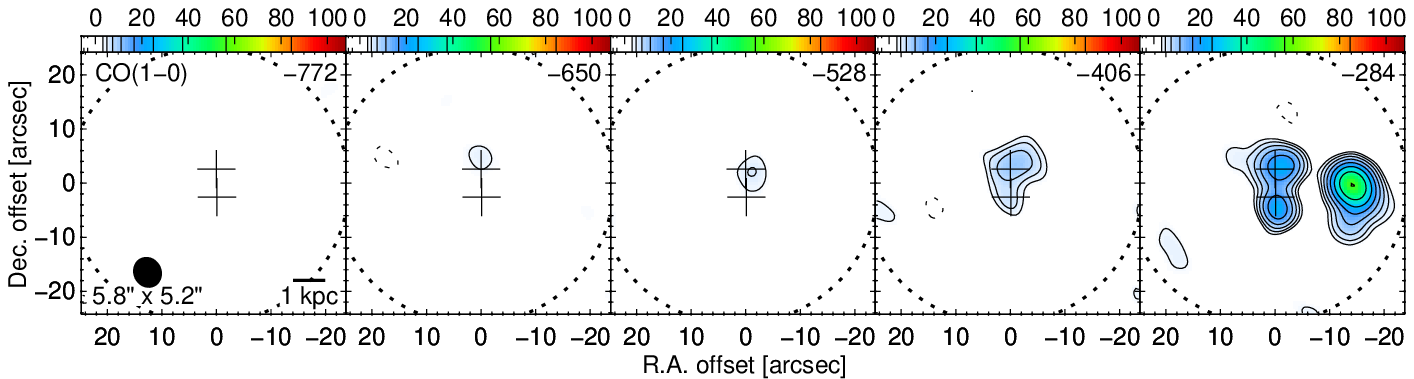} 
\epsscale{0.70}
\plotone{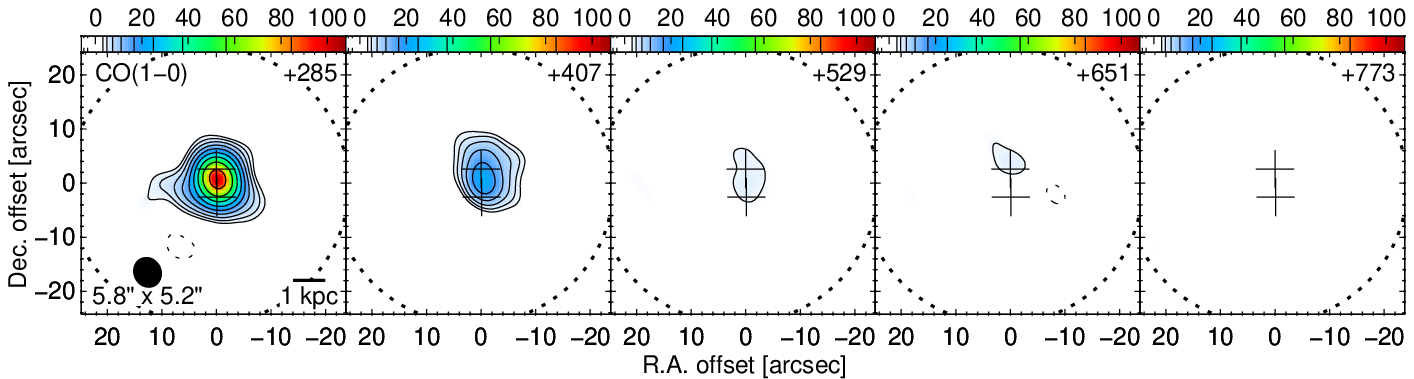} 
\caption{ \label{f.HVchans.CO10.tp.VHV}
CO(1--0) channel maps for high-velocity emission. 
Each channel is 122 \kms\ wide.
Velocity offsets from 2775 \kms\ are at the top-right corners.
Contours are at $\pm3\times1.5^{n} \sigma$ $(n=0,1,2,3,\cdots)$, 
where the rms noise is $\sigma=1.1$ mK. 
Negative contours are dashed.
The data are not corrected for the mosaicked primary-beam response, whose 50\% contours are the dotted circles.
The peak intensity of CO(1--0) in these channels is 101 mK 
while that of the 2.6 mm continuum already subtracted here is 30 mK at this resolution.
The labels of the intensity scale bars at the top are in mK.
The two plus signs in each panel are at the two nuclei. 
The black ellipse in the bottom-left corner of each leftmost panel shows the FWHM size of the observing beam.
}
\end{figure}

\begin{figure}[t]
\epsscale{0.45}
\plotone{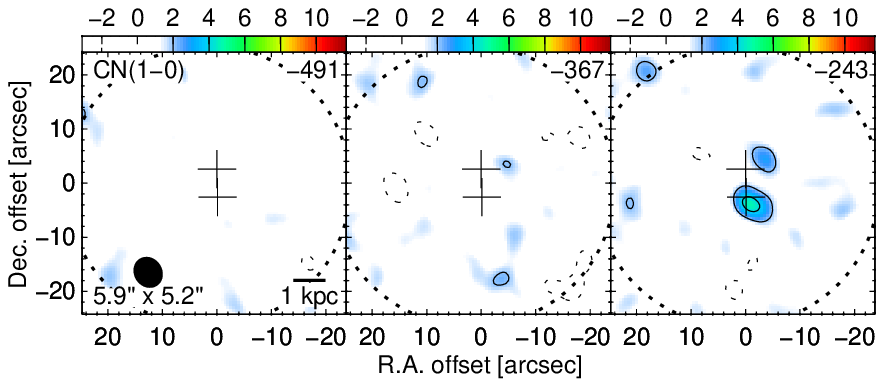} 
\epsscale{0.45}
\plotone{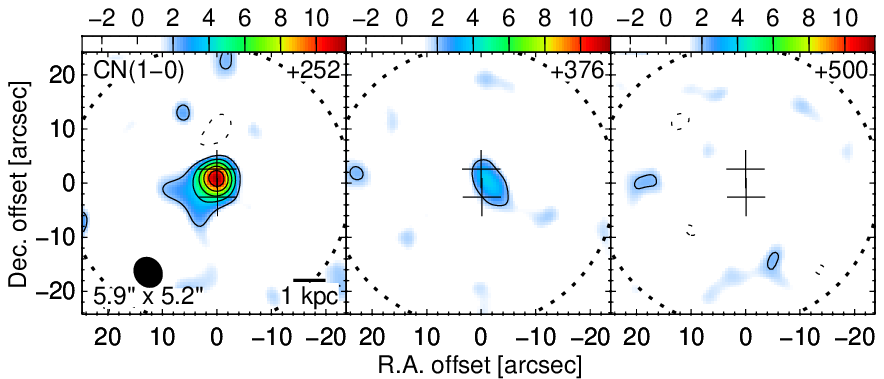} 
\caption{ \label{f.HVchans.CN10.tp.VHV}
CN(1--0, 3/2--1/2) channel maps for high-velocity emission. 
Each channel is 124 \kms\ wide.
Velocity offsets from 2775 \kms\ are at the top-right corners.
Contours are at $\pm2.5n \sigma$ $(n=1,2,3,\cdots)$, where the rms noise is $\sigma=0.80$ mK. 
Negative contours are dashed.
The data are not corrected for the mosaicked primary-beam response, whose 50\% contours are the dotted circles
(visible only at the corners).
The peak intensity of the CN(1--0) line in these channels is 11.7 mK.
The intensity scale bars at the top are labeled in mK.
The two plus signs in each panel are at the two nuclei. 
The black ellipses in the bottom-left corners show the FWHM size of the observing beam.
}
\end{figure}

\begin{figure}[h]
\begin{center}
\mbox{
\includegraphics[height=43.2mm]{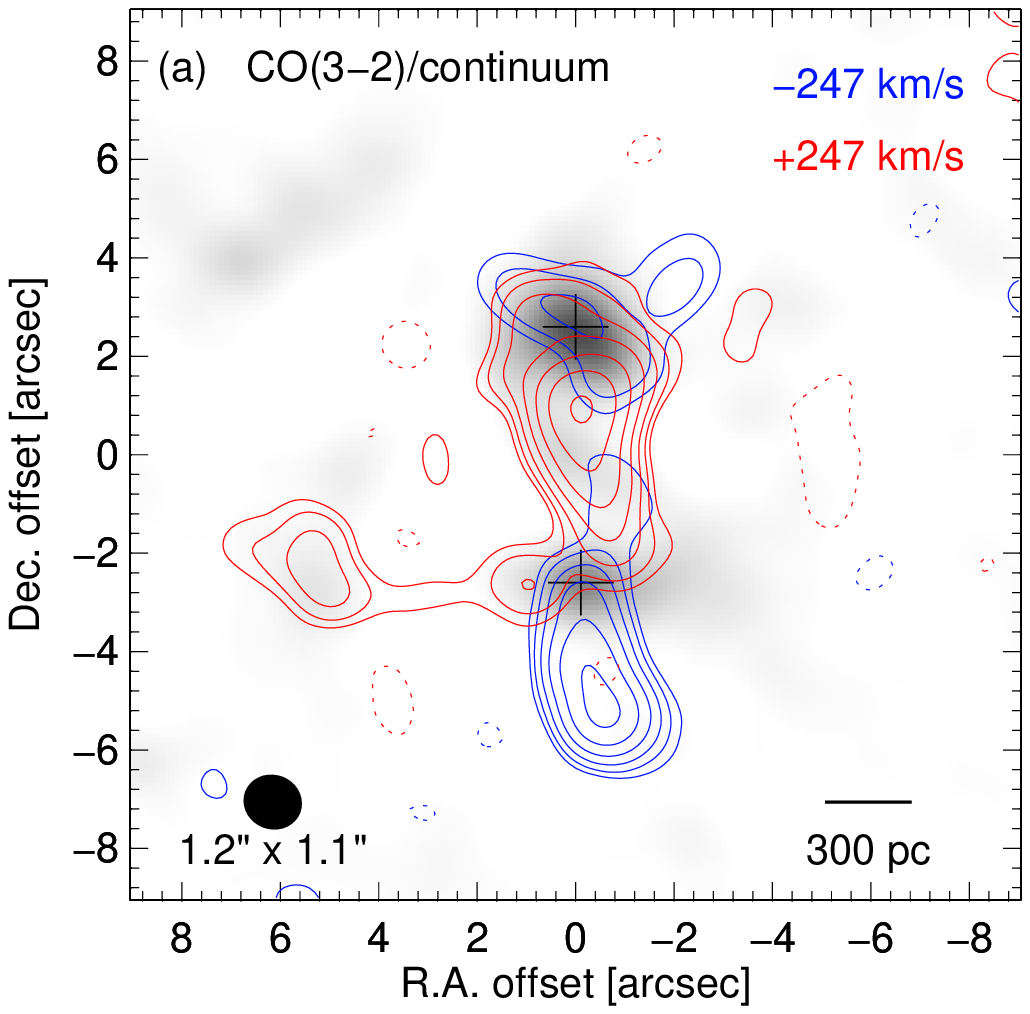} 
\includegraphics[height=43.2mm]{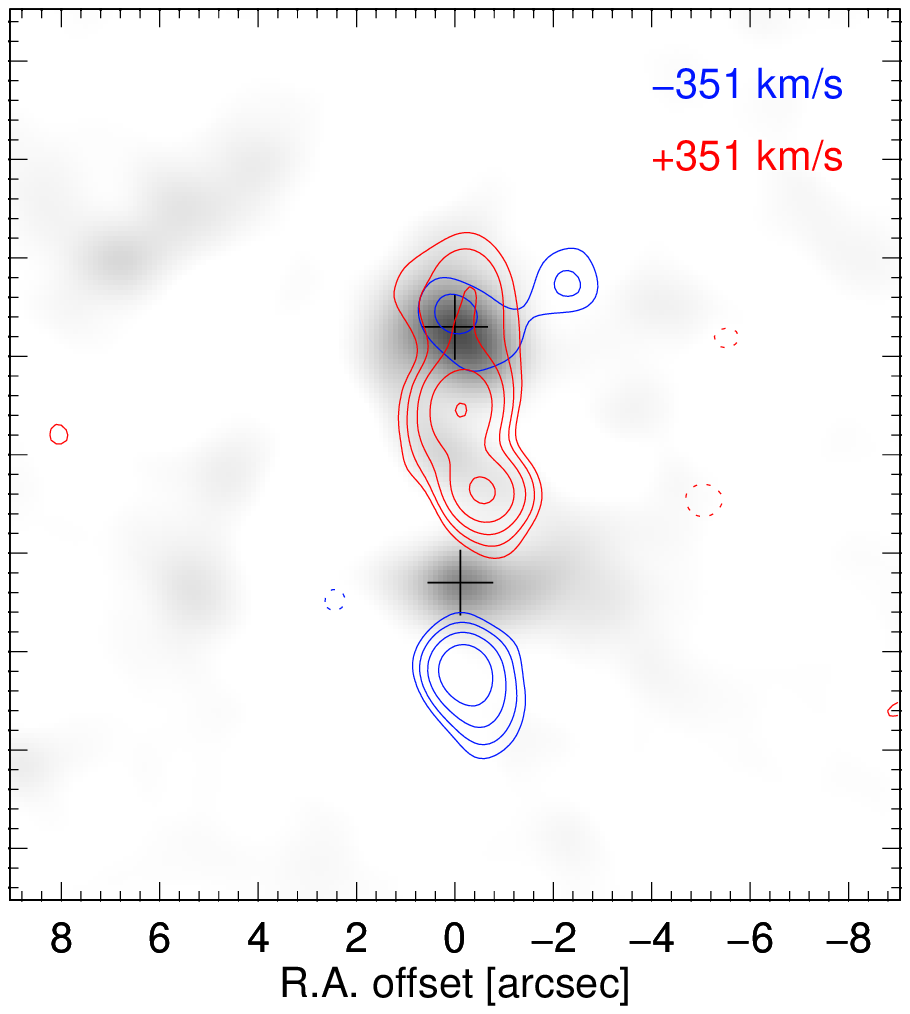} 
\includegraphics[height=43.2mm]{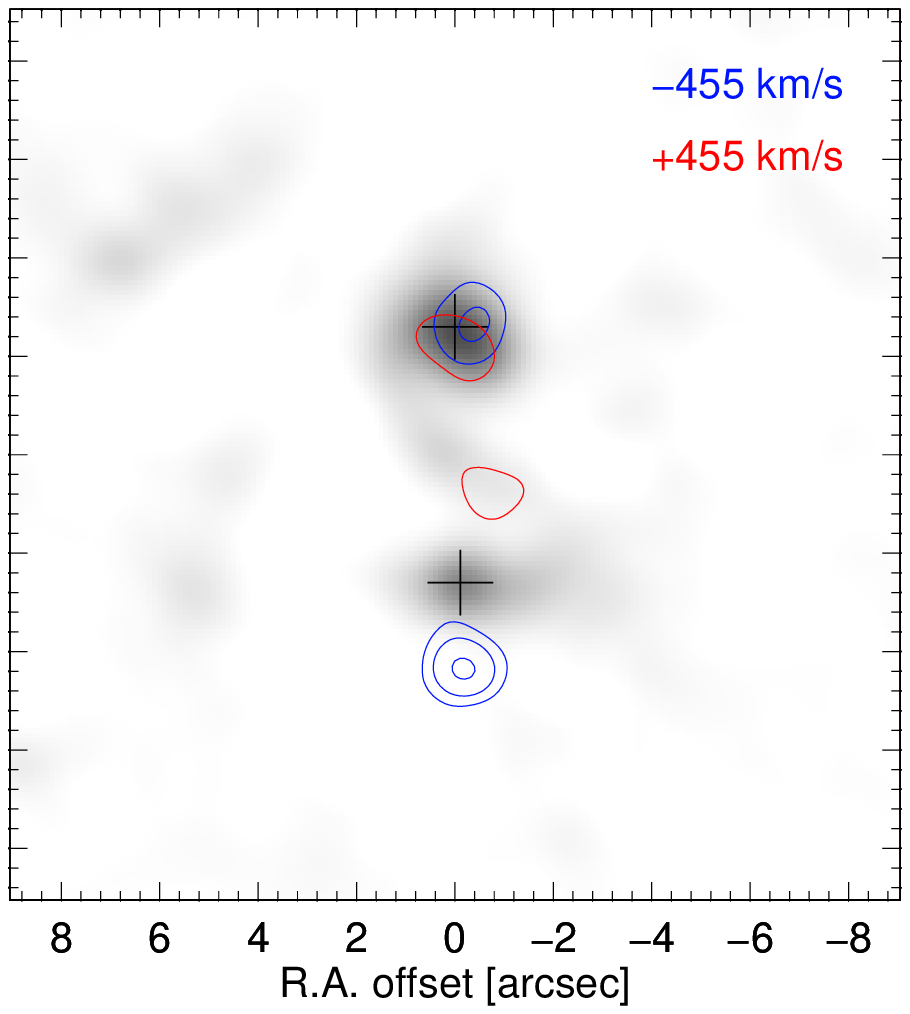} 
\includegraphics[height=43.2mm]{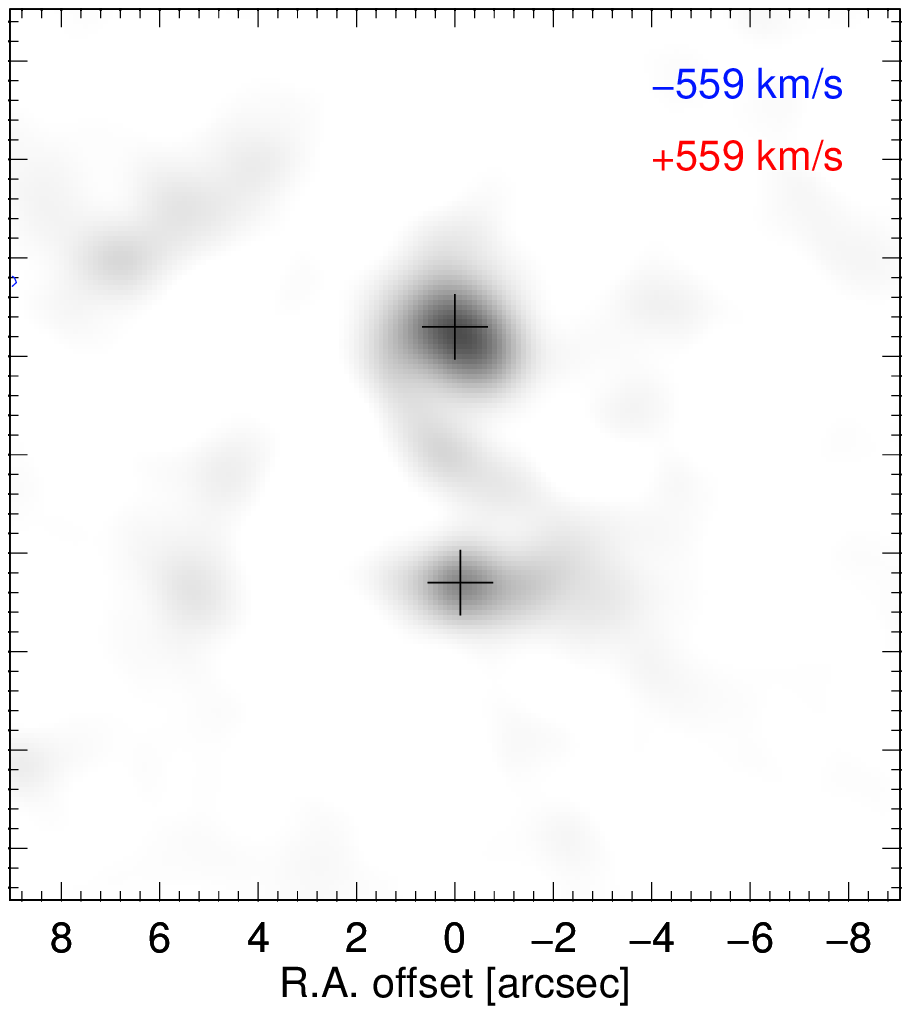} 
}
\mbox{
\includegraphics[height=35mm]{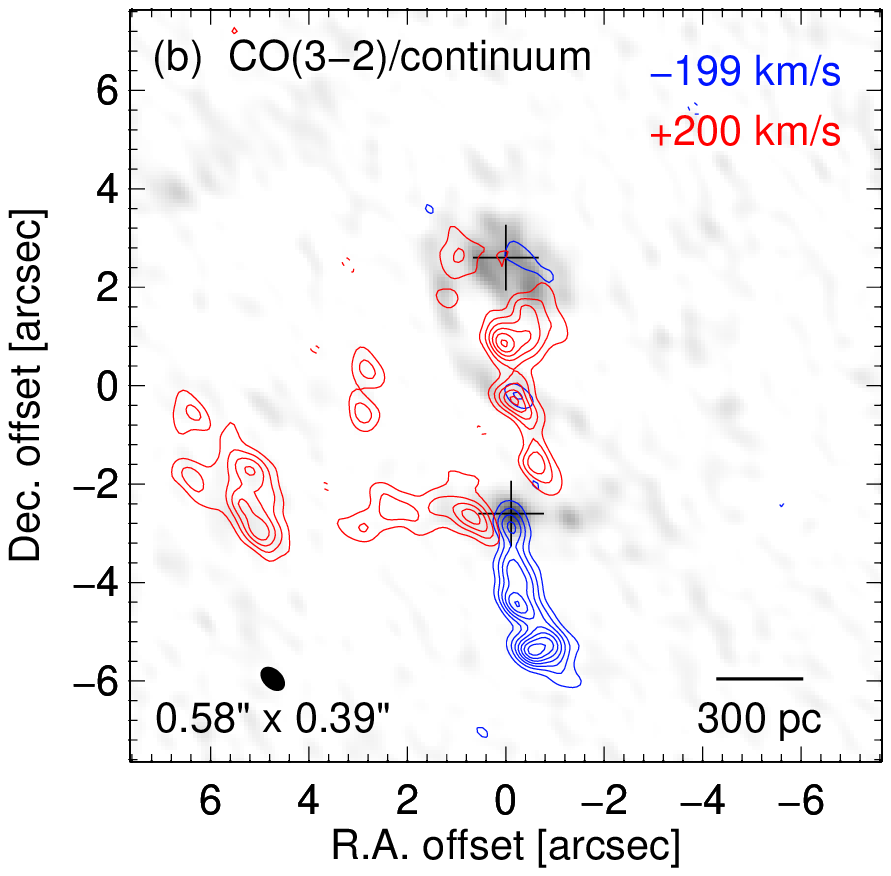} 
\includegraphics[height=35mm]{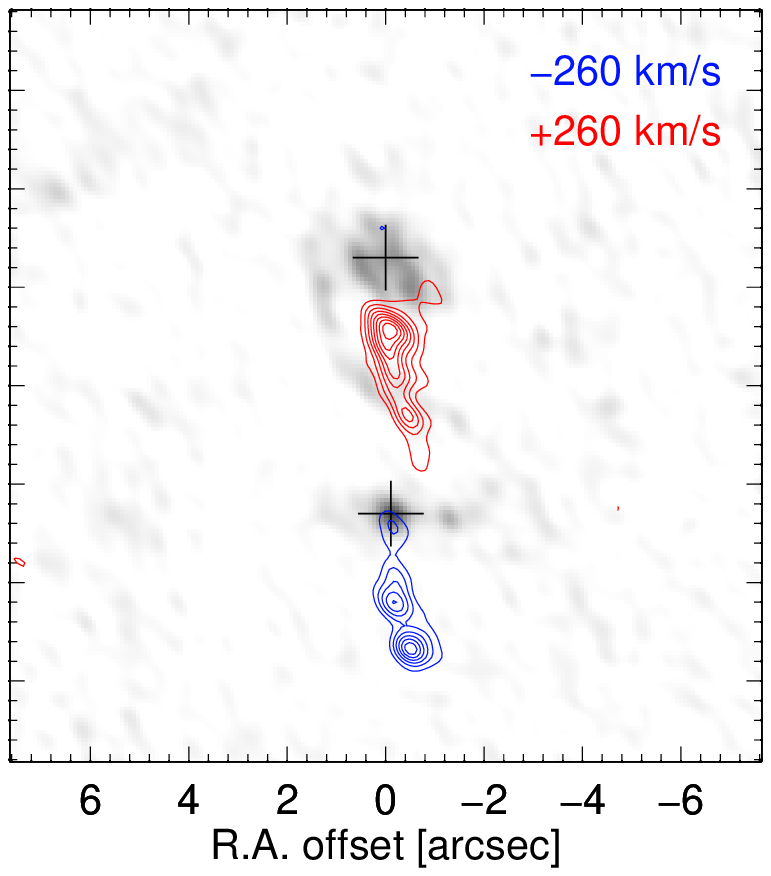} 
\includegraphics[height=35mm]{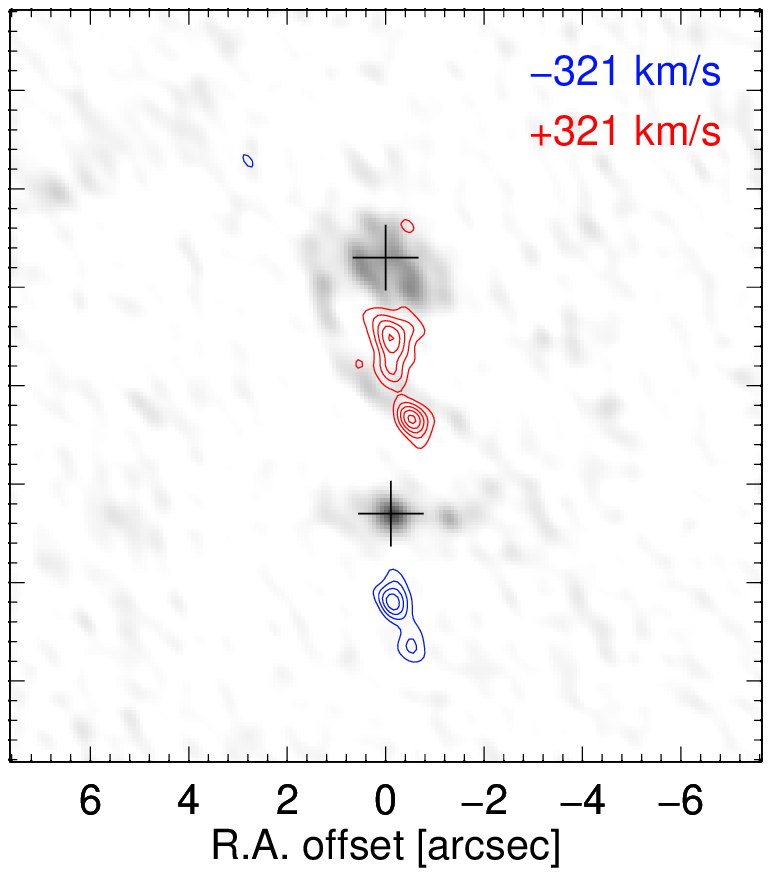} 
\includegraphics[height=35mm]{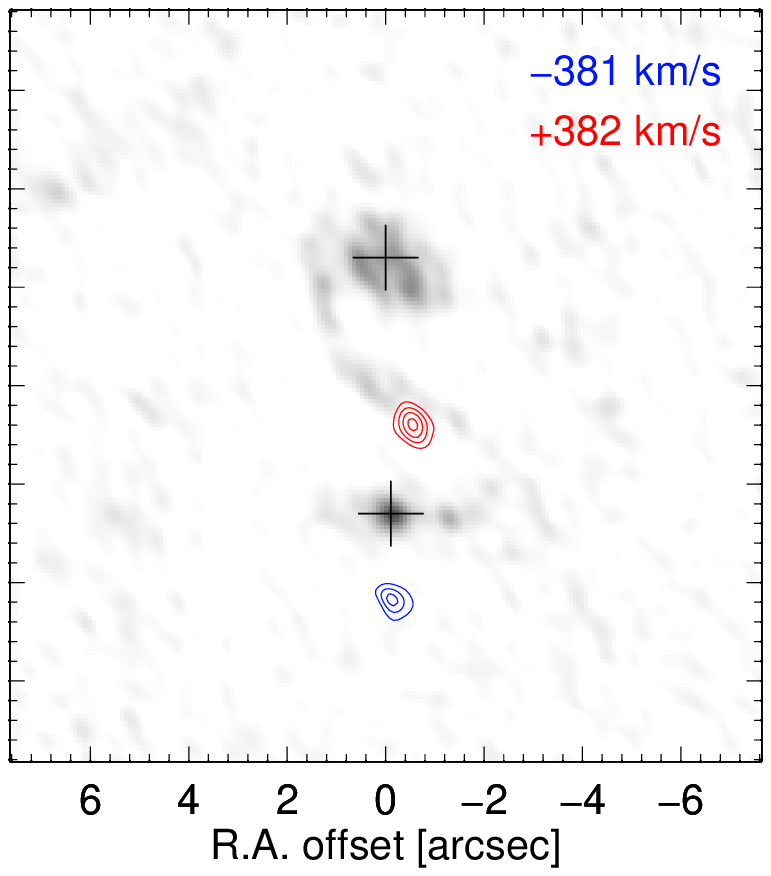} 
\includegraphics[height=35mm]{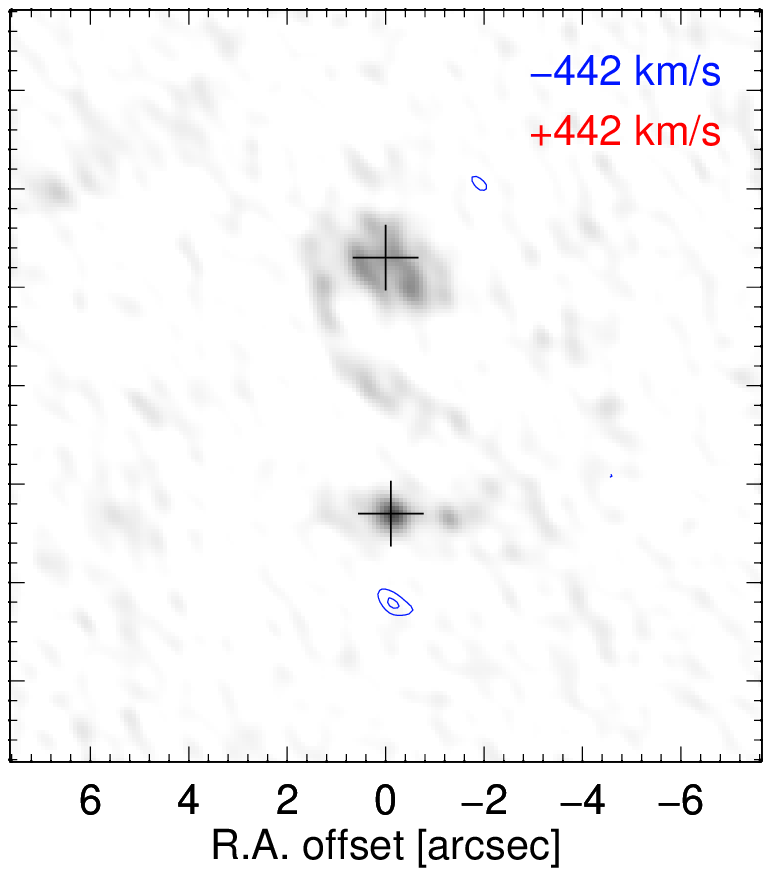} 
}
\end{center}
\caption{ \label{f.HVchans.CO32.RedBlueGray}
CO(3--2) channel maps of high-velocity emission at 1\farcs1 (a; upper row) and 0\farcs5 (b; lower row) resolutions.
Red and blue contours in each panel are respectively for a channel redshifted or blueshifted by the amount
indicated in the top-right corner from the fiducial velocity 2775 \kms.  
In (a) [(b)], each channel is 104 [61] \kms\ wide 
and contours are at $\pm4\times1.8^{n} \sigma$ $(n=0,1,2,3,\cdots)$ [$\pm4 n \sigma$ $(n=1,2,3,\cdots)$],
where the rms noise is $\sigma=5.6$ [38] mK. 
Negative contours are dashed.
The  background grayscale image is 0.86 mm continuum emission at the same resolution. 
The peak continuum intensity is 108 [237] mK.  
These data are not corrected for the mosaicked primary-beam response, whose FWHM size is about 24\arcsec.
The two plus signs in each panel show the locations of the two radio nuclei.
The leftmost panel in each row has a scale bar and an ellipse showing the FWHM size of the observing beam. 
}
\end{figure}

\begin{figure}[h]
\epsscale{0.33}
\plotone{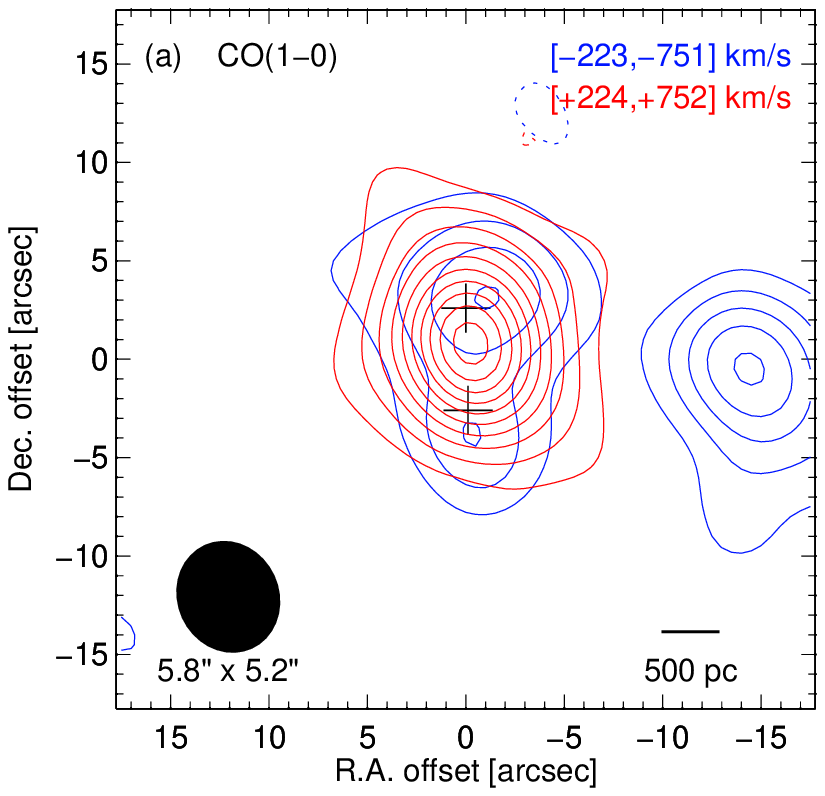} 
\plotone{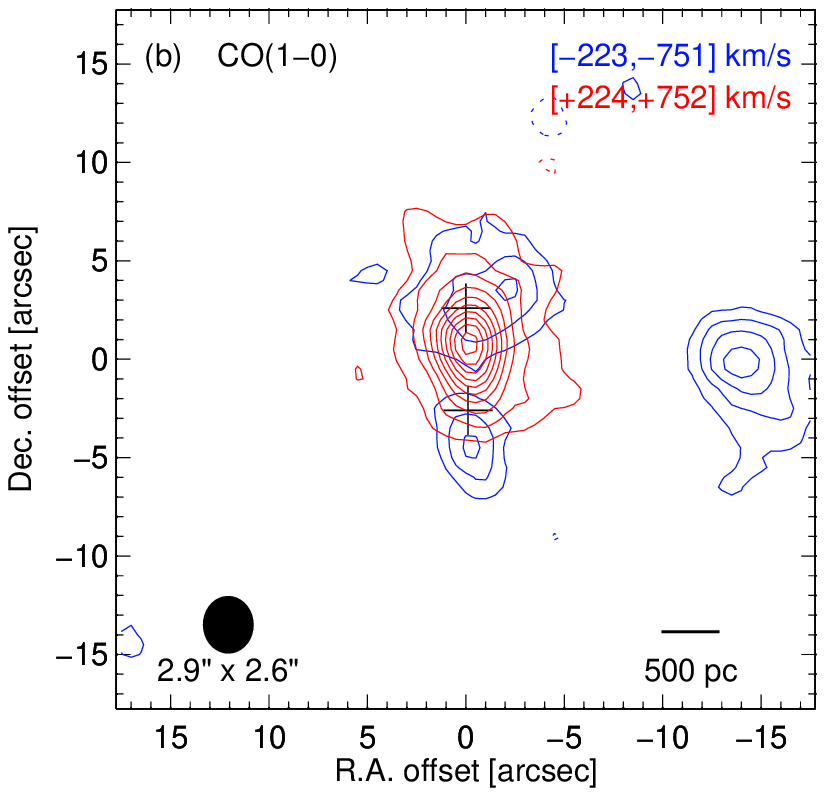} 
\caption{ \label{f.HV.CO10.RedBlues}
CO(1--0) maps of high-velocity emission at different spatial resolutions.
Red and blue contours are intensities integrated over 528 \kms\ at,  respectively,  redshifted and blueshifted
velocity ranges indicated in the top-right corner. They are offsets from the fiducial velocity 2775 \kms.  
Contours are at $\pm4 n^{p} \sigma$ $(n=1,2,3,\cdots)$ where $p=1.2$ and $\sigma=0.24$ K\kms\ in (a)
and $p=1.1, \sigma=0.79$ K\kms\ in (b).
Negative contours are dashed.
The data are not corrected for the primary-beam response whose FWHM size is 53\arcsec.
The two plus signs are at the two nuclei.
The ellipse at the bottom-left corner shows the FWHM size of the synthesized beam. 
}
\end{figure}

\begin{figure}[t]
\epsscale{0.24}
\plotone{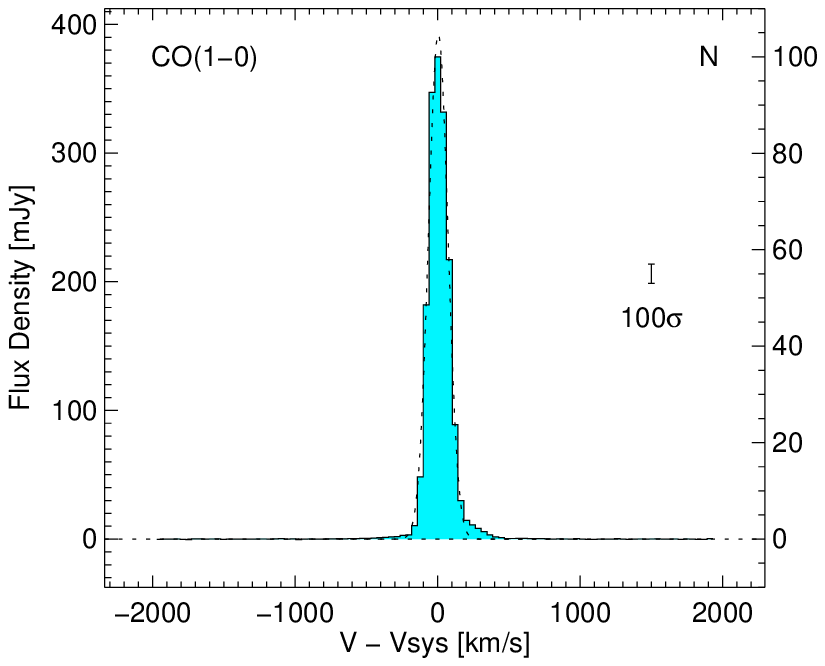} 
\plotone{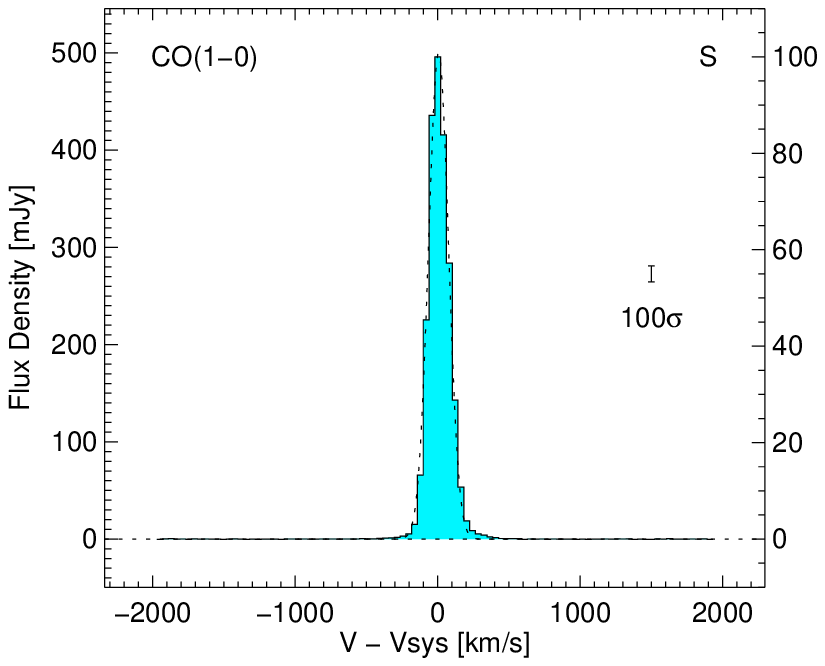} \\ 
\plotone{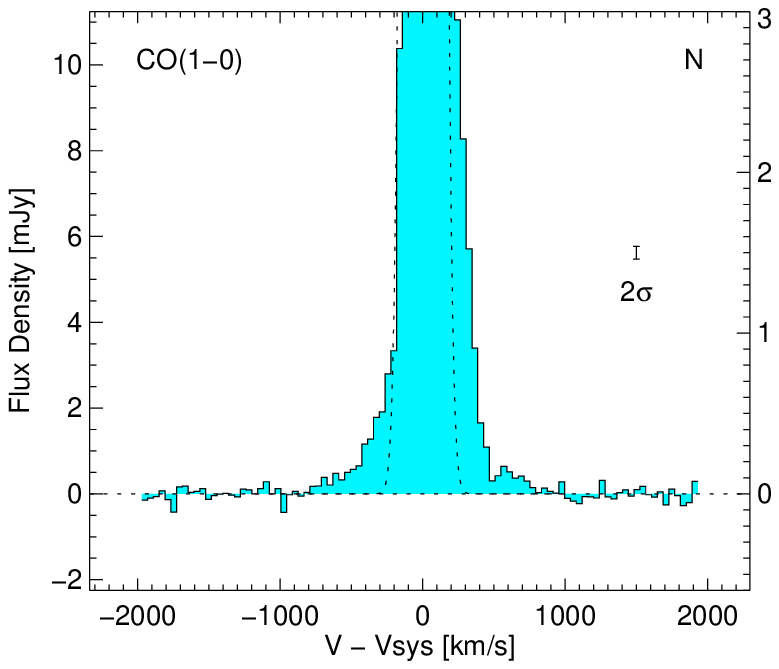} 
\plotone{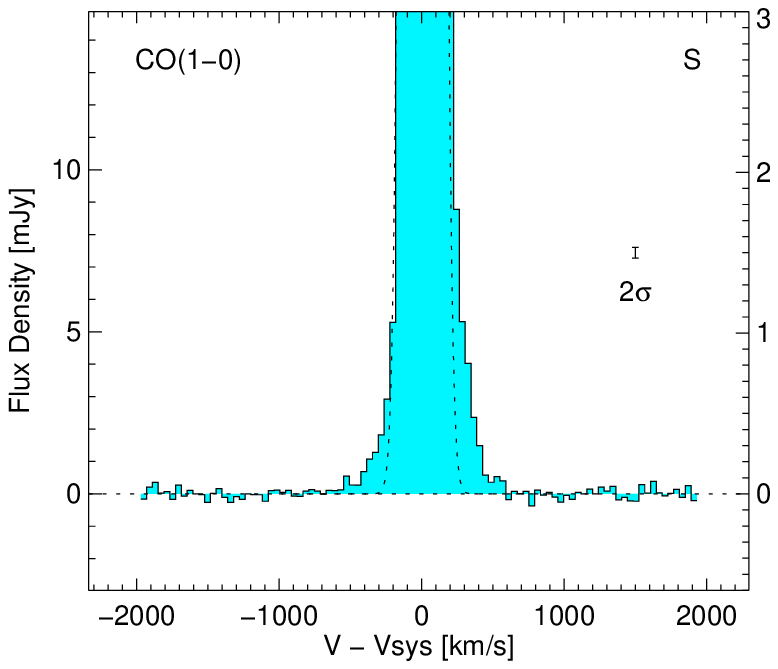} \\ 
\plotone{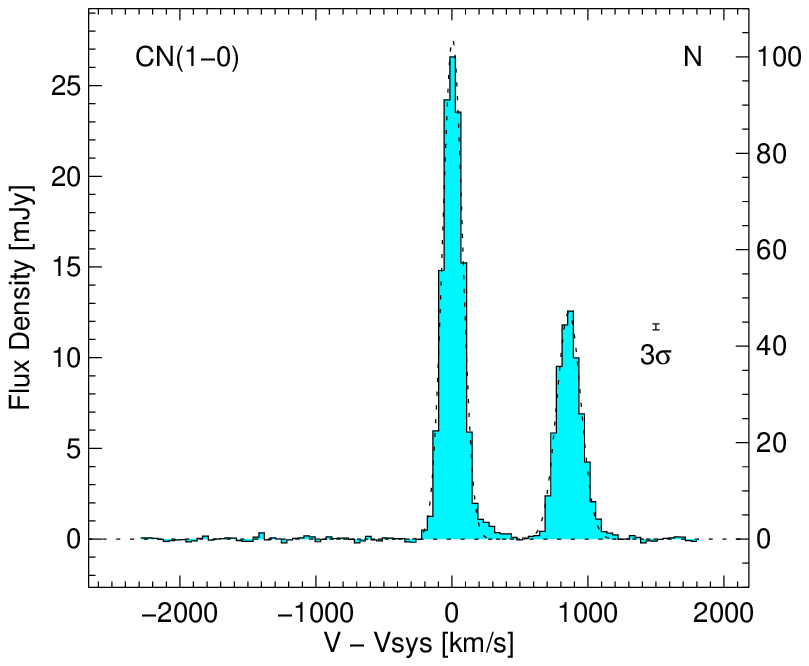} 
\plotone{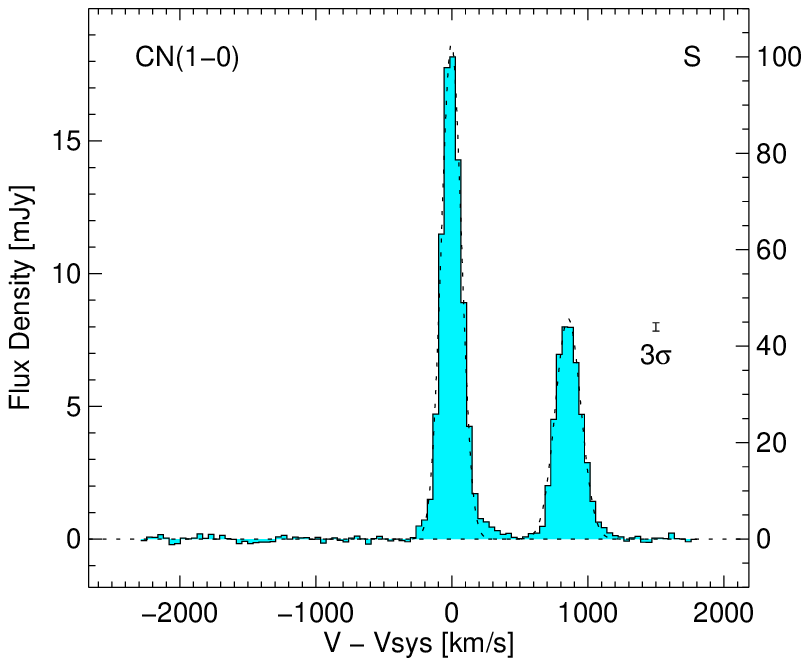} \\ 
\plotone{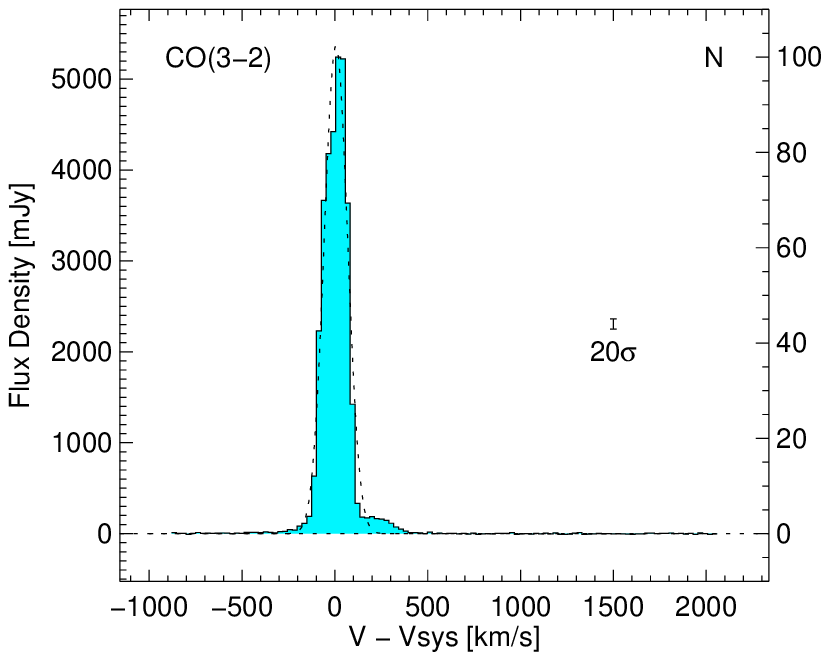} 
\plotone{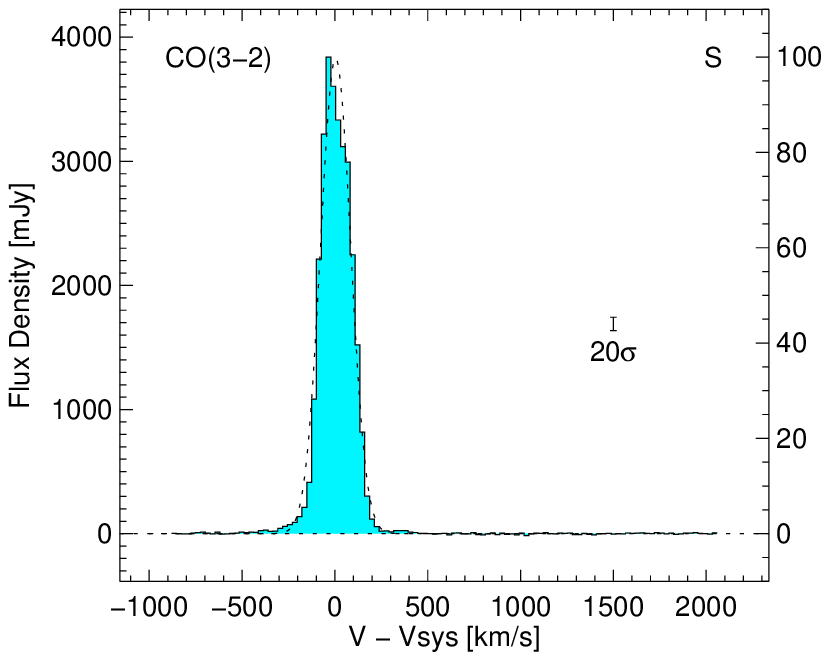} \\
\plotone{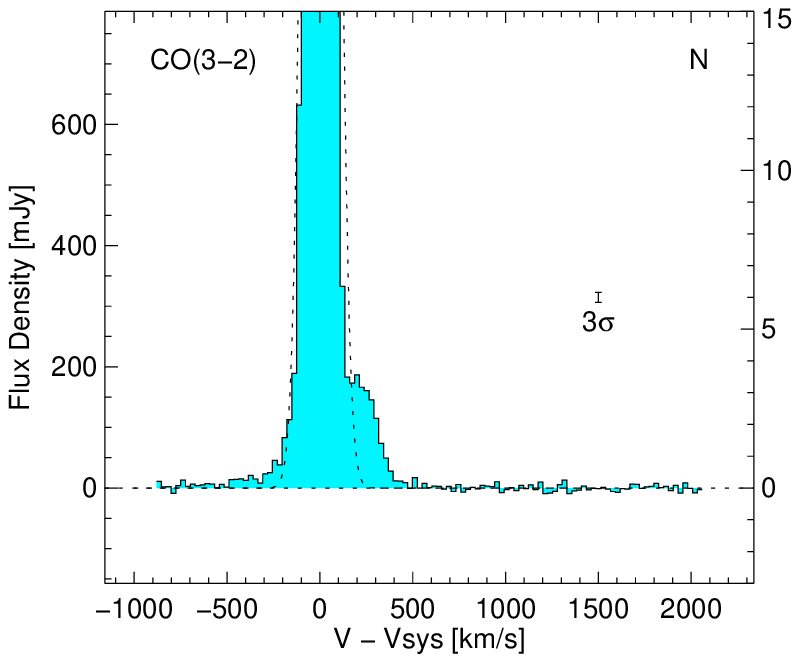} 
\plotone{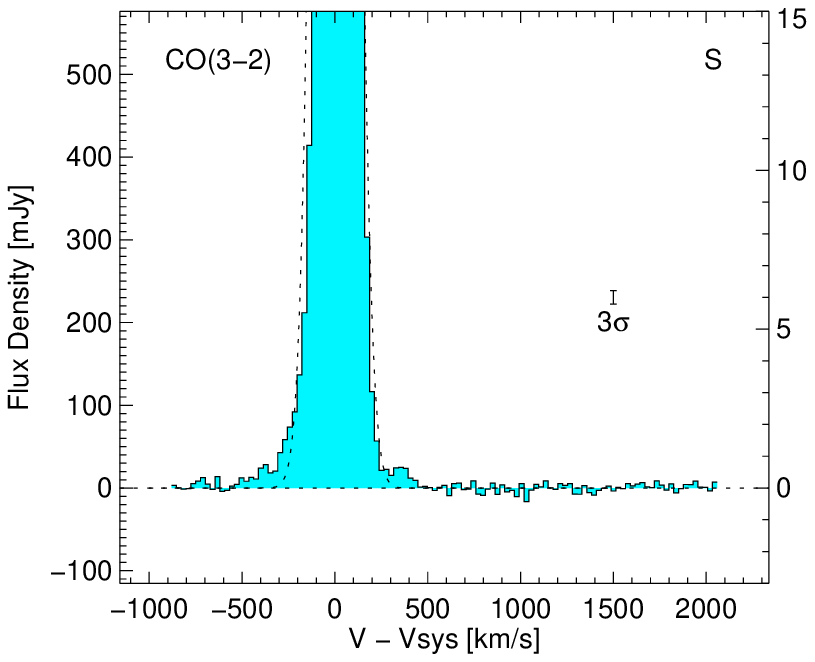} \\ 
\plotone{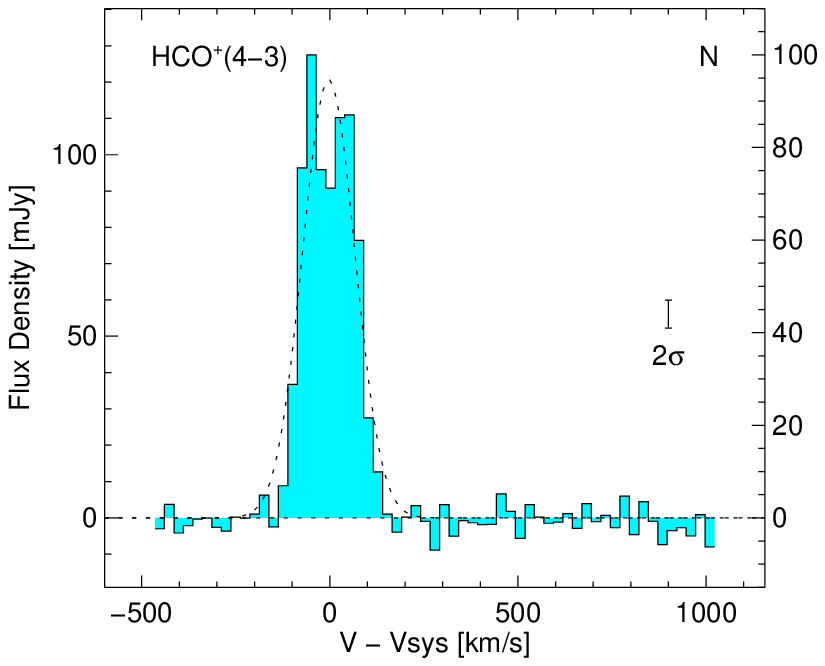} 
\plotone{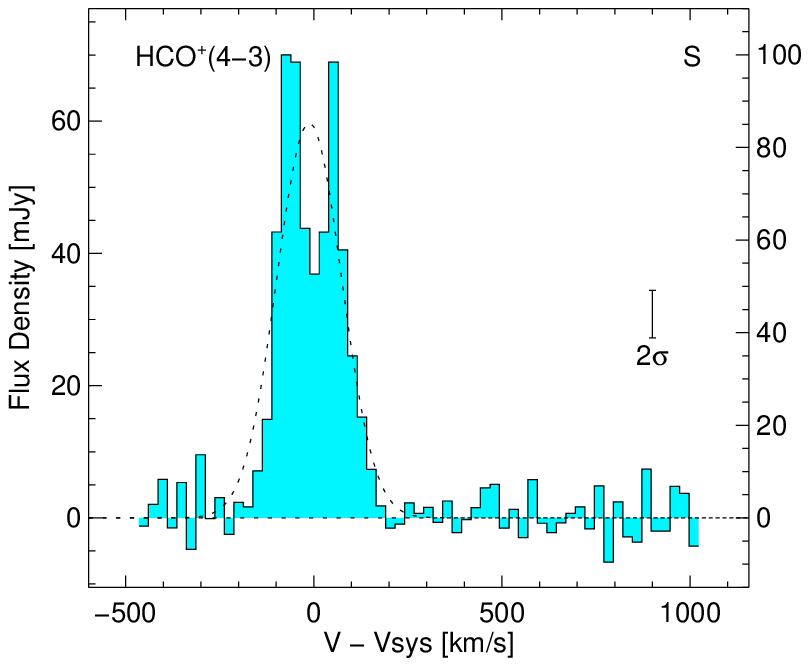} 
\caption{\scriptsize \label{f.spectra.nuclei}
Line spectra of NGC 3256 nuclei; the northern nucleus (left column) and the southern nucleus (right column). 
Line names are in individual panels.
Dotted curves are single-component Gaussian fits to the data to help identify excess emission at high velocities.
Abscissa is velocity offset from 2775 \kms.
Right ordinate is fraction in percent of the peak in the spectrum.
(Row 1 and 2) 
CO(1--0) spectra from the CSV \plus\ Cycle 0 data of \about5\arcsec\ resolution. 
Diameter of the sampling aperture is 4\arcsec.
The two rows show different ranges of ordinate.
(Row 3)
CN(1--0) spectra from the CSV \plus\ Cycle 0 data of \about5\arcsec\ resolution. 
Diameter of the sampling aperture is 4\arcsec.
The abscissa is for the brighter CN(1--0, 3/2--1/2) line.
The Gaussian fitting used two components for the two transitions.
(Row 4 and 5)
CO(3-2) spectra sampled with 4\arcsec-diameter apertures.
The spatial and spectral resolutions of the data are \about0\farcs6 and 26 \kms (30 MHz) , respectively.
(Row 6)
\HCOplus(4--3) spectra sampled with 2\arcsec-diameter apertures from \about0\farcs6 resolution data. 
}
\end{figure}

\begin{figure}[t]
\epsscale{0.27}
\plotone{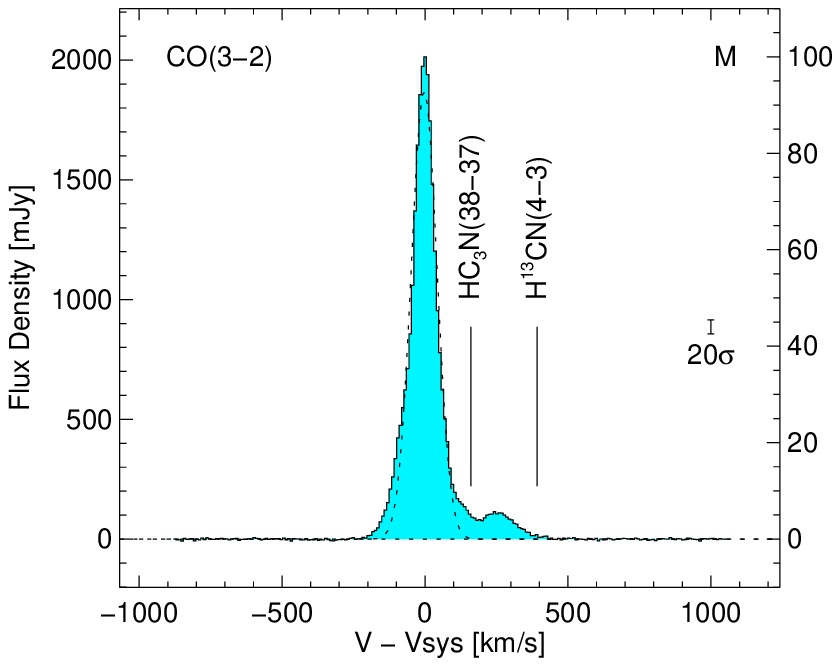} 
\plotone{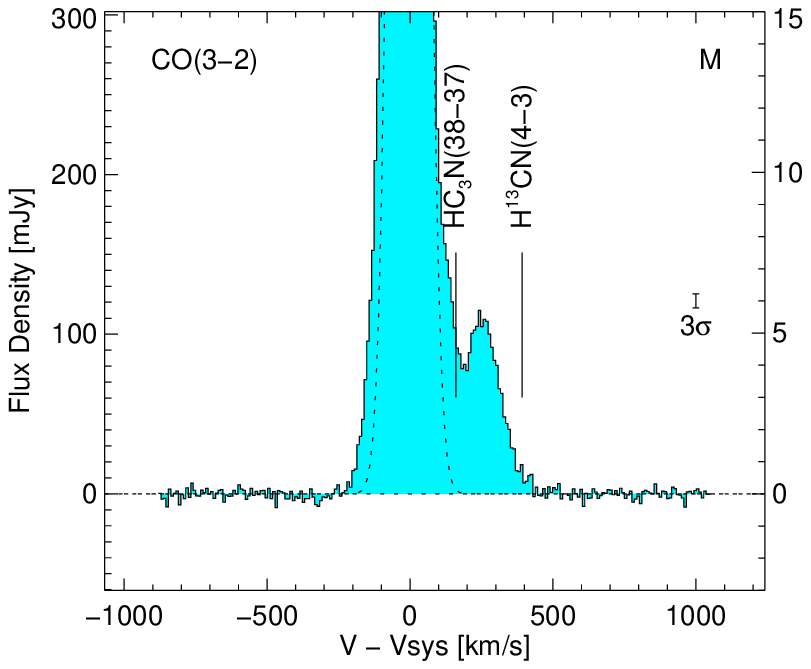} 
\caption{ \label{f.spec.noblend}
CO(3--2) spectrum sampled with a 2\arcsec-diameter aperture at the midpoint of the two nuclei
from a cube of 1\farcs15 and 8.7 \kms\ (10 MHz) resolutions.
The same spectrum is plotted twice with different y axis ranges.
\HCthreeN(38--37) and \HthirteenCN(4--3) lines should appear at the marked locations
if emitted from gas at the velocity of the CO(3--2) line peak.
}
\end{figure}

\begin{figure}[t]
\epsscale{0.32}
\plotone{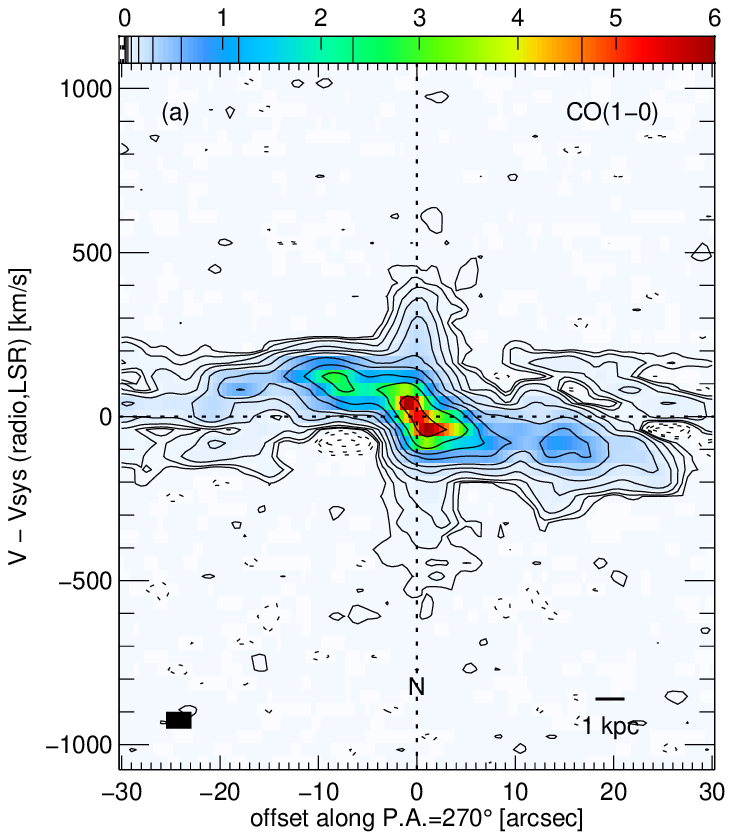} 
\plotone{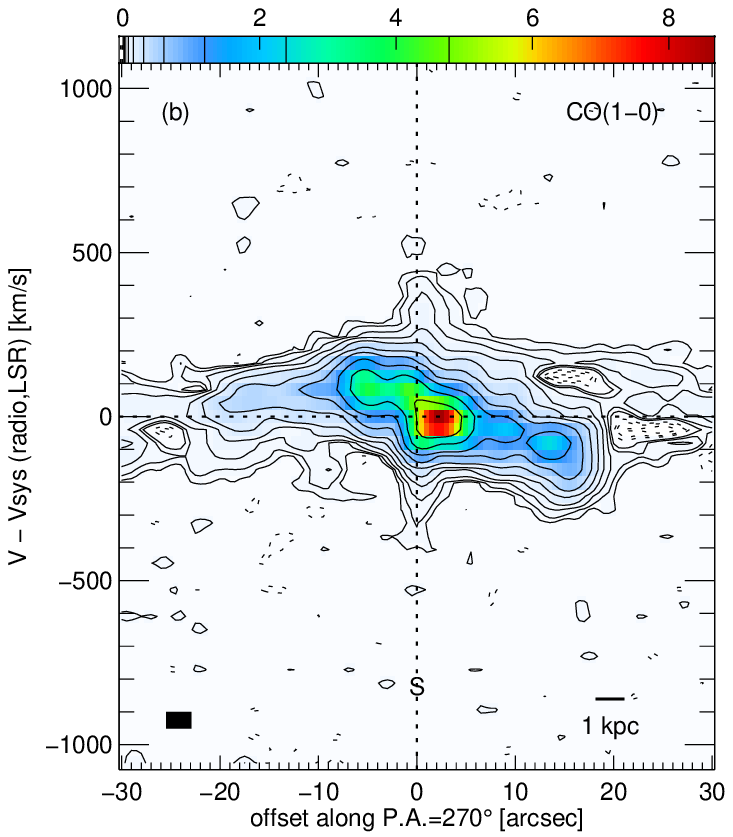} 
\plotone{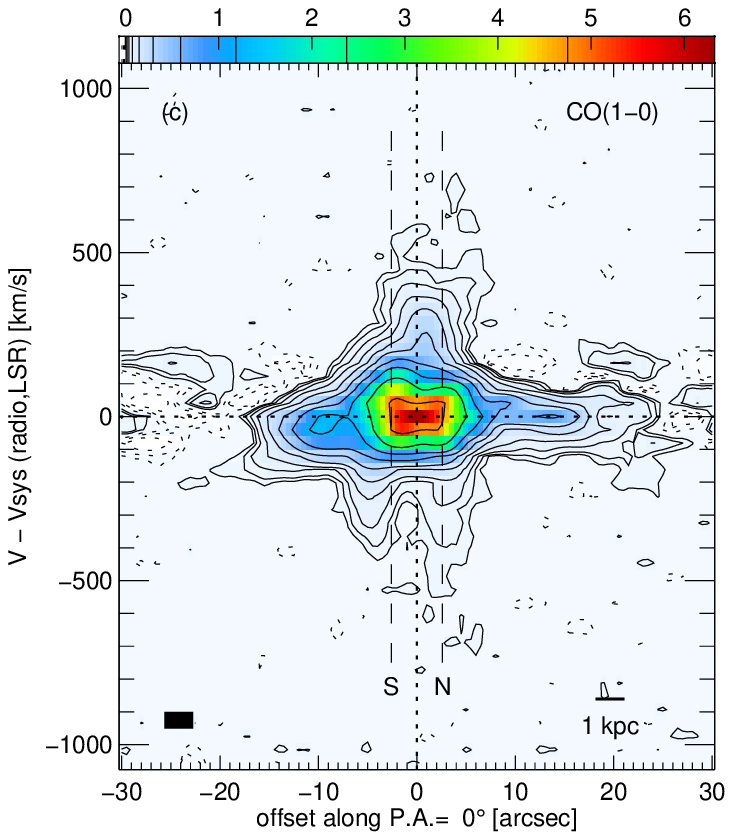}\\ 
\plotone{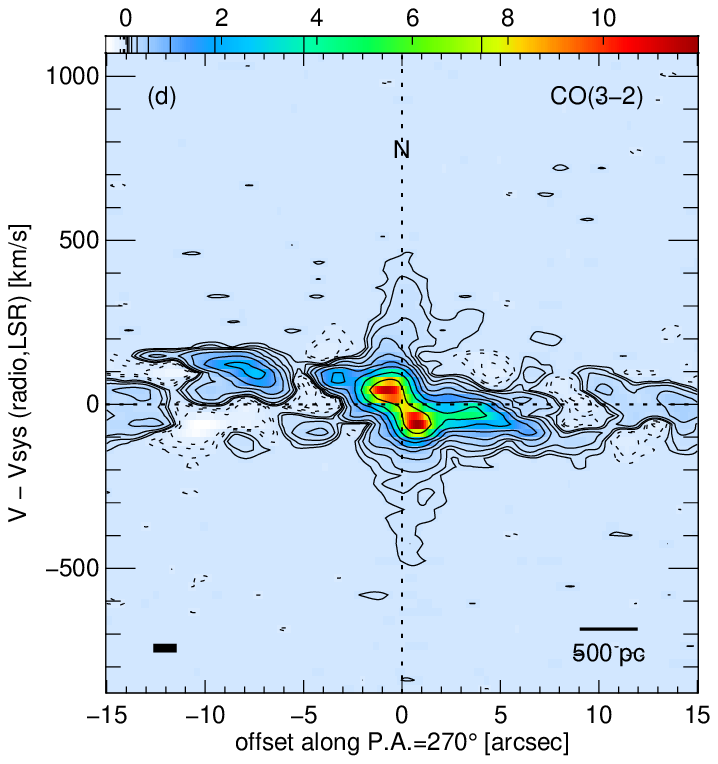} 
\plotone{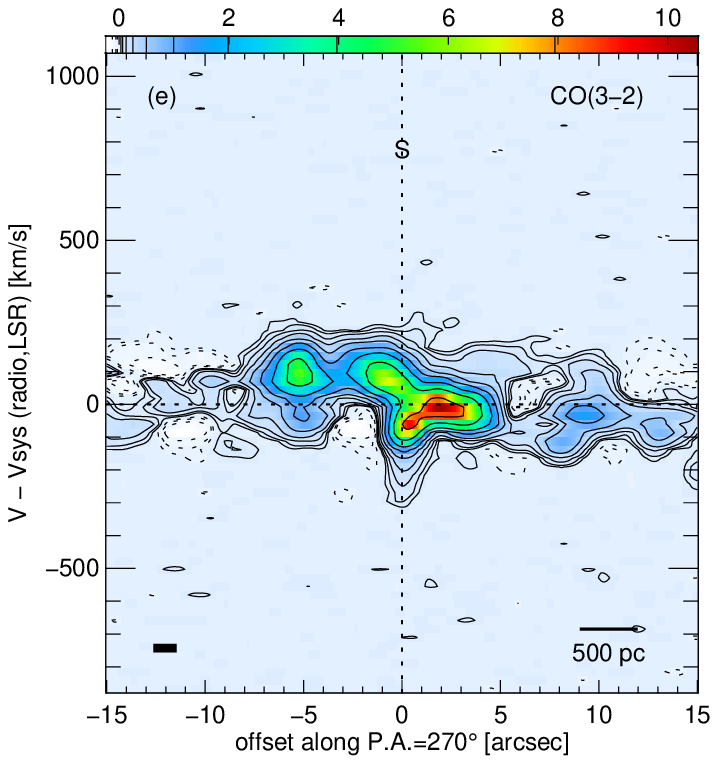} 
\plotone{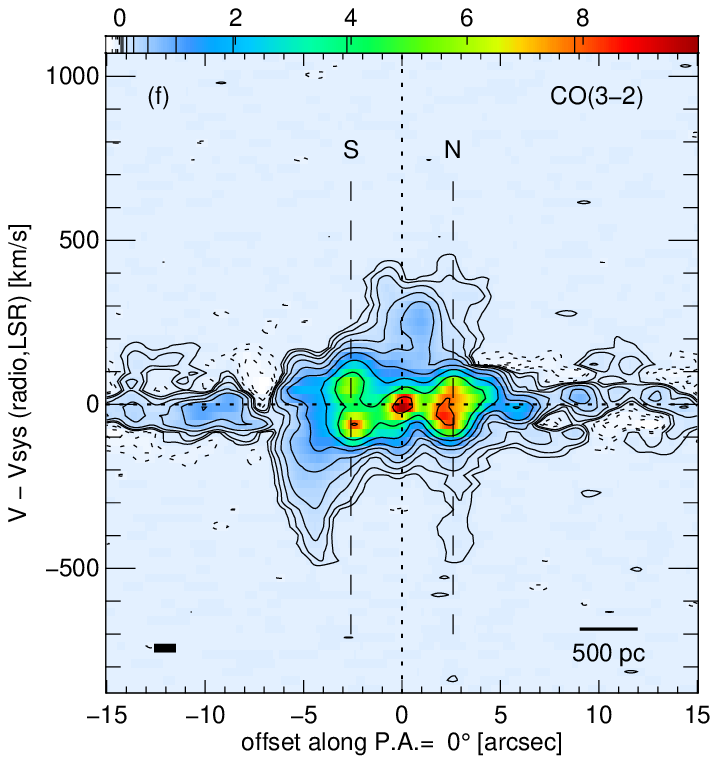} 
\caption{ \label{f.COpv}
CO  position-velocity cuts through the nuclei. 
The upper row is for CO(1--0) and lower CO(3--2).
The three columns are  
(left) p.a.=270\arcdeg\ cut through the N nucleus, 
(middle) p.a.=270\arcdeg\ cut through the S nucleus,
(right) p.a.=0\arcdeg\ cut through the midpoint of the N and S nuclei.
The locations of the two nuclei are marked with the letters N and S.
For CO(1--0), each cut is 2\farcs5 wide and the data resolution is \about 2\farcs7.
Contours are at $\pm2\times2^{n} \sigma$ $(n=0,1,2,3,\ldots)$, where $\sigma=4.6$ mK.
For CO(3--2), the slit width and the data spatial resolution are both 1\farcs1. 
Contours are at $\pm2.5\times2^{n} \sigma$ $(n=0,1,2,3,\ldots)$, where $\sigma=12$ mK.
Negative contours are dashed.
Labels of the intensity scale bars are in kelvin. 
The resolution element is shown as a black rectangle in the bottom left corner.
}
\end{figure}

\begin{figure}[t]
\epsscale{1.0}
\plotone{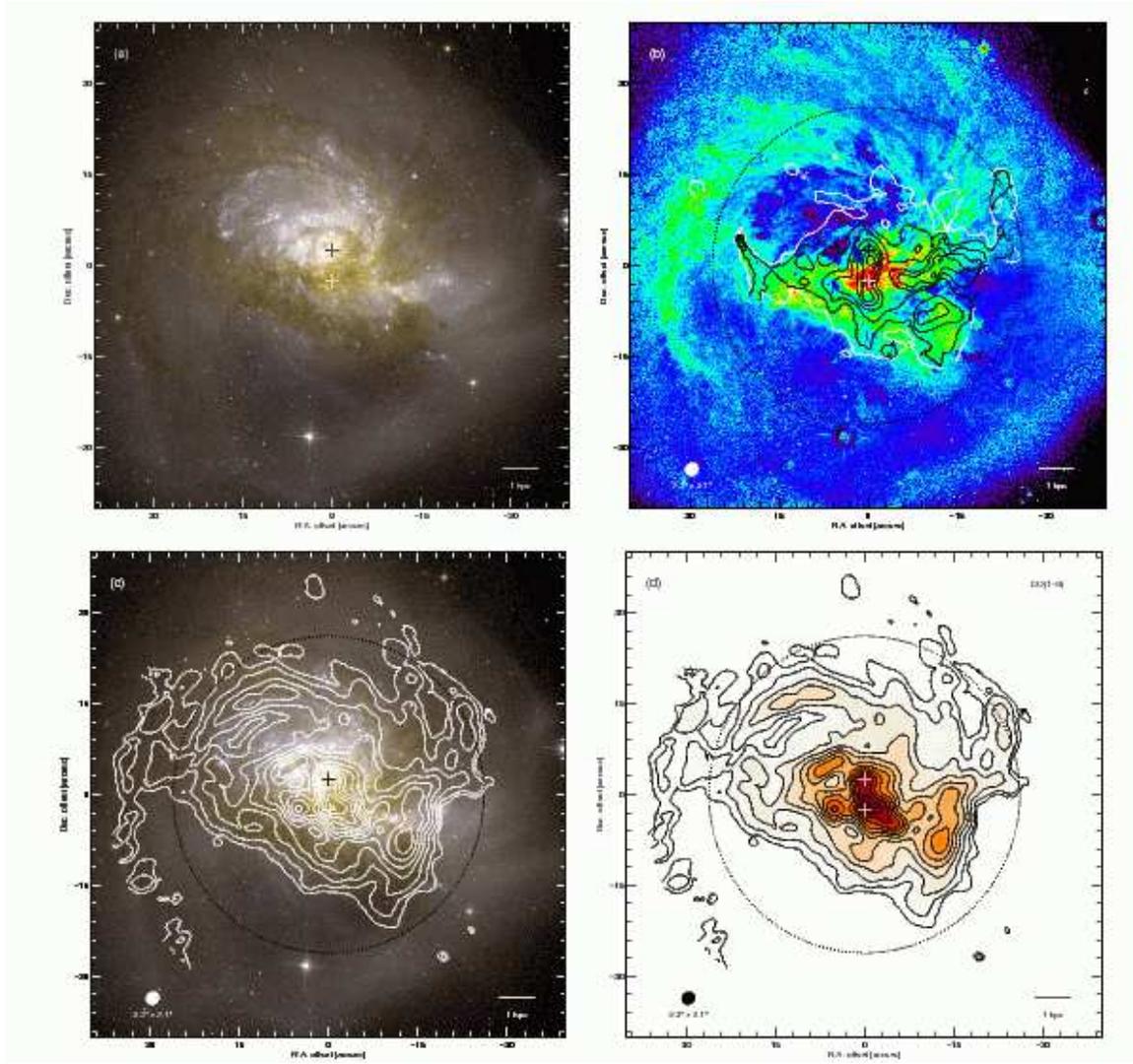}
\caption{ \label{f.hst.L}
Comparison of HST and ALMA CO(1--0) data. 
(a) A composite of F814W (\about $I$) and F435W (\about $B$).
(b) $B-I$ color index with overlaid contours of CO(1--0) line width (i.e., 2nd moment). 
Contours are in 10 \kms\ steps starting from 20 \kms\ and in white at the lowest level and black above.
(c) CO(1--0) contours on the HST composite image. 
The $n$th contours are at $0.2n^{2.5}$ \%\ of the peak integrated intensity 1340 K \kms.
(d) CO(1--0) velocity-integrated intensity with the same contours as in (c).
The two plus signs are at the radio nuclei.
The dotted circle in each panel is the 50\%\ contour of the ALMA primary beam response, for which the CO data are corrected.  
Much of CO(1--0) emission has corresponding optical dark lanes.
Regions with red optical colors generally have larger CO line widths.  
}
\end{figure}

\begin{figure}[t]
\epsscale{1.0}
\plotone{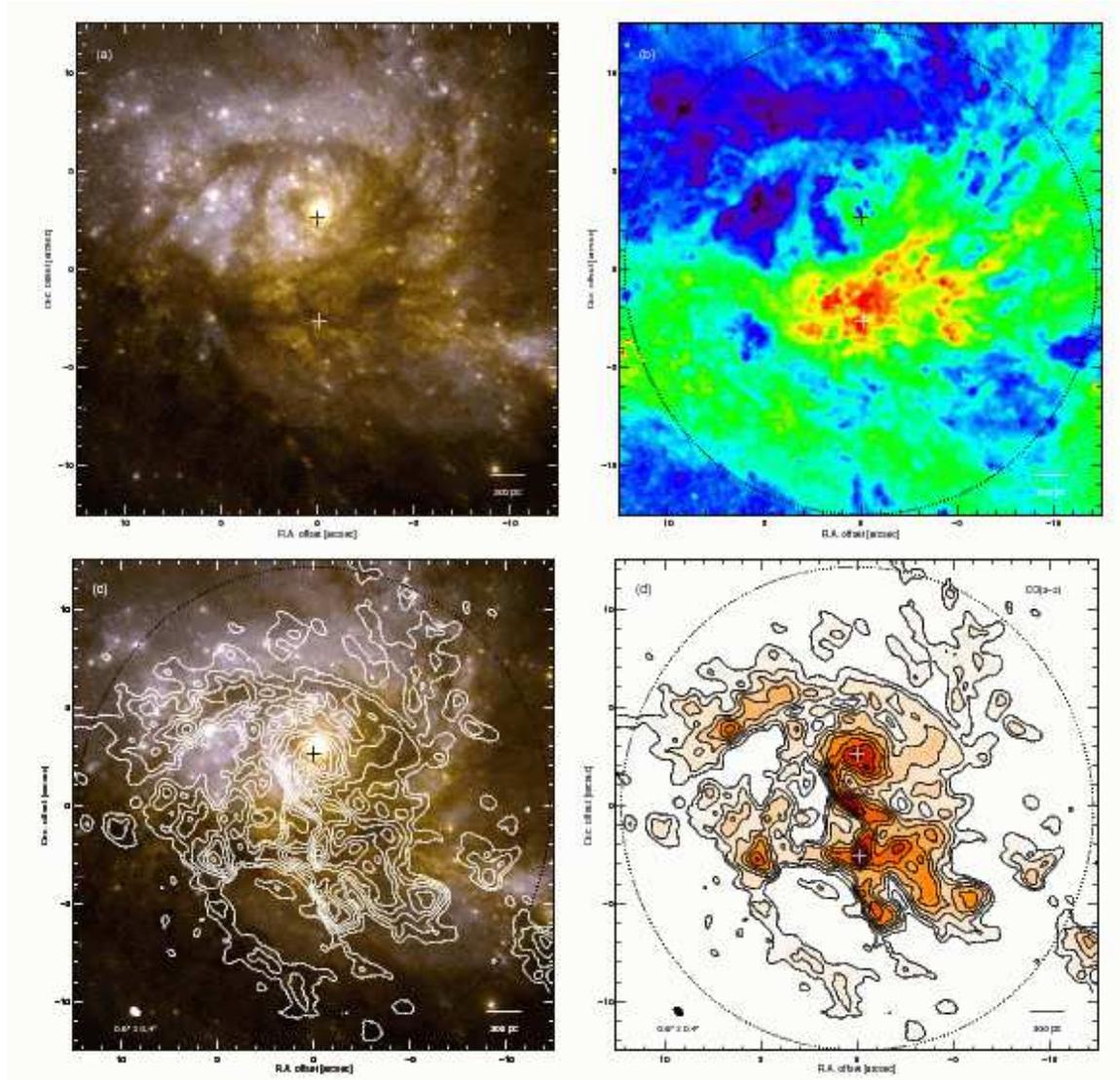}
\caption{ \label{f.hst.M}
Comparison of HST and ALMA CO(3--2) data in the central region.
(a) A composite of F814W (\about $I$) and F435W (\about $B$).
There may be a small (\about0\farcs5) astrometric offset  between the radio and optical data.
(b) $B-I$ color index.
(c) CO(3--2) contours on the HST composite image.
The $n$th contours are at $n^{2}$ \%\ of the peak integrated intensity 2730 K \kms.
(d) CO(3--2) velocity-integrated intensity with the same contours as in (c).
The two plus signs are at the radio nuclei. 
The dotted circle is the 50\%\ contour of the ALMA primary beam response, for which the CO data are corrected.  
}
\end{figure}

\begin{figure}[t]
\epsscale{0.4}
\plotone{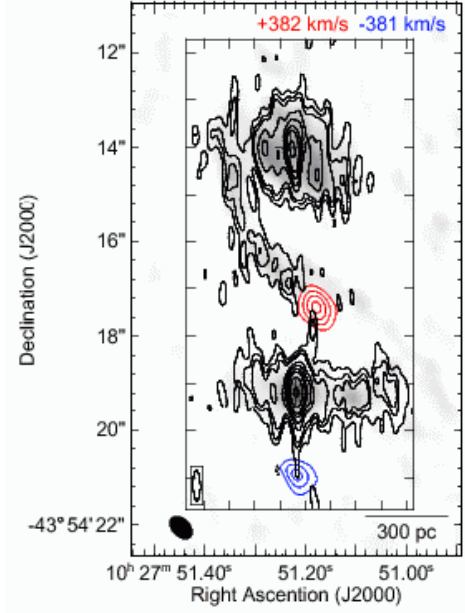} 
\caption{ \label{f.vla-almaHV}
Comparison of 3.6 cm continuum (black contours)  
with 0.86 mm continuum (gray scale) and high-velocity CO(3--2) emission (blue and red contours)
in the center of NGC 3256.
The VLA continuum image in the inset is from \citet[their Fig. 2. \copyright\ AAS. Reproduction with permission.]{Neff03} and has
a resolution of 0\farcs63 $\times$ 0\farcs15 and contours at $-0.1, 0.1, 0.15, 0.2, 0.4, 0.6, 0.9, 1.2, 2.0$ and $3.5$ mJy \perbeam\
(rms \about27 $\mu$Jy\perbeam).
Our ALMA submillimeter data are plotted in the same way as in Fig. \ref{f.HVchans.CO32.RedBlueGray} (b).
}
\end{figure}

\begin{figure}[t]
\begin{center}
\includegraphics[height=55mm]{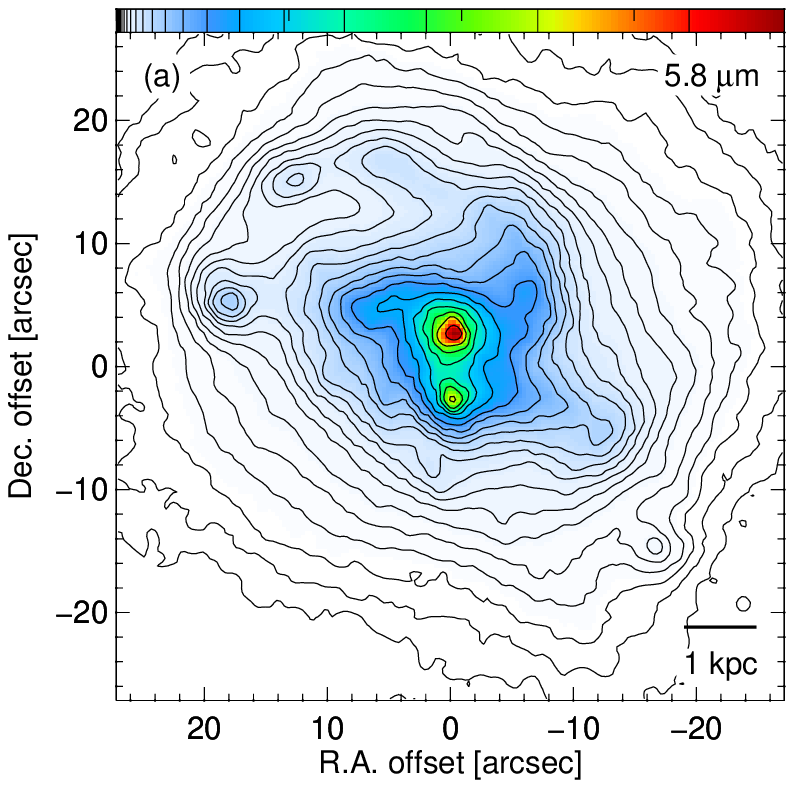} 
\includegraphics[height=55mm]{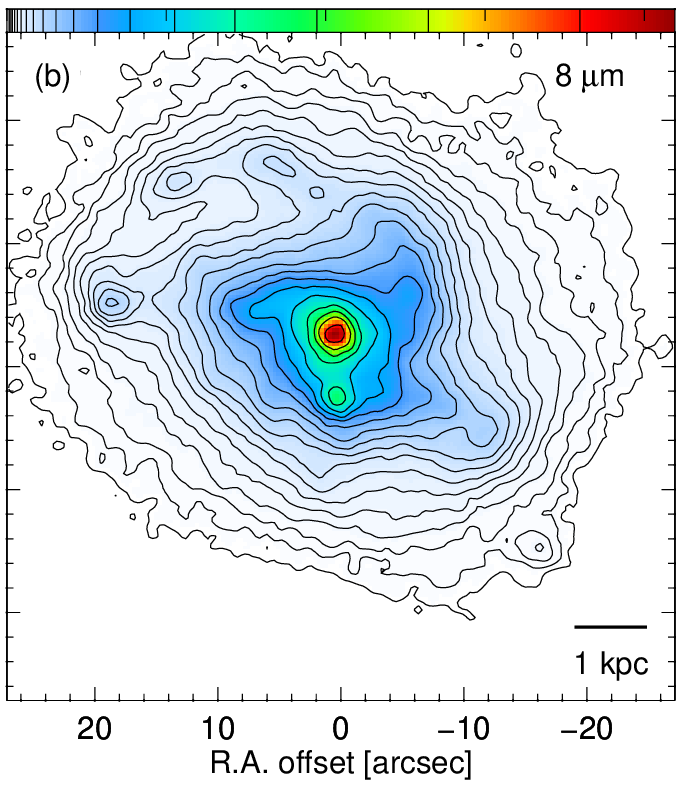} 
\end{center}
\caption{ \label{f.spitzer}
Spitzer IRAC images of the central region of NGC 3256.
Contours are in steps of \case{1}{3} mag (a factor of 1.36).
Each panel shows the same area as in Fig. \ref{f.contmaps}a for 2.8 mm continuum using the same linear intensity scale.
}
\end{figure}

\begin{figure}[t]
\begin{center}
\epsscale{0.9}
\plotone{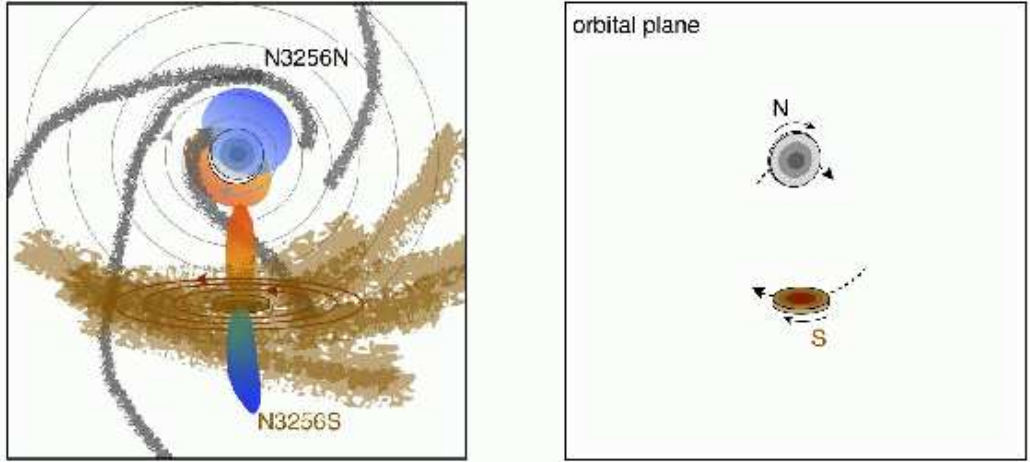} 
\end{center}
\caption{ \label{f.illust}
Illustrations of the NGC 3256 system projected to the sky plane (left) and to the orbital plane of the two nuclei (right).
The northern nucleus belongs to the merger progenitor shown in gray.
This component has a low inclination and has several spiral arms.
The southern nucleus belongs to the merger progenitor shown in brown. 
This component is close to edge-on, strongly disturbed, and is slightly foreground of the northern galaxy plane.
The orbital plane of the two nuclei is close to the sky plane but has its near side likely to the south.
Both nuclei (and progenitors) have prograde rotation with respect to the orbital motion of the two nuclei as shown in the right panel.
They have different inclinations with respect to the orbital plane and the northern nucleus is a factor of a few more massive than the southern nucleus.
Both nuclei drive their own bipolar molecular outflow. 
They are shown as blue and red lobes for redshifted and blueshifted gas, respectively.
The outflow from the northern nucleus is nearly pole-on and is wide-open. 
The southern outflow is nearly edge-on, well collimated,  
and its apparent velocity increases with distance up to about 0.3 kpc from the southern nucleus.
The blue cone of the southern outflow gradually curves toward west as it leaves the nucleus.
}
\end{figure}

The southern outflow appears highly collimated. 
It has a narrow base at the southern nuclear disk and is detected to
a projected distance of \about4\arcsec\ (0.7 kpc) from the southern nucleus
(Figs.~\ref{f.HVchans.CO32.RedBlueGray} and \ref{f.chans.CO32.br}).
Its length-to-width ratio is about 5 in Fig.~\ref{f.HVchans.CO32.RedBlueGray}.
This ratio suggests an opening angle of about 20\arcdeg\ for an edge-on cone (i.e., the flow is within $10\arcdeg$ from its central axis).
Because the outflow is well collimated along its axis to about one kpc from its origin
we can reasonably call it a bipolar molecular jet.

Looking at details, the blueshifted outflow gradually curves toward west as it goes further from the nucleus (\S \ref{s.result.highV.channelMaps}).
In our model, this is most likely due to ram pressure because the southern nucleus
is moving from west to east with respect to the northern galaxy, as inferred in \S \ref{s.configuration}.
Similar curvature is unclear in the redshifted outflow to the north in Fig. \ref{f.HVchans.CO32.RedBlueGray} 
although blueshifted emission to the north of the southern nucleus in the $-139, -113$, and $-87$ \kms\ channels 
in Fig. \ref{f.chans.CO32.br} show the expected curves.

The southern cone of the outflow is visible in the integrated intensity maps of  CO(3--2), (2--1), 
and barely in CO(1--0) in Fig. \ref{f.maps.CO102132}.
It is also hinted at in the integrated intensity map of CN(1--0, 3/2--1/2).
This feature in the integrated maps, in particular in CO(3--2), appears to have little contamination from non-outflowing ambient gas 
because the feature in the channel maps, when visible, consistently maintain its spur-like morphology.
This is expected for the almost edge-on southern nuclear disk; little non-outflowing molecular gas is expected to be at high latitudes.
The southern cone of this outflow is also visible in the 2.1 \micron\ line image 
of \HH\ 1--0 S(1) in \citet[their Fig. 2a]{Kotilainen96}.
\subsubsection{Velocity Structure}
The velocity structure of the southern outflow is noteworthy in that
the highest velocity emission at  $|V - \Vsys| \sim 400$ \kms\ is offset from the nucleus in both the blueshifted and redshifted velocities
by about 1\farcs8  (310 pc on the sky) as we noted in \S \ref{s.result.highV.channelMaps}.
About the same offsets are see at $|V - \Vsys| \sim 450$ \kms\ in CO(1--0).
This is also seen in the CO(3--2)  position-velocity diagram along ${\rm p.a.} = 0\arcdeg$ (Fig.~\ref{f.COpv} f)
in which the terminal velocity increases with distance from the nucleus until this peak.
The symmetry in the blueshifted and redshifted emission suggests this to be systematic rather than a coincidence.
The simplest model is that the molecular outflow accelerates from the nucleus to this distance.
Alternatively, it may be that only the line-of-sight velocity increases along the outflow and peaks at $d \approx 1\farcs8$.
It is possibly because of a gradual increase of the outflow opening angle, 
although at $d \approx 1\farcs8$ the high velocity emission is still compact ($< 1\arcsec$ in extent).
We regard the acceleration along the outflow to $d \approx 1\farcs8$ most likely 
but do not rule out other possible causes for the observed velocity structure.
The outflow line-of-sight velocity decreases further out and the true outflow velocity may also do so.

\subsubsection{Inclination Correction}
We estimate the most likely inclination of the southern molecular jet (and the southern nuclear disk)
to be about 80\arcdeg\ with a range of possible values between about 70\arcdeg\ and 85\arcdeg.
We already deduced in \S \ref{s.configuration.n3256s} that the southern nuclear disk is nearly edge-on; 
a conservative lower limit of the inclination is 70\arcdeg\ from the observations there. 
A sign for a larger inclination is that there is blueshifted emission at the location of the redshifted cone
and redshifted emission at the location of the blueshifted cone
(e.g., at the $-113$ and $+69$ \kms\ channels in Fig. \ref{f.chans.CO32.br}). 
The condition required to see both blueshifted and redshifted emission in a conical outflow is 
$i_{\rm S, outflow} + \theta_{\rm S, op}/2 > 90\arcdeg$ where
$i_{\rm S, outflow} $ is the inclination of the outflow axis and $\theta_{\rm S, op}$ is the full opening angle of the cone.
For the $\theta_{\rm S, op} \sim 20\arcdeg$ measured above the inclination to barely see both blue- and red-shifted emission
in both cones is  $ i_{\rm S, outflow} \gtrsim 80\arcdeg$.
On the other hand, the data do not support $ i_{\rm S, outflow} \approx 90\arcdeg$ 
because that would make both blueshifted and redshifted emission almost equally visible in each cone.
These arguments set the above-mentioned range of $i_{\rm S, outflow}$.
Its further refinement is hampered 
by uncertainties in the outflow opening angle, its spatial variation (if any), and the curvature of the outflow.

The inclination correction to the line-of-sight velocity is at least a factor of 2.9 ($= 1/\cos 70\arcdeg$) and is
5.8 and 11.4 for the $i_{\rm S, outflow}$ of 80\arcdeg\ and 85\arcdeg, respectively.
With the CO detection at least up to $\pm 450$ \kms\ along our sightline,
the maximum outflow velocity is therefore at least $\gtrsim 1000$ \kms\ even considering the
jet opening angle of about 20\arcdeg.
It is plausible, though not yet certain, that the maximum velocity is as large as 2600 \kms\
($= 450\, \kms /\cos 80\arcdeg$).
The maximum velocity is very likely larger in this southern molecular outflow than in the northern one.
The large velocity provides additional support to the description of this outflow as a molecular jet.

\subsubsection{Southern Outflow Parameters}
\label{s.outflow.southern.parameter}
The mass outflow rate from the southern nucleus is estimated to be
 $\dot{M}_{S} \approx 50 \Xtwenty $ and $25 \Xtwenty$  \Msol\ \peryr\ for
$i_{\rm S, outflow} = 80\arcdeg$ and 70\arcdeg, respectively, from the projected outflow extent of 4\arcsec,
a characteristic line-of-sight velocity of the outflow of 250 \kms,
and the gas mass estimated in \S \ref{s.result.line.hv.mass}.
The time scale for the outflow to travel 4\arcsec\ on the sky is 0.5 and 1 Myr, respectively, 
for $i_{\rm S, outflow} = 80\arcdeg$ and 70\arcdeg.
Adopting the same characteristic velocity, the kinetic luminosity of the southern outflow is on the order
of $L_{\rm kin, S} \about 9\Xtwenty \times 10^{9}  \Lsol $ (=$3\Xtwenty \times10^{36}$ W) and $1\Xtwenty \times 10^{9}  \Lsol$
for $i_{\rm S, outflow} = 80\arcdeg$ and 70\arcdeg, respectively.
This kinetic luminosity is larger than that of the northern outflow even though the northern nucleus is more luminous
in mid-infrared and presumably also in total luminosity.
It exceeds the mechanical luminosity of the southern nucleus due to supernovae and stellar winds
if $i_{\rm S, outflow} = 80\arcdeg$ and $\Xtwenty = 1$.

The gas depletion time of the southern nucleus by this outflow is only $0.6\chi_{S}$ Myr for the central 80 pc
for $i_{\rm S, outflow} = 80\arcdeg$. 
Here we used the peak gas surface density at the 80 pc resolution for the mass of gas to be depleted by the outflow.
The choice of this small size is because the base of the bipolar molecular jet appears compact.  
Again the depletion time does not depend on our choice of the CO to \HH\ conversion factor 
if the same conversion factor applies to the gas at the nucleus and in the outflow (i.e., if $\chi_{S} = 1$)
but the time scale can be ten time longer ($\chi_{S} \sim 10$) if the outflowing CO is optically thin.

The outflow rate and kinetic luminosity above are lower limits in the sense that they do not account for the mass that is in the outflow 
but has lower line-of-sight velocities than the 224 \kms\ cutoff in our flux measurement of the high-velocity gas
in \S \ref{s.result.line.hv.flux}. The gas depletion time is an upper limit for the same reason.
The omission of the low-velocity gas is more significant than for the northern outflow 
because the de-projected cutoff velocity is larger, 1300 and 650 \kms\ for  $i_{\rm S, outflow} = 80\arcdeg$ and 70\arcdeg, respectively.
The total gas mass in the outflow may well be an order of magnitude larger than the gas mass above our cutoff velocity.
In our 0\farcs6 resolution integrated intensity image in Fig. \ref{f.maps.CO102132}, 
the CO(3--2) flux in the southern cone of the southern outflow is 300 Jy \kms\ 
at distances from the southern nucleus between 1\arcsec\ and 4\arcsec\ along the outflow.
For comparison, the CO(3--2) flux at velocities above our cutoff is only 21 Jy \kms\  in this outflow cone in the same dataset.
We used the latter value for our outflow rate calculation. 
Thus the omission of low-velocity flow may cause an underestimate of the outflow rate by up to an order of magnitude.

\subsubsection{Possible Driver: Radio Jet}
\label{s.outflow.southern.driver}
The radio image in Fig. \ref{f.vla-almaHV} suggests 
that the southern molecular outflow is associated with a bipolar radio jet from the southern nucleus.
The high velocity CO(3--2) emission at $2775 \pm 380$ \kms\ are at either ends of the pair of linear radio features
that emanate to north and south from the southern nucleus, although the southern radio spur appears to go further
(see \S\ref{s.result.comparison.VLA}).
We also found that the southern cone of the molecular jet is along the southern radio spur, 
even following its westward curve.
These configurations allow a model that there is a bipolar radio jet from the southern nucleus and
the southern molecular outflow is entrained by this radio jet.
If so the apparent acceleration of molecular gas along the outflow to $d\approx 1\farcs8$ is probably due to
continuous dragging of molecular gas by the high speed plasma jet.

\subsection{Significance of the Outflows} 
Both molecular outflows are significant in the mass consumption budget of the individual nuclei
because the outflow rates are comparable to or larger than the star formation rates in the nuclei.
\citet{Lira08} estimated the star formation rates of the northern and southern nuclei to be
\about15 and \about6 \Msun\peryr, respectively, by modeling their infrared spectral energy distributions.
Our outflow rates are larger than the star formation rates at both nuclei; they are at least comparable considering
their uncertainties.
The total outflow rate, $60 X_{\rm 20, \, N\, outflow} + 50 X_{\rm 20, \,S \, outflow}$ \Msun\peryr, is also 
on the same order as the total star formation rate of NGC 3256, \about50 \Msun\peryr\
(Table \ref{t.4418param}).
This is still so when \Xtwenty\ is \about0.1 in both outflows for optically thin CO emission.  
The star formation history of the merger should be influenced by the molecular gas outflow ---
this was a conclusion of \citest{Sakamoto06} and it still holds in our new study.

Part of the outflowing molecular gas, in particular that in the southern molecular jet, will probably escape 
from their original galaxy but may not leave the merger.
The ratio of escape velocity to circular orbital velocity is 2.5--3 for extended mass distributions of galaxies \citep{Leitherer13};
it is $\sqrt{2}$ for Keplerian motion.
We estimated in \S\ref{s.configuration.Mdyn_MgasMdyn} the rotational velocities of 300 \kms\ and 200 \kms\ 
at a radius of 200 pc for the northern and southern galaxies, respectively.
Assuming a flat rotation curve in each galaxy beyond this radius, 
the ratio is 2.5 and 13, respectively, for the maximum molecular outflow velocity that we estimated
for $\gtrsim$4$\sigma$ emission in \S\ref{s.outflow.northern.parameter} and \ref{s.outflow.southern.parameter}
(i.e., 750 \kms\ for N and 2600 \kms\ for S).
The ratio is 3.3 for the \about1000 \kms\ maximum velocity obtained from the FWZI of CO(1--0) spectrum on the northern nucleus.
It is 7.5 (3.2) for our 1300 (650) \kms\ de-projected cutoff velocity used for the southern jet 
with $i_{\rm S,\, outflow} =80\arcdeg\; (70\arcdeg)$. 
A  tiny fraction of the molecular gas in the northern outflow and most of the high-velocity molecular gas in the southern
outflow are therefore above their respective escape velocities from respective galaxies.
Whether the molecular gas will escape from the merging system is a different problem and not certain.
For one thing, the escape velocity from the merger is larger than that from a constituent galaxy 
because the former is more massive.
Moreover, hydrodynamical effects on the outflowing molecular gas, already implied by the curvature of the southern molecular jet,
are likely significant and can decelerate the outflow through interaction with ambient gas in the system.

\section{Dormant? AGN in the Southern Nucleus}
\label{s.Snucleus}
The most plausible driver of the southern bipolar molecular jet is an AGN in the southern nucleus if the outflow is
entrained by a bipolar radio jet. 
It is because only AGNs are known to drive well-collimated radio jets of several 100 pc to several 100 kpc.
The contrast between the northern and southern outflows in terms of the outflow opening angle, velocity, and kinetic luminosity
also implies different driving mechanisms between them. 
Since the northern outflow is a starburst-driven superwind in all likelihood the southern outflow is left with an AGN barring
a starburst with very unusual parameters.

\subsection{Constraints on Current AGN Activities}
\label{s.Snucleus.AGNconstraints}
Despite the likely radio jet that we identified, recent searches for an  AGN in the southern nucleus as well as in NGC 3256 as a whole
have been generally negative though not unanimously so.
\citet{AlonsoHerrero12} modeled the Spitzer mid-IR spectra at \about5--38 \micron\ from the central 13\arcsec\ including both nuclei
and concluded that any AGN contribution to the bolometric luminosity of NGC 3256 is less than 1\%.
In X-rays, \citet{Lira02} detected the southern nucleus, in addition to the brighter northern nucleus, with
long (28 ks) Chandra observations but found no evidence for an AGN in either nucleus.
Their absorption-corrected X-ray luminosity for the southern nucleus in the 0.5--10 keV range was at least two orders of magnitude
below that of classical Seyfert nuclei. 
They concluded that only a low luminosity AGN comparable to that in M81 is possible.
\citet{PereiraSantaella11} analyzed 126 ks observations with XMM-Newton and concluded the absence of a luminous Compton-thick AGN.
Although they confirmed the weak 6.4 keV Fe K$\alpha$ line marginally detected by \citet{Jenkins04} 
the line equivalent width was found to be too small for a luminous AGN.
On the positive side, \citet{Neff03} found that the radio-to-X-ray ratios of both nuclei are indicative of low-luminosity AGNs.

Our ALMA observations set constraints on any hidden AGN in the southern nucleus regarding the column density
and the spatial extent of the obscuring material as well as on the AGN luminosity.
The mean absorbing column density is as high as $\log (N_{\rm H, equiv.}/{\rm cm^{-2}})  \approx 23.5$
toward the central 80 pc of the southern nucleus; here we use a half of the total column density.
Although this does not make the nucleus Compton thick this is an order of magnitude larger than the column density
that \citet{Lira02} used for absorption correction. 
The true column density toward the AGN, if any, can be much higher (or lower) than this mean value because
an AGN could be shrouded at a much smaller scale.
Regardless of the heating source behind, the very large obscuration toward the southern nucleus is consistent with
its very deep 9.7 \micron\ silicate absorption observed by \citet{MartinHernandez06} and \citet{DiazSantos10}.
The absorption index is $S_{\rm 9.7 \mu m} = \ln (f_{\rm 9.7 \mu m, obs}/f_{\rm 9.7 \mu m, cont}) < -3.0$ according to the 0\farcs36 aperture 
data in Fig. 3 of \citet{DiazSantos10}. 
This absorption index is comparable to those of Arp 220 and NGC 4418 both of which
have been suspected to host hidden Compton-thick AGNs \citep{Roche86, Spoon07}.
Their nuclei have compact and bright dusty cores with sizes of tens of parsecs, high opacities at submillimeter wavelengths,
and $\gtrsim 100$ K brightness temperatures at 860 \micron\ \citep{Sakamoto08,Sakamoto13}.
Interestingly, we did {\it not} detect such a compact and bright continuum core toward the southern (as well as northern) nucleus
nor did we detect lines from vibrationally excited molecules (Table \ref{t.linelist}) unlike toward Arp 220 and NGC 4418 \citep{Costagliola10,Sakamoto10,Martin11}.
This indicates that any Compton thick and warm absorber around an AGN in the southern nucleus must be very compact.
For example, 
a dust shroud having 860 \micron\ opacity of 0.3 (i.e., X-ray Compton opacity \about 10) and a temperature 100 K
should have the size of 6 pc (0\farcs03) so that it has the observed peak 860 \micron\ brightness temperature of 0.24 K at our 0\farcs43 resolution.
The bolometric luminosity of this core would be $2\times 10^9$ \Lsun.
It is an upper limit because 
only a part of the 860 \micron\ continuum is thermal dust emission (\S \ref{s.result-continuum}) 
and
probably only a part of  the observed 860 \micron\ dust continuum is from the central 6 pc.
The absence of a bright submillimeter core in our data is therefore consistent with 
the mid-IR estimate of the low luminosity of any AGN in NGC 3256.

\subsection{AGN Activities in Recent Past?}
\label{s.Snucleus.recent}
The molecular bipolar jet plausibly driven by an AGN combined with the absence of luminous AGN could
be explained in two ways. 
One is that the low-luminosity AGN is very efficient in driving the radio and molecular jets
and the other is that the AGN was previously active but is currently inactive possibly due to the quenching effect of the outflow.

There are indeed observations that arguably suggest a luminous AGN in NGC 3256 some $10^4$ yr ago.   
\citet{Moran99} found, in addition to sings of a several 100 \kms\ superwind,
broad \Halpha\ line emission with FWZI $\approx$ 4000--6000 \kms\ at off-center positions. 
The broad line was not detected on the northern nucleus but was detected $\gtrsim$10\arcsec\ from it in a 2\farcs5 slit along 
${\rm p.a.} = 155\arcdeg$.
No velocity shift of the broad line was found between the two sides of the nucleus.
Although the unusually large line widths and the lack of velocity shift alone could be attributed to our southern outflow, 
the locations where the broad line is detected are not in the 20\arcdeg\ opening angle of the outflow.
\citet{Moran99} determined it implausible 
that the broad line emission is reflected light of an AGN broad line region
citing the lack of a luminous AGN that can illuminate the scattering ISM several kpc away. 
However, it is possible, given our detection of high-velocity molecular jet from the southern nucleus, 
that the southern nucleus had a luminous AGN until very recently. 
If the broad line emission at least 3 kpc away from the southern nucleus is a light echo of the past activity, 
the nucleus was (much more) active  $10^4$ yr ago.
Similar variations of AGN luminosity at $10^3$--$10^5$ yr time scales have been found 
in a growing number of galaxies \citep{Keel12} and, 
in our Galaxy, the X-ray luminosity of Sgr A$^\ast$ dropped from its `high' state of the last 500 yr by 4--6 orders of magnitude within
the last 100 years \citep{Ryu13}.

A caveat for the scenario that the southern outflow was driven by a radio jet from a recently deactivated AGN is 
that AGN radio jets are not preferentially aligned with the galaxy rotation axes \citep{Kinney00,Gallimore06}
though a good alignment was recently reported for Sgr A$^\ast$ \citep{Li13}.
The southern outflow and the southern nuclear disk have apparent alignment at least in the projection onto the sky.
Unless this is another case of intrinsic galaxy-jet alignment, this probably suggests collimation by the nuclear disk.
It may be through interaction of a radio jet and the nuclear gas concentration, perhaps through which
the jet is loaded with molecular gas. 
Alternatively, the alignment might be because the outflow is not entrained by a radio jet but driven by some other mechanisms
including a compact starburst, AGN, and their combination where the nuclear disk works as a collimator.
Star formation in the southern nucleus is active, 
having one third of the star formation in the northern nucleus \citep{Lira08}
fueled by the high surface-density gas of 
$\Sigma_{\rm mol}({\rm S}) = 6\times 10^3 \Xtwenty$ \Msol \persquarepc. 
It is therefore reasonable to expect some contribution of star formation to the southern outflow.
If the southern outflow is driven mainly by starburst then the kinetic luminosity of the outflow must be much lower
than that calculated from our fiducial conversion factor and outflow inclination; 
these parameters must be lower than we assumed.
On the whole we regard that AGN jet-driven outflow is more plausible than others to be the main mechanism 
for the southern molecular jet. Some help from starburst is certain.
This model is, however, not yet proven and needs further studies for verification and 
to determine the true driving mechanism(s).

\section{Discussion and Conclusions}
\label{s.conclusions}

We have reported our ALMA and SMA observations of molecular line and continuum emission in the center of NGC 3256.
We constrained the configuration of the two merger nuclei and their nuclear molecular disks much better than before
and resolved for the first time the high-velocity molecular gas in the merger into two molecular outflows from the two nuclei.

We have suggested the southern molecular outflow from NGC 3256S to be driven by an AGN bipolar jet.
If confirmed, it joins a small group of outflows that share the same driving mechanism and have been imaged in molecular line(s).
They include the molecular outflows in M51 \citep{Matsushita04}, NGC 1266 \citep{Alatalo11}, and NGC 1433 \citep{Combes13}.
Compared with these outflows, the bipolar molecular jet of NGC 3256S is better collimated and more energetic
for a common \Xco. 
This may be because the AGN radio `jets' in the other galaxies are wider radio plumes.
Mainly because of the large outflow velocity, the kinetic luminosity of the southern outflow approaches
that of local ultraluminous infrared galaxies and quasar hosts observed by \citet{Cicone14}, who obtained 
outflow kinetic luminosities on the orders of $10^{36}$--$10^{37}$ W with a conversion factor 3 times lower than ours.
The large maximum velocity of the southern outflow is also comparable to or larger than those in their survey 
but this is probably because ours is helped much by the high ALMA sensitivity and the proximity of NGC 3256.

The overall significance of AGN-driven, jet-entrained molecular outflows is an open question.
AGN time variability similar to the one we suggested for NGC 3256S may reduce the apparent AGN contribution to 
galactic molecular outflows.
Regarding jet-entrained outflows, on one hand, radio jets have been found only in minority of AGNs. 
For instance, \citet{Ho01} found ``linear'' structures of radio continuum in 14/52 = 27\% of optically selected, nearby Seyfert galaxies.
On the other hand, the parameters of our southern outflow imply that a jet-entrained outflow can be more powerful and efficient
than other outflows when normalized by the source bolometric luminosity.
It is possible therefore that the small number and/or short lifetime of the outflows driven by AGN radio jets are offset to some extent
by their efficiencies and luminosities. 
The two molecular outflows in NGC 3256 are excellent targets for such assessment
because we can simultaneously study properties and driving mechanisms of two powerful molecular outflows of different natures.

Our observations have added two similarities between NGC 3256 and Arp 220
in addition to both being late stage mergers with large infrared luminosities.
One is the presence of outflows from both of the two merger nuclei; for Arp 220
blueshifted molecular line absorption indicative of outflow has been detected toward both nuclei
\citep{Sakamoto09}.
The other is that the two merger nuclei with less than 1 kpc projected separation still retain
their nuclear gas disks with misaligned rotational axes; for Arp 220 this was first imaged by \citet{Sakamoto99}.
Our submillimeter observations also revealed a clear difference between the two mergers.
Namely, the nuclei of NGC 3256 are less obscured than the Arp 220 nuclei in terms of
gas and dust column density averaged at 100 pc scale.
This is most clearly seen in the submillimeter continuum emission whose opacity due to dust is almost unity at 860 \micron\
toward the nuclei of Arp 220 but  about two orders of magnitude lower toward the nuclei of NGC 3256.
In order for NGC 3256 to evolve into Arp 220, therefore, significant gas accretion is needed to the nuclei 
despite the ongoing strong molecular outflows that would deplete the gas in the nuclei in Myrs.
Such evolution may indeed occur because Arp 220 is probably more advanced as a merger than NGC 3256 judging from their nuclear separations.
NGC 3256 may become more luminous in that process, perhaps as luminous as Arp 220, because
there is a statistical trend for larger nuclear obscuration (i.e., more gas funneling to the nuclei) and
larger total luminosities in more advanced mergers \citep{Haan11,Stierwalt13}.
Further studies on NGC 3256 are warranted also for the purpose of tracing the late evolutionary path of a merger 
that is plausibly about to become ultraluminous.

Finally we reemphasize our caution on \Xco\ in particular for the high-velocity molecular outflows.
The large line widths of the outflow gas reduce the CO column density per line width and hence
may well result in optically thin CO emission. 
The conversion factor for that case is $\Xtwenty \sim 0.1$.
Such a low conversion factor for optically-thin CO has been adopted, for example, for the molecular outflow in NGC 1266
on the basis of multi-line CO excitation analysis \citep{Alatalo11}.
The outflows in NGC 3256 may have a similar situation and \Xco.
Alternatively, the outflowing gas may consist of an ensemble of optically-thick (in CO) clouds that spread in a wide velocity range.
In its partial support is our detection of CN(1--0) lines, with likely enhancement relative to CO(1--0), 
in the high-velocity gas (\S\ref{s.result.line.hv.spectra}).
Although CN may be subthermally excited, the detection of a line with a $10^6$ \percubiccm\ critical density 
implies gas clumping for dense gas to exist in the high velocity outflows.
Even if individual clumps are not virialized as assumed for the standard \Xco, the conversion factor for optically thick
clumps will be larger than that for optically thin CO (and lower that that for virialized CO-thick clouds).
Similar clumping and presence of dense gas in a galactic molecular outflow have been deduced for Mrk 231
by \citet{Aalto12} from their detection of broad line wings in HCN, \HCOplus, and HNC lines.
Because most outflow parameters in Table~\ref{t.4418measured.param} depend on \Xtwenty,
followup studies on the physical and chemical properties of the high-velocity gas are highly desired.

%
Our primary findings are: 

1. Each of the two merger nuclei has its own nuclear disk where molecular line and continuum emission peak.
The northern nuclear disk is nearly face-on ($i$ \about30\arcdeg), 
has a \about200 pc characteristic radius,
and clearly rotates around the northern nucleus.
The southern nucleus has  a more compact emission peak and a linear structure extending \about200 pc on either side.
It is deduced to be a nearly edge-on nuclear disk rotating around the southern nucleus.
The mean molecular gas surface densities of both nuclei is about 3$X_{20} \times10^4$ \Msol\persquarepc\
at 240 pc resolution, where $X_{20}$ is the CO-to-\HH\ conversion factor in the unit of $10^{20}$ \unitofX.
The peak gas surface density is $6 X_{20} \times10^4$ \Msol\persquarepc\ at the southern nucleus at 80 pc resolution.

2. The high velocity molecular gas previously found at the center of the merger is resolved to two molecular outflows
associated with the two nuclei.
We detected not only CO but also CN lines with enhancement in these outflows.
The CN detection in a galactic outflow is for the first time to our knowledge.
The total molecular outflow rate of the two outflows is on the same order of the total star formation rate in NGC 3256.

3. The molecular outflow from the northern nuclear disk is a bipolar flow with a wide opening angle
and  a nearly pole-on viewing angle. 
It has de-projected outflow velocities up to 750 \kms\ at $\gtrsim$4$\sigma$ 
and an outflow time scale (crossing time) of 1 Myr.
Its molecular gas mass is $6 X_{20} \times10^7$ \Msol,
mass outflow rate $60 X_{20}$ \Msol\peryr, 
and kinetic luminosity on the order of $4 X_{20} \times 10^8$ \Lsun.
The last three are for the gas at de-projected velocities above 260 \kms.
At the current rate the outflow would deplete molecular gas in the northern nuclear disk in 3 Myr 
if the same conversion factor applies to the nuclear disk and the outflow.
Most of the outflow/superwind signatures found so far at other wavelengths in NGC 3256 
must be from this outflow.

4. The molecular outflow from NGC 3256S is a well collimated bipolar jet
with a \about$20\arcdeg$ opening angle and is nearly edge on. 
It has a de-projected maximum velocity 2600 \kms\ for a favored inclination angle 80\arcdeg\ 
or 1300 \kms\ for $i=70\arcdeg$.
The line-of-sight outflow velocity increases with distance up to 300 pc from the nucleus.
This molecular jet has a 0.5 Myr crossing time,
a mass of $2.5 X_{20} \times10^7$ \Msol, 
a mass outflow rate $50 X_{20}$ \Msol\peryr, and a kinetic luminosity
on the order of $90 X_{20} \times 10^8$ \Lsun\ for $i=80\arcdeg$. 
These are for gas at projected velocities above 220 \kms\ and the lower velocity gas in the outflow
may be an order of magnitude larger in mass.
The gas depletion time for the central 80 pc is \about0.6 Myr
under the same assumption about the conversion factor as above and ignoring the lower velocity flow.

5. The northern outflow is a starburst driven superwind in all likelihood.
The southern outflow is most likely entrained by a radio jet from a weak or recently dimmed AGN in the southern nucleus.
Pieces of evidence for the latter outflow driver are the large differences in the outflow parameters from the northern superwind, 
off-nuclear broad \Halpha\ lines in NGC 3256, 
and a pair of radio spurs from the southern nucleus that matches in shape the southern molecular bipolar jet.

6. Continuum spectral indexes are negative at 3 mm and positive at 0.86 mm for both nuclei.
The index is lower, in particular at 0.86 mm, for the southern nucleus, suggesting
significant synchrotron and/or free-free emission even at 860 \micron.
Neither nucleus has a bright ($\Tb > 10$ K) dust continuum core of several 10 pc size at 860 \micron\ 
such as those found in Arp 220 and NGC 4418.
This disfavors presence of a highly Compton-thick and currently luminous AGN in the nuclei of NGC 3256.

The new observations presented in this paper contain more information than we could fit in a single paper. 
Further analysis will be reported elsewhere.

\vspace{5mm}
\acknowledgements
We are grateful to the people who worked or supported to make ALMA a reality. 
We also thank the ALMA and SMA staff who carried out our observing runs or made the data assessments.
This paper made use of the following ALMA data: ADS/JAO.ALMA\#2011.0.00002.SV and ADS/JAO.ALMA\#2011.0.00525.S. 
ALMA is a partnership of ESO (representing its member states), NSF (USA) and NINS (Japan), 
together with NRC (Canada) and NSC and ASIAA (Taiwan), in cooperation with the Republic of Chile. 
The Joint ALMA Observatory is operated by ESO, AUI/NRAO and NAOJ.
This paper also uses observations made with the Submillimeter Array, which is a joint project 
between the Smithsonian Astrophysical Observatory and the
Academia Sinica Institute of Astronomy and Astrophysics, and is
funded by the Smithsonian Institution and the Academia Sinica.
This research is also partly based on observations made with the NASA/ESA Hubble Space Telescope, 
and obtained from the Hubble Legacy Archive, 
which is a collaboration between the Space Telescope Science Institute (STScI/NASA), 
the Space Telescope European Coordinating Facility (ST-ECF/ESA) 
and the Canadian Astronomy Data Centre (CADC/NRC/CSA).
This research has made use of the NASA/ IPAC Infrared Science Archive, which is operated by the Jet Propulsion Laboratory, 
California Institute of Technology, under contract with the National Aeronautics and Space Administration.
The authors also made use of the NASA/IPAC Extragalactic Database (NED),
NASA's Astrophysics Data System (ADS),
and
the splatalogue database for astronomical spectroscopy.
KS was supported by the Taiwanese NSC grants 99-2112-M-001-011-MY3 and 102-2119-M-001-011-MY3.

{\it Facilities:} \facility{ALMA, SMA, HST, Spitzer}

\clearpage

\clearpage
\begin{deluxetable}{lcc}
\tablewidth{0pt}
\tablecaption{NGC 3256 parameters  \label{t.4418param} }
\tablehead{
	\colhead{Parameter}  &
	\colhead{Value}  &
	\colhead{note}
}
\startdata  
R.A. (J2000)         & 10\hr27\mn51\fs23       & (1) \\ 
Dec. (J2000)       & \minus43\arcdeg54\arcmin16\farcs6 & (1) \\
$V_{\rm sys}$ [\kms] &  2775 &  (2) \\
$D$ [Mpc] & 35 & (3) \\
Scale. 1\arcsec\ in pc & 170  \\
$L_{\rm 8-1000\mu m}$ [\Lsol] & $10^{11.56}$ & (4) \\
SFR	[\Msun\peryr]	&  50		& (5) \\	
P.A.  [\arcdeg] & \about70 & (6) 
\enddata
\tablecomments{
(1)
The middle point of the two radio nuclei in \citet{Neff03}.
We used this as the center for the ALMA Cycle 0 and SMA observations and for all figures 
in offset coordinates.
The ALMA pointing position for the CSV observations was 
R.A.=10\hr27\mn51\fs60 Dec.=\minus43\arcdeg54\arcmin18\farcs0.
(2) Systemic velocity of the merger in radio-definition with respect to the LSR \citesp{Sakamoto06}.
(3) Adopted galaxy distance \citep{Sanders03}.
(4) From the IRAS flux data \citep{Sanders03}.
(5) Star formation rate calculated from the bolometric luminosity using the calibration of \citet{Murphy11} 
and assuming no AGN contribution. 
(6) The major-axis position angle of the bulk molecular gas motion in the central  \about6 kpc
\citesp{Sakamoto06}.
}
\end{deluxetable}

\begin{deluxetable}{clcccccccc}
\tabletypesize{\scriptsize}
\tablewidth{0pt}
\tablecaption{Log of ALMA observations  \label{t.obslog} }
\tablehead{ 
         \colhead{ID.} &
	\colhead{UT date}  &
	\colhead{config.} &
	\colhead{$N_{\rm ant}$} &	
	\colhead{BL range} &		
	\colhead{$\langle T_{\rm sys}\rangle$} &
	\colhead{$S_{\rm gain}$} &
	\colhead{$\alpha_{\rm gain}$} &
	\colhead{$\alpha_{\rm bp}$} &		
	\colhead{$T_{\rm gal}$} 
	\\
	\colhead{ }  &	
	\colhead{ }  &
	\colhead{ } &
	\colhead{} &
	\colhead{[m]} &	
	\colhead{[K]} &
	\colhead{[Jy]} &
	\colhead{ } &	
	\colhead{ } &		
	\colhead{[min]} 
	\\
	\colhead{(1)}  &
	\colhead{(2)}  &
	\colhead{(3)} &
	\colhead{(4)} &
	\colhead{(5)} &
	\colhead{(6)} &
	\colhead{(7)} &
	\colhead{(8)} &	
	\colhead{(9)} &		
	\colhead{(10)} 	
}
\startdata
Band 3: \\
B3-c1 & 2011-04-16 & CSV  &  7  & 12--90  &  66 & 1.94 & $-0.25$ & $-0.25$ &  90 \\  
B3-c2 & 2011-04-17 & CSV  &  8  & 14--89  &  61 & 1.94 & $-0.25$ & $-0.25$ &  90 \\ 
B3-1 & 2011-12-29 & COM & 14  & 16--259 & 74 & 1.35 & $-0.50$ & $-0.34$ & 31 \\
B3-2 & 2011-12-30 & COM & 13  & 15--253 & 72 & 1.35 & $-0.50$ & $-0.34$ & 31 \\
B3-3 & 2012-01-27 & COM & 17  & 17--269  & 72 & 1.33 & $-0.67$ & $-0.336$ &34  \\
B3-4 & 2012-01-27 & COM & 17  &  17--264  & 69 & 1.33 & $-0.67$ & $-0.336$ & 34  \\ 
B3-5 & 2012-03-27 & EXT   & 14  & 20--374  & 64 & 1.49 & $-0.56$ & $-0.43$ & 34  \\ 
B3-6 & 2012-07-28 & EXT  & 23  &  16--364  & 78 & 1.46 & $-0.68$ & $-0.67$ & 29  \\ 
Band 7: \\
B7-1 & 2012-01-24 & COM & 16  & 15--256 & 181 & 0.54 & $-1.03$ & $-0.61$ & 30   \\      
B7-2 & 2012-05-21 & EXT  & 14  &  16--360 & 186 & 0.75 & $-1.18$ & $-0.67$ & 30  \\ 
B7-3 & 2012-06-04 & EXT  & 20  &  16--360  & 149 & 0.67 & $-0.89$ & $-0.86$ & 30
\enddata
\tablecomments{
(2) date of observations. 
(3) ALMA array configuration. Cycle 0 had two configurations, COM=compact and EXT=extended.
(4) Number of useable antennas after flagging bad data. Some of them are partly flagged.
(5) The range of projected baselines toward NGC 3256.
(6) Median single-side-band system temperature toward NGC 3256.
(7) and (8) Flux density and spectral index of the gain calibrator, 
J1037\minus295 in the CSV observations and 
J1107\minus448 in our Cycle 0 observations. 
The flux densities in Band 3 are the ones at 100 GHz and those in Band 7 are at 348 GHz.
Primary flux calibrators were Mars or Titan.
(9) Spectral index of the bandpass calibrator, 
J1037\minus295 and 3C279 in the CSV and Cycle 0 observations, respectively.
(10) Total integration time on NGC 3256.
The calibration parameters (7)--(9) for the data B3-c2 are taken from the measurements in B3-c1.
Those for B3-1 and B3-2 as well as B3-3 and B3-4 are respectively from combined fitting of the two
calibration datasets. 
}
\end{deluxetable}

\begin{deluxetable}{lcccc}
\tablewidth{0pt}
\tablecaption{Frequency Coverage  \label{t.freqCoverage} }
\tablehead{ 
         \colhead{obs.} &
	\colhead{band} &
	\colhead{SB} &	
	\colhead{$f$(LSRK)} &
	\colhead{\frest}
	\\
	\colhead{ }  &	
	\colhead{} &
	\colhead{} &
	\colhead{GHz} &
	\colhead{GHz} 		
	\\
	\colhead{(1)}  &
	\colhead{(2)}  &
	\colhead{(3)} &
	\colhead{(4)} &
	\colhead{(5)} 
}
\startdata
Cycle 0 & B7 & U &  352.220--355.619 & 355.511--358.941 \\
Cycle 0 & B7 & L &  340.226--343.629 & 343.405--346.839 \\
CSV & B3 & U &  111.693--115.110 & 112.737--116.185 \\
Cycle 0 & B3 & U &  111.598--114.962 & 112.641--116.036 \\
CSV & B3 & L &  99.622--102.823 & 100.553--103.784 \\
Cycle 0 & B3 & L & 99.606--102.910 & 100.537--103.871 
\enddata
\tablecomments{
(3) U= upper sideband, L=lower sideband.
(4) The range of LSRK frequencies covered in all executions for a source in the direction of NGC 3256.
The CSV data has a \about0.1 GHz gap in the middle of USB. 
The execution B7-2 in Table \ref{t.obslog} does not have the upper half of the USB. 
(5) The frequency coverage in the rest frame of NGC 3256 at $V({\rm radio, LSRK})= 2775 \kms$.
}
\end{deluxetable}

\begin{deluxetable}{llrlccccc}
\tablewidth{0pt} 
\tablecolumns{9}
\tablecaption{Lines Imaged toward NGC 3256  \label{t.linelist} }
\tablehead{ 
         \colhead{species, transition} &
	\colhead{\frest} &
	\colhead{$\Eu/k$} &	
	\colhead{note}
	\\
	\colhead{ }  &	
	\colhead{[GHz]} &
	\colhead{[K]} &
	\colhead{} 	
	\\
	\colhead{(1)}  &
	\colhead{(2)}  &
	\colhead{(3)} &
	\colhead{(4)} 
}
\startdata
\HCOplus(J=4--3) & 356.7342 & 42.8 & B7 USB \\ 
CO(J=3--2)             & 345.7960 & 33.2 & B7 LSB \\  
CO(J=2-1) & 230.5380 & 16.6 & SMA USB \\
\thirteenCO(J=2--1) & 220.3987 & 15.9 & SMA LSB \\
CO(J=1--0)             & 115.2712 & 5.5 & B3 USB \\
CN(N=1--0; J=3/2--1/2) & 113.4949 & 5.4 & B3 USB \\ 
CN(N=1--0; J=1/2--1/2) & 113.1688 & 5.4 & B3 USB  \\ 
\propyne($J_K$=$6_0$--$5_0$) & 102.5480 & 17.2 & B3 LSB \\   
\sidehead{non-detections: } 
\HCOplus($v_2=1,J=4$--$3,l=1f$)   & 358.2424 & 1236.7 & B7 USB \\
HCN($v_2=1,J=4$--$3,l=1f$)   & 356.2556 & 1067.1 & B7 USB 
\enddata
\tablecomments{
The data are taken from splatalogue. 
(2) Rest frequency.  
For CN the \frest\ frequencies are mean values of five and four transitions (of almost identical \Eu), 
respectively, for the one group at around 113.4949 GHz and the other at around 113.1688 GHz.
The ranges of \frest\ in each group are 32 and 68 MHz, respectively, for the first and the second group.
\propyne\ (propyne or methyl acetylene) is a symmetric top molecule 
and has several transitions around this frequency with different \Eu\ for different $K$. 
Listed above is the one with $K=0$ involving the lowest \Eu. 
Intensity-weighted mean frequency can be several MHz lower than this for excitation temperatures on the order of 50 K. 
(3) The upper level energy divided by the Boltzmann constant.
}
\end{deluxetable}

\begin{deluxetable}{ccccclclrl}
\tablewidth{0pt}
\tablecaption{Continuum Data Properties  \label{t.data_cont_properties} }
\tablehead{ 
         \colhead{Source} &
         \colhead{Band} &
	\colhead{SB} &
	\colhead{wt} &
	\colhead{beam} &	
	\colhead{rms} & 
 	\colhead{max} &
	\colhead{rms} &
 	\colhead{max} &
	\colhead{Fig.}
	\\
	\colhead{ }  &	
	\colhead{ }  &	
	\colhead{ }  &	
	\colhead{ }  &		
	\colhead{[\arcsec, \arcsec]} &
	\multicolumn{2}{c}{[mJy \perbeam]} &
	\multicolumn{2}{c}{[mK]} &
	\colhead{}
	\\
	\colhead{(1)}  &
	\colhead{(2)}  &
	\colhead{(3)} &
	\colhead{(4)} &
	\colhead{(5)}  &
	\colhead{(6)}  &
	\colhead{(7)} &
	\colhead{(8)} & 
	\colhead{(9)} &
	\colhead{(10)}
}
\startdata
Cyc0 & B7 & D & br & $0.52\times0.36$ & 0.16 & 4.42 & 8.6 & 237 & \ref{f.contmaps}c, \ref{f.HVchans.CO32.RedBlueGray}b, \ref{f.vla-almaHV} \\ 
Cyc0 & B7 & D & na  & $0.65\times0.49$ & 0.13 &  5.80  &  4.1 &   183 & \ref{f.contmaps}b \\
Cyc0 & B7  & U & na  &  $0.63\times0.48$ & 0.17 &  5.76  & 5.5 &  185 & \\
Cyc0 & B7  & L & na  & $0.67\times0.50$ &  0.15  &  5.71  & 4.7 &  179 & \\
Cyc0 & B7  & D & tp  &  $1.19\times1.11$ & 0.24 & 14.21 &  1.8 & 108 & \ref{f.HVchans.CO32.RedBlueGray}a \\ 
 CSV\plus Cyc0 & B3 & D & na & $2.77\times2.36$ &   0.017 &  5.09 & 0.28 &   84 & \ref{f.contmaps}a \\
 CSV\plus Cyc0 & B3 & U & na &  $2.63\times2.40$ &  0.037 & 5.00  & 0.56 &    75 & \\  
 CSV\plus Cyc0 & B3 & L & na  & $2.81\times2.35$  &  0.018 & 5.13  & 0.32 &    91 &
\enddata
\tablecomments{
(1) Cyc0 = ALMA Cycle 0, CSV\plus Cyc0 = ALMA CSV and Cycle 0 combined.
(3) sideband. D=DSB, U=USB, L=LSB.
(4) visibility weighting for imaging.   br= Briggs with {\tt robust = 0}, na = natural, tp = tapered.
(5) Size of the synthesized beam in FWHM.
(6)--(7) rms noise and maximum in the continuum image in the unit of mJy \perbeam. 
These are measured before correction for the primary-beam attenuation.
(8)--(9) The same rms noise and maximum in the unit of milli-kelvin.
}
\end{deluxetable}

\begin{deluxetable}{cccccrrrrc}
\tabletypesize{\scriptsize}
\tablewidth{0pt}
\tablecaption{Line Data Properties  \label{t.data_line_properties} }
\tablehead{ 
         \colhead{Source} &
         \colhead{line} &
	\colhead{wt} &
	\colhead{$\delta V$} &
	\colhead{beam} &	
	\colhead{rms} &
	\colhead{max} & 
 	\colhead{rms} &
	\colhead{max } &
	\colhead{Fig.}
	\\
	\colhead{ }  &	
	\colhead{ }  &	
	\colhead{ }  &	
	\colhead{\kms}  &		
	\colhead{[\arcsec $\times$\arcsec]} &
	\multicolumn{2}{c}{[mJy \perbeam]} &
	\colhead{[mK]}  &	
	\colhead{[K]}  &
	\colhead{}
	\\
	\colhead{(1)}  &
	\colhead{(2)}  &
	\colhead{(3)} &
	\colhead{(4)} &
	\colhead{(5)}  &
	\colhead{(6)}  &
	\colhead{(7)} &
	\colhead{(8)} & 
	\colhead{(9)} &
	\colhead{(10)}
}
\startdata
Cyc0 & \HCOplus(4--3) & na & 25.2 &  $0.65\times0.49$   &  1.53  &  42.9  & 47 & 1.3 &  \ref{f.chans.HCOp43.na30MHz}, \ref{f.spectra.nuclei} \\ 
Cyc0 & \HCOplus(4--3) & na &  \phn8.4  &  $0.65\times0.49$  &  2.55 &  46.5  & 78 &   1.4  & \ref{f.maps.nonCO}  \\ 
Cyc0 & CO(3--2) & br & 26.0 & $0.58\times0.39$ &   1.92  & 470.9 & 88 &  21.7  &   \ref{f.chans.CO32.br} \\  
Cyc0 & CO(3--2) & na &  26.0 & $0.68\times0.50$  &  1.20 & 624.2& 37 & 19.0  & \ref{f.spectra.nuclei}  \\ 
Cyc0 & CO(3--2) & tp &  26.0 &  $1.20\times1.10$ & 1.67   & 1725.7& 13 &  13.6  &    \ref{f.COpv} \\   
Cyc0 & CO(3--2) & br &  \phn8.7 & $0.58\times0.39$ &   3.31 & 479.2 & 152 & 22.1  & \ref{f.maps.CO102132}, \ref{f.maps.CO32nuc}, \ref{f.HVchans.CO32.RedBlueGray}, \ref{f.hst.M}, \ref{f.vla-almaHV} \\ 
Cyc0 & CO(3--2) & tp &   \phn8.7 &   $1.20\times1.10$ &   2.48 & 1777.3 & 20 &  14.0  &    \ref{f.HVchans.CO32.RedBlueGray}, \ref{f.spec.noblend} \\ 
SMA & CO(2--1)                & na & 10.0 & $1.03\times0.56$ & 13.0\phn & 480.0  & 526 & 19.4  & \ref{f.maps.CO102132}  \\
SMA & \thirteenCO(2--1) & tp & 20.0 & $2.14\times1.69$ & 14.1\phn & 138.5  & 100 & 0.98  & \ref{f.maps.nonCO}  \\                
Cyc0 & CO(1--0) & br &   10.4 &  $1.61\times1.24$ &   1.46 & 321.2 & 69 & 15.1  & \ref{f.maps.CO102132}  \\ 
Cyc0 & CO(1--0) & na &  10.4 & $2.32\times2.06$  &  0.96 & 589.8  & 19 & 11.6  & \ref{f.hst.L}   \\ 
Cyc0 & CN(1--0, \case{3}{2}--\case{1}{2} ) & br & 26.4 & $1.69\times1.28$ & 0.72 & 26.6 & 32 & 1.19 & \ref{f.maps.nonCO}, \ref{f.chans.CN10.br10MHz} \\ 
Cyc0 & CN(1--0, \case{1}{2}--\case{1}{2} ) & br & 26.5 & $1.69\times1.28$ & 0.72 & 12.5 & 33 & 0.56 & \ref{f.maps.nonCO}, \ref{f.chans.CN10.br10MHz} \\ 
Cyc0 & \propyne(6--5) & na &  29.2 &  $2.62\times2.14$ &   0.34 &   3.7 &  7.2 &  0.078  &  \ref{f.maps.nonCO}, \ref{f.chans.CH3C2H.Na10MHz} \\ 
CSV\plus Cyc0 & CO(1--0) & br & 40.6 & $1.69\times1.31$ &  0.69 &  314.9 &  29 & 13.3 & \ref{f.chans.CO10.br} \\ 
CSV\plus Cyc0 & CO(1--0) & na & 40.6 & $2.92\times2.57$ &  0.43 & 736.6 &  5.4 & 9.2  &   \ref{f.HV.CO10.RedBlues}, \ref{f.COpv} \\ 
CSV\plus Cyc0 & CO(1--0) & tp & 40.6 & $5.78\times5.17$ &   0.52 & 1732.2 &  1.6 &  5.4  & \ref{f.HVchans.CO10.tp.VHV}, \ref{f.HV.CO10.RedBlues}, \ref{f.spectra.nuclei}  \\ 
CSV\plus Cyc0 & CN(1--0, \case{3}{2}--\case{1}{2}) & tp  & 41.3 &  $5.90\times5.16$  & 0.41 &   87.4  & 1.3 &   0.28  &  \ref{f.HVchans.CN10.tp.VHV}, \ref{f.spectra.nuclei}   
\enddata
\tablecomments{
(1) Cyc0 = ALMA Cycle 0, CSV\plus Cyc0 = ALMA CSV and Cycle 0 combined.
(2) Emission line. See Table \ref{t.linelist} for more line information.
(3) visibility weighting for imaging. br= Briggs with {\tt robust = 0}, na = natural, tp = tapered.
(4) Velocity resolution for the line.
(5) Size of the synthesized beam in FWHM.
(6)--(7) rms noise and maximum in the channel maps in the unit of mJy \perbeam. 
These are measured before correction for the primary-beam attenuation.
(8)--(9) The same rms noise and maximum in brightness temperature.
}
\end{deluxetable}

\begin{deluxetable}{cccccccccc}
\tablewidth{0pt}
\tablecaption{Continuum Flux Densities and Spectral Indexes  \label{t.contFluxSpix} }
\tablehead{ 
         \colhead{band} &
	\colhead{SB} &
	\colhead{$\nu_{\rm mean}$} &	
	\colhead{$S_\nu(20\arcsec)$} &
	\colhead{$S_\nu$(N)} &
	\colhead{$S_\nu$(S)} &
	\colhead{$\theta$} &
	\colhead{$\alpha(20\arcsec)$} &
	\colhead{$\alpha$(N)} &
	\colhead{$\alpha$(S)} 	
	\\
	\colhead{ }  &	
	\colhead{} &
	\colhead{GHz} &
	\colhead{mJy} &
	\colhead{mJy} &
	\colhead{mJy} &
	\colhead{\arcsec} &
	\colhead{} &
	\colhead{} &
	\colhead{} 				
	\\
	\colhead{(1)}  &
	\colhead{(2)}  &
	\colhead{(3)} &
	\colhead{(4)} &
	\colhead{(5)}  &
	\colhead{(6)}  &
	\colhead{(7)}  &
	\colhead{(8)}  &
	\colhead{(9)}  &
	\colhead{(10)}			 
}
\startdata
B7 & U & 354.236 & 122.2 & 13.2 & 9.6 & 1 & $+3.00$ & $+3.72$ & $+1.56$ \\
B7 & L & 341.203  & 109.2 & 11.5 & 9.1 & 1 & \nd & \nd & \nd  \\
B3 & U & 113.313 &  \phn28.4 & 5.1 & 5.0 & 2.7 & $-0.10$ & $-0.12$ & $-0.19$ \\
B3 & L &  101.222 & \phn28.8  & 5.2 & 5.1 & 2.7 & \nd & \nd & \nd   
\enddata
\tablecomments{
(2) U= upper sideband, L=lower sideband.
(3) Mean frequency of the single side band continuum.
(4) Flux density of the continuum emission integrated over a 20\arcsec\ diameter aperture centered at
the midpoint of the two nuclei. Only the Cycle 0 data are used. 
(5) and (6) Flux densities at the radio positions of the northern and southern nuclei, respectively, within
the Gaussian beams whose FWHM sizes are in (7).
Data in columns (4), (5), and (6) are corrected for the primary beam responses but not for any missing flux.
More digits are shown than allowed for their absolute accuracies because the ratio between USB and LSB
are free from the common sources of error for the two sidebands.
(8), (9), and (10)  Spectral indexes for the 20\arcsec\ aperture, northern nucleus, and southern nucleus
calculated between the USB and LSB of each receiver band.
}
\end{deluxetable}

\clearpage
\begin{deluxetable}{lcc}
\tablewidth{0pt} 
\tablecolumns{3}
\tablecaption{NGC 3256 Measured and Estimated Parameters  \label{t.4418measured.param} }
\tablehead{
	\colhead{Parameter}  &
	\colhead{Value}  &
	\colhead{note}
}
\startdata
\sidehead{northern galaxy/nucleus: }
inclination  [\arcdeg] $i_{\rm N}$         & 30      & (1) \\ 
major axis p.a.  [\arcdeg]  & 75   (90) & (2) \\
$\max \Sigma_{\rm mol}({\rm N, 240\,pc})$ [\Msol \persquarepc] & $4\Xtwenty \times 10^3$ & (3) \\
$\Mmol({\rm HV, N})$ [\Msol] & $6\Xtwenty \times10^{7} $  & (4) \\
$\max(v_{\rm outflow, N}) $ [\kms] & 750 (1000) & (5) \\
$\langle v_{\rm outflow, N} \rangle$ [\kms] & 300 & (6) \\
$l_{\rm outflow, N} $ [pc] & $1\farcs8/\sin i_{\rm N} = 820$ & (7) \\
$t_{\rm outflow, N}$ [Myr]  & 1.1 & (8) \\
$\dot{M}_{\rm N}$ [ \Msol\ \peryr ] &  $ 60\Xtwenty$ & (9) \\
$L_{\rm kin, N}$ [\Lsol] & $ 4\Xtwenty \times 10^{8} $ & (10) \\
$t_{\rm dip, N}$(300 pc) [Myr]  & $3\chi_{N}$ & (11) \\
\sidehead{southern galaxy/nucleus: }
inclination   $i_{\rm S}$ [\arcdeg]    & 80    & (1) \\ 
major axis p.a. [\arcdeg]                   & 90    & (2) \\
$\max \Sigma_{\rm mol}({\rm S, 240\,pc})$ [\Msol \persquarepc] & $3\Xtwenty \times 10^3$ & (3) \\
$\max \Sigma_{\rm mol}({\rm S, 80\,pc})$ [\Msol \persquarepc] & $6\Xtwenty \times 10^3$ & (3) \\
$\Mmol({\rm HV, S})$ [\Msol] & $2.5\Xtwenty \times10^{7}$  & (4) \\
$\theta_{\rm S, op}$  [\arcdeg] & $20$ & (12) \\
$\max(v_{\rm outflow, S})$ [\kms]  & $2600 c_{80}$ & (5) \\
$\langle v_{\rm outflow, S} \rangle$ [\kms]  & $1400 c_{80}$ & (6) \\
$l_{\rm outflow, S}$ [pc]  & $4\arcsec/\sin i_{\rm S} = 690 s_{80}$  & (7) \\
$t_{\rm outflow, S}$ [Myr]  & $0.5 t_{80}$ & (8) \\
$\dot{M}_{\rm S}$ [\Msol\ \peryr] & $50\Xtwenty t_{80}^{-1} $   & (9) \\
$L_{\rm kin, S}$ [\Lsol] & $ 9\Xtwenty c_{80}^2 t_{80}^{-1} \times 10^{9}  $ & (10) \\
$t_{\rm dip, S}$(80 pc) [Myr]  & $0.6 \chi_{S} t_{80}$ & (11) \\
\sidehead{merger system: }
$V_{\rm sys}$ [\kms] &  2775 &  (13) \\
$\Mdyn({\rm N})/\Mdyn({\rm S})$ & $2.5 \pm 1$  & (14) \\
$\Mmol(r \leq 10\arcsec)$ [\Msol] & $7\Xtwenty \times10^{9}$  & (15) \\
$\Mmol(r \leq 20\arcsec)$ [\Msol] & $1\Xtwenty \times10^{10}$  & (15) 
\enddata
\tablecomments{
 We define $\Xtwenty \equiv X_{\rm CO}/(10^{20}\, \unitofX)$, which is the ratio of the true \Xco\ for the gas in consideration to 
 our fiducial conversion factor.
 The parameter $\chi$ is $\Xco({\rm nucleus})/\Xco({\rm outflow})$, the ratio of \Xco\ between the nuclear disk and
 the outflow in consideration.
 We also define, for the southern nucleus, the ratio of trigonometric values for our most favored inclination
 to those for the true inclination of the southern nucleus.
 Namely, 
 $s_{80} \equiv \sin 80\arcdeg/\sin i_{\rm S}$, 
 $c_{80} \equiv \cos 80\arcdeg/\cos i_{\rm S}$,
 and
 $t_{80} \equiv \tan 80\arcdeg/\tan i_{\rm S}$.
 For $i_{\rm S} = $70\arcdeg\ and 85\arcdeg, $t_{80}$ is 2.1 and 0.5, respectively.
(1) Inclination of the nuclear disk.
(2) Major axis position angle of the nuclear disk. 
The number in the parenthesis for the northern galaxy is the one for the very center of the nuclear disk.
(3) Peak molecular gas surface density of the nucleus at the given resolution.
(4) The mass of high-velocity molecular gas associated with the nucleus. 
This only includes the gas whose line-of-sight velocity is more than 224 \kms\ offset from the systemic velocity.
(5) The maximum outflow velocity after the correction for inclination.
For the northern outflow this corresponds to the $> 3\sigma$ CO(1--0) emission in Fig.~\ref{f.HVchans.CO10.tp.VHV} and the number in parenthesis
is from the CO(1--0) FWZI in Fig.~\ref{f.spectra.nuclei}.
For the southern outflow this corresponds to the $> 4\sigma$ CO(3--2) emission in Fig.~\ref{f.HVchans.CO32.RedBlueGray}a.
(6) Characteristic outflow velocity that we use to calculate kinematical luminosity.
(7) Extent of the outflow on each side of the nucleus. This is corrected for the outflow inclination.
(8) Outflow timescale, i.e., travel time.
(9) Molecular mass outflow rate.
(10) Kinetic luminosity of the molecular outflow.
(11) Timescale for the outflow to deplete to deplete molecular gas from the nuclear region of the given diameter. 
(12) Opening angle of the molecular outflow.
(13) The systemic velocity of the merger. We also use this as the fiducial velocity for individual nuclei.
(14) The ratio of the dynamical masses in the central kpc of the two nuclei.
(15) Mass of molecular gas within the indicated radius on the sky from the midpoint of the two nuclei.
}
\end{deluxetable}


\clearpage

\clearpage

\clearpage

\clearpage

\clearpage

\clearpage

\clearpage

\clearpage

\clearpage

\clearpage

\clearpage

\clearpage

\clearpage

\clearpage

\end{document}